\documentclass{aa}  
\usepackage{subfigure}
\usepackage{graphicx}
\usepackage{txfonts}
\usepackage{color}
\usepackage[none]{hyphenat}

\usepackage[]{hyperref}
\hypersetup{colorlinks=true,citecolor=blue,  linkcolor=blue,urlcolor=black}

\def\Lsun{L_\odot}

\def\Mp{M_{\rm p}}
\def\Rp{R_{\rm p}}
\def\Sp{S_{\rm p}}

\def\Ls{L_{\star}}

\def\Mearth{M_\oplus}
\def\Rearth{R_\oplus}
\def\Searth{S_\oplus}

\newcommand{\eq}[1]{Eq.\,(\ref{#1})}

\newcommand{\fig}[1]{Fig.\,\ref{#1}}

\newcommand{\sect}[1]{Sect.\,\ref{#1}}

\begin{document}

   \title{The habitability of Proxima Centauri~b}
   \subtitle{II. Possible climates and observability}

   \author{Martin Turbet\inst{1}
          \and
          J\'er\'emy Leconte \inst{2}
          \and
          Franck Selsis\inst{2}
          \and
          Emeline Bolmont\inst{3}
          \and
          Francois Forget\inst{1}
          \and
          Ignasi Ribas\inst{4}
          \and
          Sean N. Raymond \inst{2}
          \and
          Guillem Anglada-Escud\'e\inst{5}
          }
          
   \institute{Laboratoire de M\'et\'eorologie Dynamique, Sorbonne Universit\'es, UPMC Univ Paris 06, CNRS, 4 place Jussieu, 75005 Paris, France; \email{mturbet@lmd.jussieu.fr}
         \and
Laboratoire d'astrophysique de Bordeaux, Univ. Bordeaux, CNRS, B18N, all\'ee Geoffroy Saint-Hilaire, 33615 Pessac, France
         \and
NaXys, Department of Mathematics, University of Namur, 8 Rempart de la Vierge,
5000 Namur, Belgium 
\and 
Institut de Ci\`encies de l'Espai (IEEC-CSIC), C/Can Magrans, 
s/n, Campus UAB, 08193 Bellaterra, Spain
\and
School of Physics and Astronomy, Queen Mary University of London, 327 Mile End 
Rd, London E1 4NS, UK}

   \date{Received; accepted}

  \abstract
  {
Radial velocity monitoring has found the signature of a $M \sin i = 1.3$~M$_\oplus$
planet located within 
the Habitable Zone (HZ) of Proxima Centauri \citep{Anglada16}. 
Despite a hotter past and an active host star the planet Proxima~b could have
retained enough 
volatiles to sustain surface habitability \citep{Ribas2016}. 
Here we use a 3D Global Climate Model (GCM) to simulate Proxima b's
atmosphere 
and water cycle for its two likely rotation modes (1:1 and 3:2 spin-orbit resonances) 
while varying the unconstrained surface water inventory and atmospheric greenhouse
effect. \\
Any low-obliquity low-eccentricity planet within the HZ of 
its star should be in one of the climate regimes discussed here. We find that a
broad range of atmospheric compositions allow surface liquid water. 
On a tidally-locked planet with sufficient surface water inventory, liquid water is
always present, at least in the substellar region. 
With a non-synchronous rotation, this requires a minimum greenhouse warming
($\sim$10~mbar of CO$_2$ and 1~bar of N$_2$).
 If the planet is dryer, $\sim$0.5~bar/1.5~bars of CO$_2$ (respectively for
asynchronous/synchronous rotation) 
suffice to prevent the trapping of any arbitrary small water inventory into
polar/nightside ice caps. 
 \\
  We produce reflection/emission spectra and phase curves for the simulated climates.
We find that atmospheric characterization will be possible by direct imaging with
forthcoming large telescopes. The angular 
separation of $7 \lambda/D$  at 1~$\mu$m (with the E-ELT) and a contrast of
$\sim$10$^{-7}$ will enable high-resolution spectroscopy and the search for
molecular signatures, including H$_2$O, O$_2$, and CO$_2$. \\
  The observation of thermal phase curves can be attempted with JWST, thanks to a
contrast 
of $2\times10^{-5}$ at 10~$\mu$m. Proxima~b will also be an exceptional target for
future IR interferometers.
  Within a decade it will be possible to image Proxima~b and possibly determine
whether this exoplanet's surface is habitable.
  }

\keywords{Stars: individual: Proxima Cen --- Planets and satellites:individual: Proxima b --- Planets and satellites: atmospheres --- Planets and satellites: terrestrial planets --- Planets and satellites: detection --- Astrobiology}

   \maketitle
%

\section{Introduction}

Proxima Centauri~b, recently discovered by \cite{Anglada16}, is not only the closest known extrasolar planet but also the closest potentially habitable terrestrial world, located at only $\sim$~4.2 light years from the Earth  \citep{leeuwen2007}.

Proxima Centauri~b, also called Proxima~b, receives a stellar flux of $\sim 950$~W~m$^{-2}$ \citep[$0.65-0.7~\Searth$ at 0.05~AU based on the bolometric luminosity from][]{Demory2009,Boyajian2012} that places it undoubtedly well within the so called Habitable Zone (HZ) of its host star ($M_{\star}$~=~0.123~M$_{\odot}$), defined as the range of
orbital distances within which a planet can possibly maintain liquid water on its surface \citep{Kasting1993,Kopparapu2013,Leco:13nat,Yang2013,Kopparapu2014,Kopparapu2016}. Indeed, for the effective temperature of Proxima \citep[3050~K,][]{Anglada16} climate models locate the inner edge between 0.9 and 1.5~$\Searth$ depending on the planet rotation \citep{Kopparapu2016} and the outer edge at $\sim 0.2~\Searth$ \citep{Kopparapu2013}. Nonetheless, surface habitability requires the planet to be endowed with a sufficient amount of water and atmospheric gases able to maintain a surface pressure and possibly a greenhouse effect (typically with CO$_2$).

Quantifying this last statement is the main goal of this study.
While most previous studies on climate and habitability focused on estimating the edges of the Habitable Zone, here we rather investigate the variety of necessary atmospheric compositions and global water content to ensure surface liquid water.  Using the limited amount of information available on Proxima~b, we can already provide some constraints on its possible climate regimes as a function of a key parameter: the volatile inventory---which includes the amount of available water above the surface as well as the amount and type of greenhouse/background gases in the atmosphere. Investigating extreme inventory scenarios is especially important in our specific case because Proxima is an active M dwarf. This means that the planet's atmospheric content has probably been dramatically influenced by various types of escape, especially during the pre-main-sequence phase where the planet underwent a runaway greenhouse phase. See the companion paper by \citet{Ribas2016} for a detailed discussion.

Assuming a circular orbit, Proxima~b should be in synchronous rotation with permanent dayside and nightside (1:1 resonance). However, \citet{Ribas2016} showed that the orbit of Proxima~b might not have had time to circularize and that an eccentricity above $\sim$0.06 would be sufficient to capture the planet into a 3:2 spin-orbit resonance similar to Mercury. At higher eccentricities, higher resonances such as the 2:1 become possible as well. The climate on a tidally locked (synchronous) planet can dramatically differ from the asynchronous case. For a given volatile inventory, we will thus systematically try to infer the difference in behavior between a planet in a 1:1 and 3:2 resonance. The choice of the specific resonance order, however, has a much more subtle impact on the climate, so that the investigations of higher order resonances will be left for further studies.

Guided by various works on previously observed terrestrial exoplanets \citep{Word:11ajl,Pierrehumbert2011,Leco:13},
this study thus explores the climate regimes available for Proxima~b as a function of its spin state, atmospheric composition and thickness, and total amount of water available in the system. For this purpose, we use the LMD Generic Global Climate Model whose implementation for this specific study is detailed in \sect{lmd_gcm}.

\begin{figure*}
\centering
\includegraphics[scale=0.23]{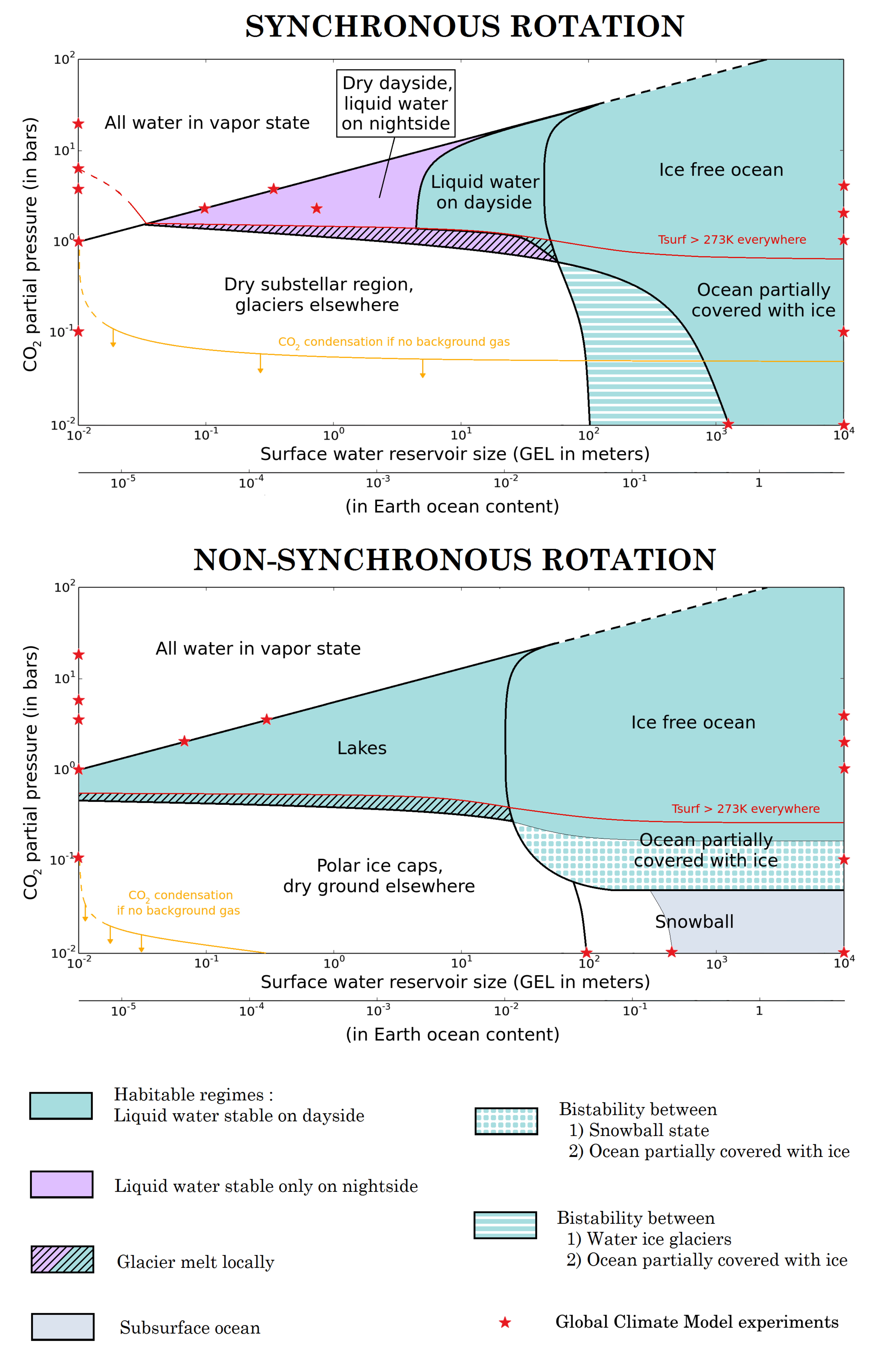}
\caption{Schematic diagrams of the possible climate regimes reached as function of the CO$_2$ atmospheric content (in bar) and the H$_2$O content available at the surface (in Global Equivalent Layer -- GEL, in meters). The top and bottom panels describe respectively the case of synchronous and asynchronous spin states. The red stars indicate the parts of the diagram that have been probed in this work using GCM simulations. The stars that lie on top of the 
left y-axis correspond to the case of a dry surface and atmosphere while those on the right y-axis consider planets completely covered by water. We note that the presence of a background gas like N$_2$ could modify significantly the lower part of the diagram (pCO$_2$~$<$~1~bar typically). It would favor the heat redistribution, which in turn could (1) prevent the CO$_2$ atmospheric collapse and (2) reduce the amount of ice possibly trapped in water ice glaciers.}
\label{proxima_diagrams}%
\end{figure*}

For further reference,  Figure~\ref{proxima_diagrams} summarizes this attempt to quantify the possible climates of Proxima~b for the two most likely spin states (1:1 and 3:2 spin-orbit resonance), as a function of the total water inventory\footnote{The total water inventory is expressed in Global Equivalent Layer (GEL) which is the globally averaged depth of the layer that would result from putting all the available water in the system at the surface in a liquid phase. } and the greenhouse gas content (CO$_2$ here). It will serve as a guide throughout the various sections of this work.
 
 To add a twist, Proxima~b---probably being our closest neighbor---should be amenable to further characterization by direct imaging in the near future. With its short orbital period, multi-epoch imaging could then rapidly yield a visible and NIR phase curve of the planet. It could be one of our first opportunity at characterizing a temperate terrestrial planet and its climate. We thus put a particular emphasis on to quantifying observable signatures for the various type of atmospheres discussed here.

 After presenting the details about the physical parameterizations used to model Proxima~b, sections \ref{dry_section} to \ref{large_water_reservoir} contain our major findings about the climate regimes achievable on Proxima~b. They are ordered following the global water inventory available from completely dry (\sect{dry_section}), water limited planets (\sect{small-res}) to water-rich worlds (\sect{large_water_reservoir}). Finally, in \sect{observability}, we highlight potential observable signatures of these various climate regimes, and discuss how direct imaging with upcoming facilities could help us to constrain the actual climate of Proxima~b.

\section{Method - The LMD Generic Global Climate Model}
\label{lmd_gcm}

This model originally derives from the LMDz 3-dimensions Earth Global Climate Model \citep{Hour:06}, which solves the primitive equations of geophysical fluid dynamics using a finite difference dynamical core on an Arakawa C grid. The same model has been used to study very diverse atmospheres of terrestrial planets, ranging from: 
\begin{enumerate}
\item Low irradiated planets as Early Mars \citep{Forg:13,Word:13,Word:15,Turb:16sub1}, Archean Earth \citep{Char:13}, Snowball Earth-like planets \citep{Turb:16sub2} or exoplanets like Gliese~581d \citep{Word:11ajl};
\item Planets receiving stellar flux similar to the Earth (\citealt{Bolm:16aa}, this paper);
\item Highly irradiated planets such as future Earth \citep{Leco:13nat} or tidally locked exoplanets like Gliese~581c~/~HD85512b \citep{Leco:13}.
\end{enumerate}

Our simulations were designed to represent the characteristics of Proxima~b, which includes the stellar flux it receives (956~W~m$^{-2}$~/~$0.7~\Searth$), its radius (7160~km~/~1.1~R$_\oplus$) and gravity field (10.9~m~s$^{-2}$) calculated assuming a mass of $1.4~\Mearth$ \citep{Anglada16} and the density of Earth (5500~kg~m$^{-3}$), a flat topography, and various rotation speeds, namely 6.3$\cdot$10$^{-6}$, 9.7$\cdot$10$^{-6}$, and 1.3$\cdot$10$^{-5}$~rad~s$^{-1}$, for 1:1 3:2, and 2:1 orbital resonances, respectively. 
All the simulations were performed assuming a circular orbit. Even if the maximum possible eccentricity of Proxima~b is 0.35 \citep{Anglada16}, for dynamical reasons \citep{Ribas2016}, the upper limit of 0.1 would be more realistic. Therefore, the mean flux approximation seems here reasonable \citep{Bolm:16aa}.
We also worked with an obliquity of 0$^{\circ}$, as expected for such a planet influenced by gravitational tides (see \citealt{Ribas2016} for more details).

\begin{table*}
\centering
\caption{Adopted stellar and planetary characteristics of the Proxima system. We also show the adopted parameters for various GCM parametrizations.}
\begin{tabular}{lcc}
\hline
Parameter & Value & Unit \\
\hline
$\Ls$    & 0.0017 &$\Lsun$  \\
$T_{\rm eff}$           & 3000  & K    \\
Age                  & 4.8  &Gyr     \\
\hline
$\Mp \sin i$   & 1.27    & $\Mearth$\\
$\Mp$           & 1.4    & $\Mearth$ \\
$\Rp$   & 1.1  & $\Rearth$     \\
Semi-major axis   & 0.0485 &AU      \\
$\Sp$     & 0.7  &$\Searth$     \\
Spin-orbit resonance                  & 1:1~/~3:2~/~2:1     \\
$\Omega_p$  & 6.3$\cdot$10$^{-6}$~/~9.7$\cdot$10$^{-6}$~/~1.3$\cdot$10$^{-5}$  &rad~s$^{-1}$ \\
Stellar Day          &  $\infty $~/~22.4~/~11.2 &Earth days \\
Obliquity   & 0 & $^{\circ}$\\
Eccentricity            & 0 \\
\hline
Surface types                &  Rock~/~Liquid water~/~Ice   &  \\
Thermal inertia &  1000~/~20000~/~2000  &J~m$^{-2}$~K$^{-1}$~s$^{-\frac{1}{2}}$  \\
Albedo          & 0.2~/~0.07~/~wavelength-dependant (Figure~\ref{spectrum_proxima}) & \\
\hline
\end{tabular} 
\tablefoot{The values for the stellar and planetary parameters are derived from \citet{Anglada16}.}
\label{table_param} 
\end{table*}

The simulations presented in this paper were all carried out at a horizontal resolution of 64~$\times$~48 (e.g., 5.6~$^{\circ}~\times$~3.8$^{\circ}$) in longitude~$\times$~latitude. In the vertical direction, the model is composed of 26 distinct atmospheric layers that were built using hybrid $\sigma$ coordinates and 18 soil layers. These 18 layers are designed to represent either a rocky ground (thermal inertia I$_{\text{rock}}$~=~1000~J~m$^{-2}$~K$^{-1}$~s$^{-\frac{1}{2}}$), 
an icy ground (I$_{\text{ice}}$~=~2000~J~m$^{-2}$~K$^{-1}$~s$^{-\frac{1}{2}}$) or an ocean (I$_{\text{ocean}}$~=~20000~J~m$^{-2}$~K$^{-1}$~s$^{-\frac{1}{2}}$ to take into account the vertical mixing) depending on the assumed surface. Oceanic heat transport is not included in this study. Each of theses configurations is able to capture the diurnal waves for the non-synchronous orbital configurations. The planet day maximum explored duration is 22.4 Earth days for the 3:2 resonance orbital configuration.
Table~\ref{table_param} summarizes all the parameterizations adopted in this work.

\begin{figure}
\begin{center}
\includegraphics[width=\linewidth]{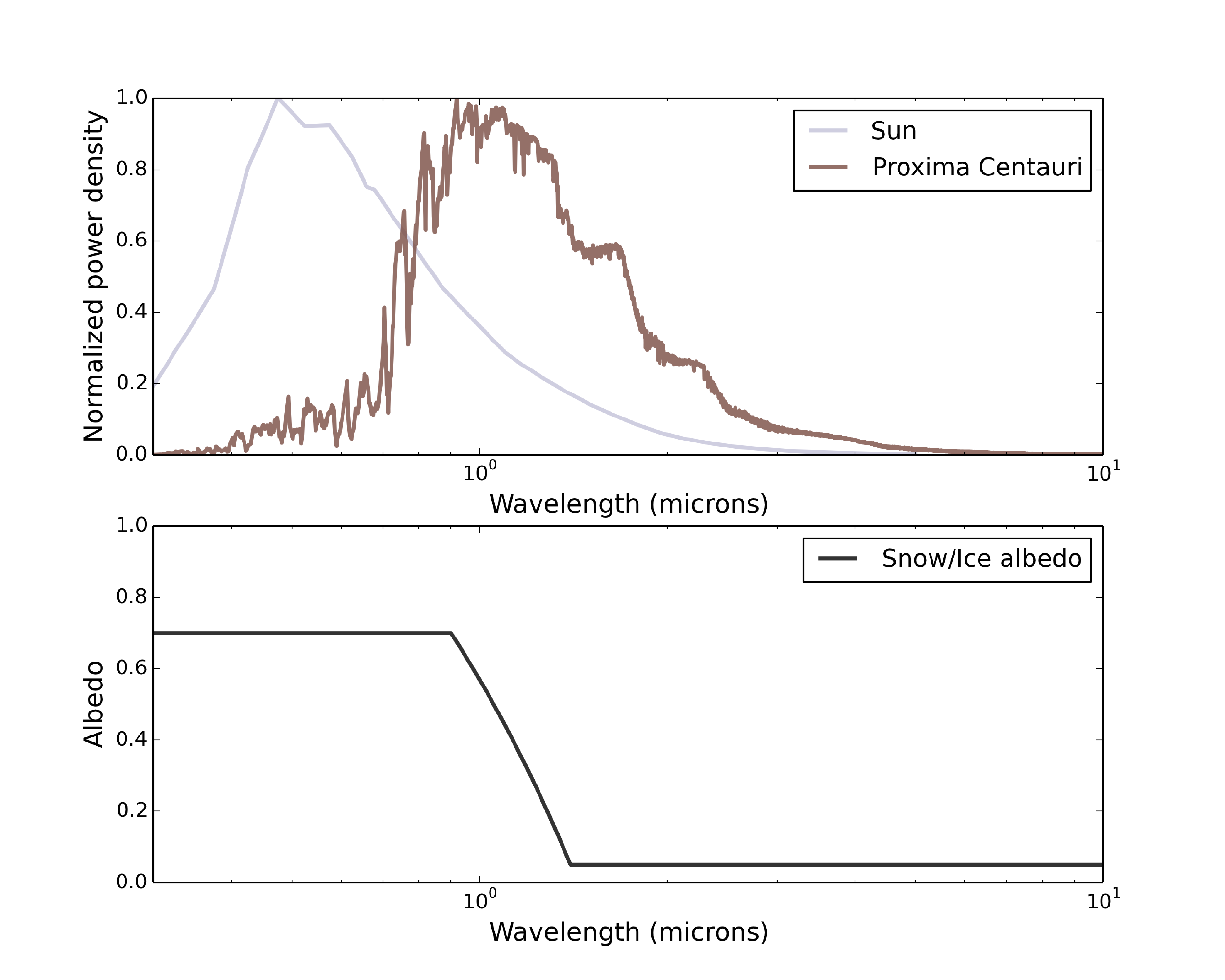}
\caption{Top panel: Synthetic emission spectrum of a Proxima Centauri-like star (normalized by the peak value) used as input for the GCM calculations. For comparison, we put the spectrum of the Sun as usually computed in the LMD-GCM radiative transfer. 
Bottom panel: Spectral distribution of snow/ice surface albedo as computed in the GCM. 
Integrated snow/ice albedo values are respectively 0.55~/~0.27 for the Sun/Proxima Centauri.}
\label{spectrum_proxima}
\end{center}
\end{figure}

The GCM includes an up-to-date generalized radiative transfer \citep{Rothman2009,Wordsworth2010,Richard2012} for variable gaseous atmospheric compositions made of various cocktails of CO$_{2}$, N$_2$ and H$_{2}$O, using the correlated-k method \citep{Fu1992,Eymet2016}. Processes such as the radiative effect of clouds or Rayleigh scattering are taken into account. 
The emission spectrum of Proxima Centauri (see Figure~\ref{spectrum_proxima}, top panel) was computed using the synthetic BT-Settl spectrum\footnote{Downloaded from  https://phoenix.ens-lyon.fr} \citep{Rajpurohit2015} of a M5.5 star, with $T_{\text{eff}}$~=~3000~K, $g$~=~10$^3$~m~s$^{-2}$ and [M/H]~=~0~dex.

Around a red dwarf like Proxima~b, the bolometric albedo of ice~/~snow is significantly reduced \citep{Joshi2012} due to the shape of its reflectance spectrum. To account for this effect, the GCM computes the bolometric albedo of ice from a simplified law of the spectral albedo of ice~/~snow calibrated to get ice~/~snow bolometric albedo of 0.55 around a Sun-like star (\citealt{Warren1980,Warren1984,Joshi2012}; see bottom panel of Figure~\ref{spectrum_proxima}). 
Around Proxima~b, our average bolometric albedo for ice~/~snow is 0.27. 
Yet, because of the varying spectral transmission of the atmosphere (due to variable water vapor and clouds), the bolometric albedo can locally reach values as high as 0.55.

Melting, freezing, condensation, 
evaporation, sublimation, and precipitation of H$_2$O are included in the 
model. Similarly, we take into account the possible condensation/sublimation 
of CO$_2$ in the atmosphere (and on the surface) but not the radiative effect of CO$_2$ ice clouds because their scattering greenhouse effect \citep{Forg:97} should not exceed 10~Kelvins in most cases \citep{Forg:13,Kitz:16}.

\section{The case of a completely dry planet}
\label{dry_section}

Because Proxima~b probably lost a massive amount of water during its early evolution around a pre-main sequence, active star  \citep{Ribas2016}, we need to consider the possibility that the planet may now be rather dry.
When water is in very limited supplies, its state is mostly determined by the temperatures of the coldest regions of the surface, the so-called cold traps (\citealt{AAS11,Leco:13}; see next section).

To get some insight into the location and properties of these cold-traps, we first consider the simple case of a completely dry planet. In this section, we will focus on the greenhouse gas content needed to prevent the formation of a cold-trap where ice could accumulate. Necessary conditions for the atmosphere to be stable are also discussed.

To evaluate this situation, we performed GCM simulations of rocky planets (surface albedo of 0.2) enveloped by pure CO$_2$ atmospheres with pressures ranging from 0.1~bar to 20~bars. Around a red dwarf like Proxima, CO$_2$ is a powerful greenhouse gas because (1) it has stronger absorption lines in the near-infrared than in the visible and (2) it does not contribute much to the stellar reflection by the Rayleigh scattering. 

Figure~\ref{dry_case} shows that the temperatures are quite high across the surface even for CO$_2$ atmospheres of moderate thickness, and despite the low insolation compared to Earth ($\Sp$~=~$0.7~\Searth$). 
In fact, GCM simulations show that, whatever the atmospheric pressure or the orbital configuration, surface temperatures are always greater than 273~K somewhere, although it is not the most relevant factor for the stability of liquid water (and therefore habitability), as indicated by the case of current day Mars \citep{Read2004,Millour2015}.

\subsection{Synchronous rotation}

 For the synchronous orbit case, surface pressures of 6~bars of CO$_2$ are required to warm the entire surface above the melting point of water. However, for lower atmospheric content, the surface temperature contrasts can be very high, for example as much as 150~K difference between the substellar and the coldest points in the 1~bar pure CO$_2$ simulation (see Figure~\ref{dry_case}, left bottom corner). In the synchronous configuration, the planet has 2 cold points located at symmetric positions around longitude $\pm$180$^{\circ}$ and latitude $\pm$60$^{\circ}$. The existence of these two cold traps persist in all the tidally locked simulations (dry, wet, down to 1~bar thin or up to 20~bars thick atmospheres, etc.) explored in this work, but their position can slightly vary, as a result of planetary-scale equatorial Kelvin and Rossby wave interactions \citep{Showman2011}. In particular, our simulations show that thick atmospheres tend to move these two cold gyres towards the west direction~/~higher latitudes (see also Figure~\ref{aquaplanet_case}). 

GCM simulations also tell us that CO$_2$-dominated atmospheres thinner than $\sim$~1~bar are not stable at all because the surface temperature at the two cold points is lower than the temperature of condensation of CO$_2$ (see Figure~\ref{dry_case}, left column~/~third row -- blue dashed line). We identified here a positive feedback: when CO$_2$ starts to collapse due to the decrease of the total gas content, the heat redistribution becomes less efficient, which increases the temperature contrast between the substellar point and the gyres and therefore favors the CO$_2$ condensation at the cold points. In this case, the atmosphere would inevitably collapse until reaching an extremely low CO$_2$ atmospheric content, in a regime of temperatures~/~pressures not well described by our model parametrizations.

In the process, CO$_2$ ice could (1) be trapped for eternity but also (2) form glaciers that could flow efficiently to warm regions and resupply continuously the atmosphere in gaseous CO$_2$ \citep{Turb:16sub2}.
Moreover, the scattering greenhouse effect of CO$_2$ ice clouds \citep{Forg:97,Turb:16sub2} that would form preferentially in the coldest regions of the planet could drastically limit the CO$_2$ atmospheric collapse.

In any case, this shows that having enough atmospheric background gas (main agent of the heat redistribution + additional pressure broadening) may favor the stability of the atmosphere and therefore the habitability of Proxima~b. 
For example, for an atmosphere of 1~bar of N$_2$ - as could be more or less expected on an Earth-sized planet of $\Rp$~$\sim$~1.1~$\Rearth$ \citep{Kopparapu2014} - and 376~ppm of CO$_2$, the dayside has mean surface temperatures above 273~K and the atmosphere does not collapse (see Figure~\ref{dry_case}, left bottom corner).

\subsection{Asynchronous rotation}

\begin{figure*}
\centering
\includegraphics[width=\linewidth]{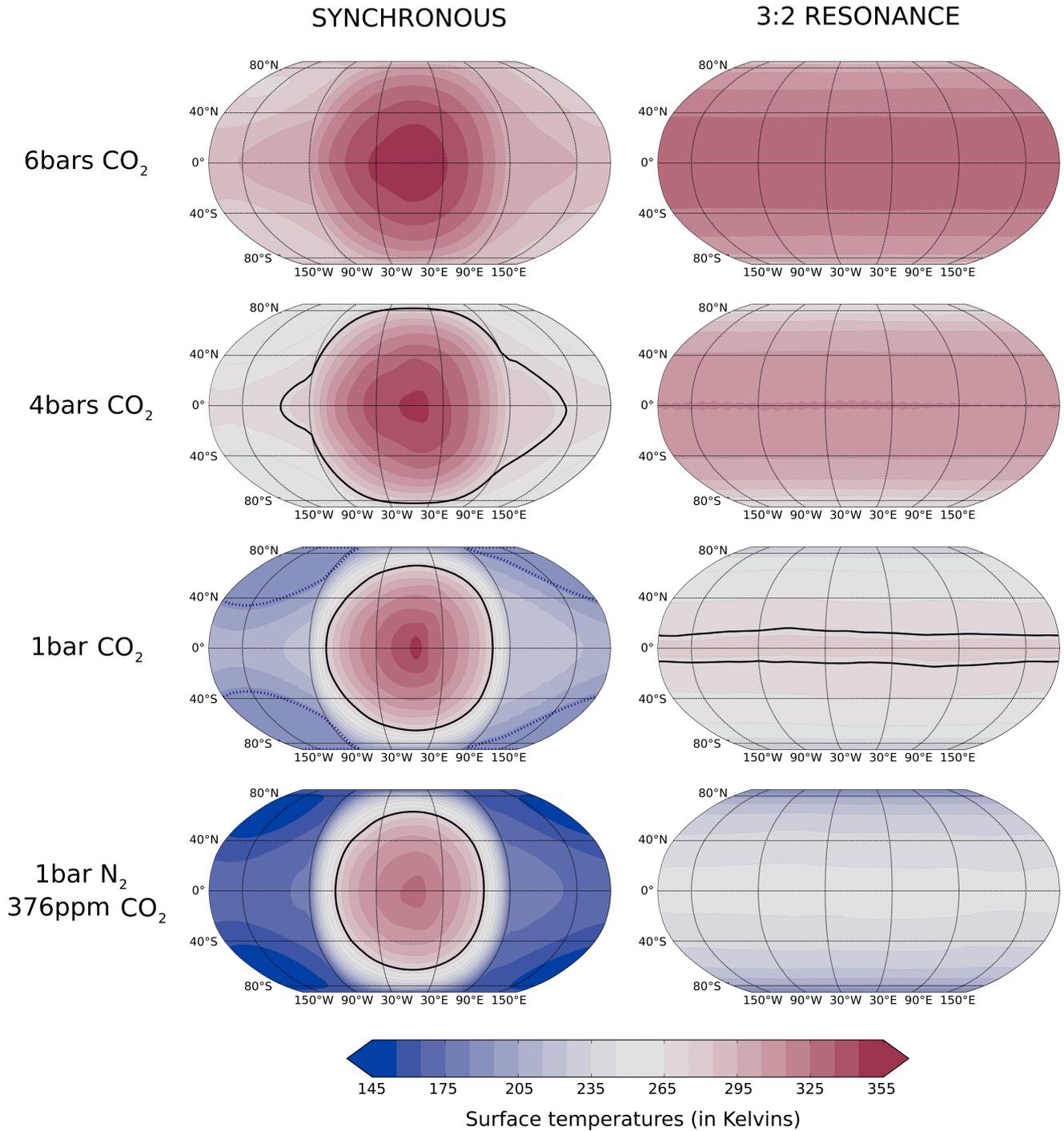}
\caption{Biennal mean surface temperatures of completely dry atmospheres, for 2 orbital configurations (synchronous and 3:2 orbital resonance) and 4 atmospheric compositions/pressures (6~bars of pure CO$_2$, 4~bars of pure CO$_2$, 1~bar of pure CO$_2$ and 1~bar of N$_2$~+~376~ppm of CO$_2$ -- Earth-like atmosphere). The solid black line contour corresponds to the 273.15~K isotherm; the dashed blue line contour indicates the regions where the atmospheric CO$_2$ collapses permanently into CO$_2$ ice deposits.  Note: the 1~bar pure CO$_2$ (synchronous) simulation is not stable in the long term since CO$_2$ would collapse at the two cold points.}
\label{dry_case}
\end{figure*}

For non-synchronous cases, the substellar temperature "eye" pattern disappears and 
the atmospheric pressure at which surface temperatures are all strictly above 273~K is slightly 
lower because the stellar radiation is now distributed equally among the longitudes. 
For the 3:2 resonance case, Figure~\ref{dry_case} shows that this condition is reached for atmospheric pressures greater 
or equal to 4~bars. 3:2 and 2:1 spin-orbit resonance configurations do not exhibit significant differences in term of 
surface temperature maps. The shorter stellar day in 2:1 (11.2 Earth days compared to 22.4) which weakens day/night 
contrasts is compensated by the higher rotation rate, which weakens equator-to-pole heat redistribution \citep{Word:11ajl,Kasp:15}.

Another crucial consequence of the efficiency of the heat redistribution relates to the CO$_2$ collapse, which occurs now at CO$_2$ atmospheric pressure as low as 0.1~bar in the 3:2 orbital resonance GCM simulations (this is a factor 10 lower than for the tidally locked configuration). Therefore, in such configuration, asynchronous rotation would favor the stability of the atmosphere and thus the habitability of Proxima~b.

\section{Limited water reservoir}
\label{small-res}

Another possibility is that Proxima~b may have a limited, but non-zero, water inventory. In this case, the question is to know where this water is stored. To answer that question, \sect{sec:water-atm} first makes an attempt at quantifying how much water vapor can be stored in the atmosphere without ever condensing at the surface. Then, in \sect{sec:water-surf}, we discuss what happens to the water reservoir when it condenses at the surface. In particular, we try to estimate how much water can be stored before it forms planetary scale oceans (the latter being discussed in \sect{large_water_reservoir}).

\subsection{Maximum amount of water stored in the atmosphere}\label{sec:water-atm}

 The amount of water that can be maintained in a planetary 
atmosphere before it starts to condense at the surface depends primarily on the atmospheric temperatures. For example, Venus has a water Global Equivalent Layer of $\sim$2~cm \citep{Bougher1997,Bezard2009} in its atmosphere that may form liquid water droplets in the atmosphere but never condenses at the surface. But it could store much more.


To quantify this possibility in the case of Proxima-b, we perform a simulation with a 4~bars CO$_2$ dominated atmosphere\footnote{Hereafter, water vapor will always be included in the atmospheric calculations, expect when otherwise specified. When referring to a specific atmospheric composition, only the amount of background gases will be specified.}, for a synchronous rotation, in aquaplanet mode (static ocean model, I~=~20000~J~m$^{-2}$~K$^{-1}$~s$^{-\frac{1}{2}}$, surface albedo of 0.07), and until equilibrium is reached (after $\sim$~1000 orbits). Suddenly, we change the properties of the planet: (1) we remove all the water available at the surface, (2) we use a rocky surface (I~=~1000~J~m$^{-2}$~K$^{-1}$~s$^{-\frac{1}{2}}$, surface albedo of 0.2), in place of the removed water at the surface. Then, every orbit, we remove from the planet all the extra water that condenses. We repeat the process until precipitations completely stop for 50 consecutive orbits. 
Up to 45~cm GEL of water could be trapped in the atmosphere before it starts to condense (not shown). In this case, surface temperatures reach 370-450~K (a 45~K increase on average compared to the aquaplanet simulation).

Repeating this procedure for atmospheric pressures of 1~bar~/~2~bars~/~4~bars, we respectively obtained maximum amounts of water vapor (before it condenses permanently at the surface) of 1/10/45~cm GEL. The corresponding simulations were reported on Figure~\ref{proxima_diagrams} and define the black curve separating the region ``All water in vapor state'' from the rest of the diagram.
The same GCM simulations (when taking into account the greenhouse effect of water vapor) show that surface pressures as low as 1.5~bars (and even lower for the 3:2 orbital resonance) are now sufficient to get surface temperatures above 273~K everywhere on the planet.

Thick, Venus-like atmospheres (pCO$_2$ typically $>$~10~bars) could potentially store very large amounts of water, although we did not perform the necessary simulations (they would require a dedicated radiative transfer model). Interestingly, for atmospheres with a huge greenhouse effect, there is no reason for the partial pressure of water not to exceed that of the critical point. In this case, the temperature would also exceed the critical point so that discussing the transition to a liquid phase at the surface would not be very meaningful. The habitability of such environment seems largely unexplored

\subsection{Maximum amount of water stored on the surface}\label{sec:water-surf}

\subsubsection{Cold climates: limits on glaciers.}
\label{thin_co2}

As soon as water is available in sufficient amount (which depends on the atmospheric gas content, as detailed above), it can start to condense permanently at the surface. 
For a synchronous rotation, CO$_2$-dominated atmospheres with pressure typically lower than 1.5~bars exhibit surface temperatures lower than 273~K at their coldest points. 
Therefore, all the extra water available at the surface gets trapped at the two cold gyres forming inevitably stable water ice deposits. 
The range of water inventory for which such unhabitable climate regimes subsist depends on the atmospheric gas content and composition, but also possibly on the internal heat flux of Proxima Centauri b.

As the water inventory increases, the ice deposits thicken and start to form water ice glaciers that can at some point flow from the coldest regions of the planet toward warmer locations. 
There are in fact two distinct processes that can limit the growth of water ice glaciers \citep{Leco:13,Meno:13}:

\begin{enumerate}
\item The gravity pushes the glaciers to flow in the warm regions where ice can be either sublimated or melted. This limit depends mostly on (1) the gravity of the planet and (2) the mechanical properties of water ice (e.g. viscosity).

\item The internal heat flux of the planet causes the basal melting of the water ice glaciers. In such condition, glaciers would slip and flow to warmer regions where, once again, ice could melt and sublimate. This limit depends primarily on (1) the geothermal heat flux of the planet and (2) the thermodynamical properties of water ice (e.g. thermal conductivity).
\end{enumerate}

\citet{Meno:13} has shown that for tidally locked terrestrial planets with Earth-like characteristics,
the basal melting should be the condition that limits the thickness of the water ice glaciers. 
He finds maximum global equivalent ice thicknesses typically ranging from 320~m (for a 1~bar N$_2$, 3.6~$\%$ CO$_2$ atmosphere) and 770~m (0.3~bar N$_2$, 360~ppm CO$_2$ atmosphere).

In the same vein, we compute the maximum ice thickness before basal melting for four of our dry GCM simulations from \fig{dry_case}: we use 2 distinct atmospheric  compositions (1~bar pure CO$_2$ atmosphere -- below that, the CO$_2$ can collapse permanently for synchronous rotation; 1~bar of N$_2$~+~376~ppm of CO$_2$) and 2 orbital configurations (synchronous and 3:2 resonance). This thickness, $h_{\text{ice}}^{\text{max}}$, is given by 
\citep{Abbot2011}:
\begin{equation}\label{ice_thickness}
\centering
h_{\text{ice}}^{\text{max}}~=~\frac{A}{F_{\text{geo}}}~\ln~\left(\frac{T_{\text{melt}}}{T_{\text{surf}}}\right),
\end{equation}
where $F_{\text{geo}}$ is the internal heat flux, and $T_{\text{melt}}$ is the melting temperature of ice at the base of the glacier. The latter being a function of the pressure below the ice, it implicitly depends on $h_{\text{ice}}^{\text{max}}$ so that the above equation must be solved numerically, once the local surface temperature is known (\citealt{Leco:13}; see appendix \ref{appendix:melting curve} for details). This temperature is taken from the GCM outputs (see the third and fourth rows of Figure~\ref{dry_case}).

Assuming that it scales roughly with $\Mp^{1/2}$ \citep{Abbot2011}, the geothermal flux can be extrapolated based on the Earth value ($\sim$90~mW~m$^{-2}$; \citealt{Davies2010}), yielding $F_{\text{geo}}$~=~110~mW~m$^{-2}$.
Of course, this estimate holds only because the Earth and Proxima have similar ages \citep{Bazot2016}. 
However, tidal heating could also take place. \citet{Ribas2016} showed that an initial eccentricity would not be damped significantly over the lifetime of the system. They also argue that, assuming the planet is alone in the system, it would be difficult to excite the orbital eccentricity above 0.1. This would correspond to an extra tidal heat flux of $\sim$~70~mW~m$^{-2}$ for a tidal dissipation 10 times lower than Earth. Therefore, for non-synchronous orbits, we arbitrarily set the geothermal heat flux $F_{\text{geo}}$ to be equal to 110+70~=~180~mW~m$^{-2}$. An upper limit on the tidal heating can also be derived from observations of Proxima~b that put an upper limit of 0.35 on the eccentricity of the planet \citep{Anglada16}.
This configuration produces a tidal dissipation heat flux of $\sim$~2.5~W~m$^{-2}$, which is similar to Io \citep{Spencer2000}.
Yet, in this case, most of the heat would probably be extracted through convection processes (e.g. volcanism) instead of conduction, as on Io. Thus only a (unknown) fraction of this flux should be used in \eq{ice_thickness}. We thus decided to use a geothermal flux of ~180~mW~m$^{-2}$ for our baseline scenario, but the reader should keep in mind that an order of magnitude change could be possible.


\begin{figure*}
\centering
\includegraphics[width=\linewidth]{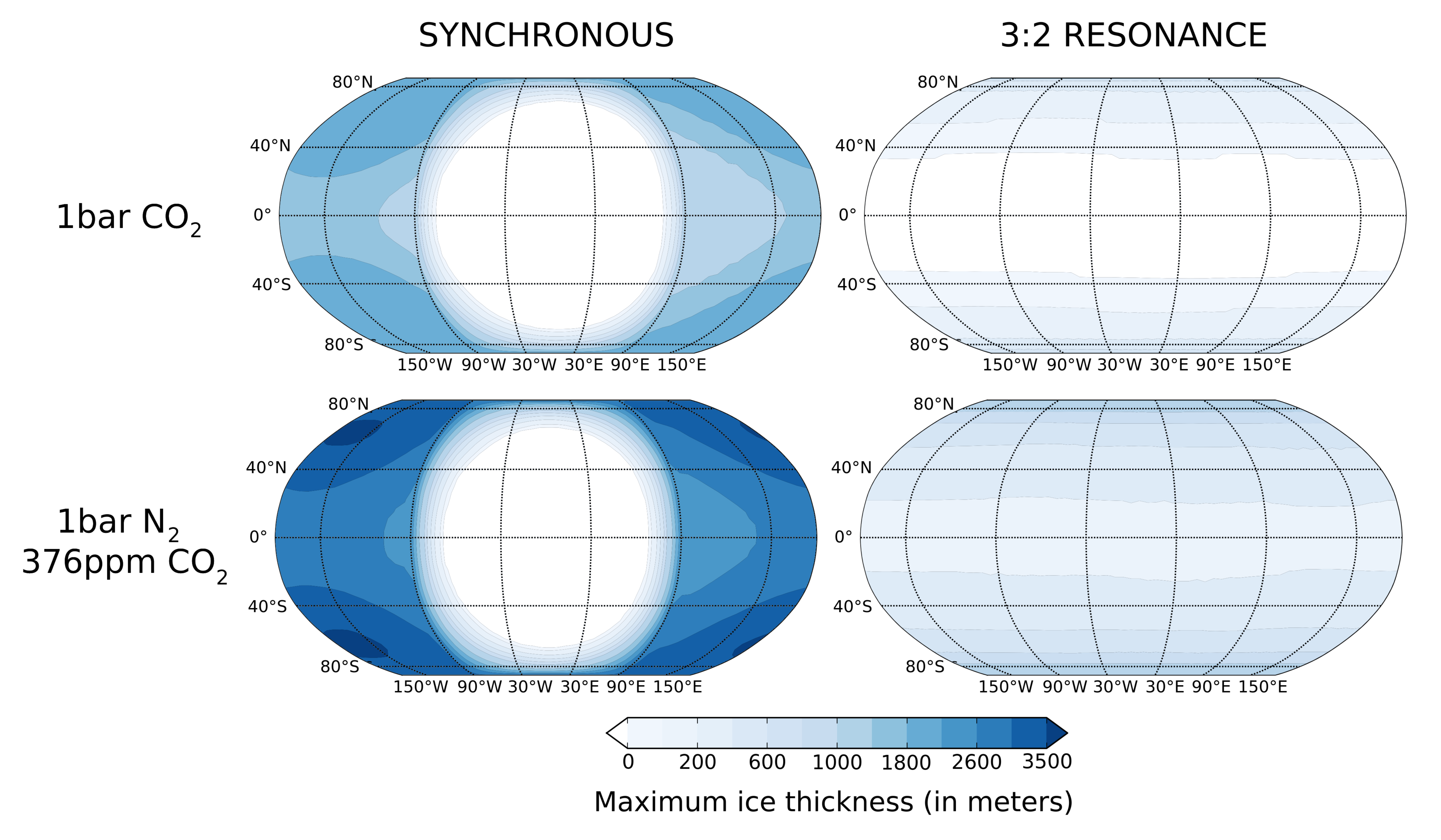}
\caption{Maximum ice thickness calculated from 4 completely dry GCM simulations made of 2 different atmospheric composition (1~bar pure CO$_2$ atmosphere, 1~bar N$_2$-dominated atmosphere with 376~ppm of CO$_2$) and 2 orbital configurations (synchronous and 3:2 resonance). The maximum ice thickness is calculated from the basal melting condition derived from the mean surface temperature from Figure~\ref{dry_case}. The internal heat flux is assumed to be 110~mW~m$^{-2}$ for the synchronous orbit cases and 180~mW~m$^{-2}$ for the asynchronous cases. The global equivalent thickness of ice is 940~m (1bar of CO$_2$) and 1650~m (Earth atm.) for the synchronous rotation and 115~m~/~490~m for the 3:2 resonance, respectively. Note: the amount of ice calculated from the 1~bar CO$_2$ simulation is probably a lower estimate since CO$_2$ would collapse, making the cold points even colder.}
\label{proxima_glacier_map}
\end{figure*}

Figure \ref{proxima_glacier_map} shows the water ice maximum thickness maps derived for our 4 simulations. After spatial averaging, this yields maximum equivalent global thicknesses of ice of 940~m~/~115~m for the 1~bar pure CO$_2$ atmosphere (respectively for sync.~/~async. rotations) and of 1650~m~/~490~m for an Earth-like atmosphere.

\begin{itemize}
\item[$\bullet$] On the one hand, very large amounts of ice (up to 61~$\%$ of Earth ocean content in the Earth-like atmosphere config.) can be trapped in the tidally locked case due to the high contrasts of temperature throughout the surface. \textit{This is not an upper limit}. Thinner atmospheres (due to CO$_2$ collapse for example, see section \ref{dry_section}) could entail much more extreme surface temperature contrasts. Such a \textit{Pluto-like} planet could potentially trap tremendous amounts of water in the form of ice.

\item[$\bullet$] On the other hand, much more limited quantities of ice can be trapped in asynchronous simulations due to both (1) a better heat redistribution and (2) a higher geothermal heat flux. We note that, for an eccentricity of 0.35, the amount of trapped ice would be probably much less than 115~m~/~490~m  due to the increased internal heat flux. 
\end{itemize}

We mention that these GCM simulations were performed with a dry atmosphere: the lack of water vapor---a powerful greenhouse gas---leads to an overestimate of the amount of ice possibly trapped.

Eventually, once the water ice glaciers start to spill, they can possibly melt at their edge, either on dayside or nightside. It has been shown by \citet{Leco:13} that such configuration could be long-lived.
We roughly put in Figure~\ref{proxima_diagrams} (hatched region) the range of CO$_2$~/~water inventory for which such scenario would happen. 
This is an exotic form of habitability.

\subsubsection{Warm climates: appearance of lakes}
\label{intermediate_co2_content}

As discussed in \sect{dry_section}, if the greenhouse effect of the atmosphere is sufficient, the coldest temperatures at the surface are above the freezing point of water. Although we might intuitively think that this situation is very different from above, from the point of view of the atmosphere, it is not. The latter will always tend to transport water from the hottest to the coldest regions. The fact that the cold-trap temperature is above the freezing point is irrelevant as long as there is not enough water at the surface to redistribute water more rapidly than it is brought in.

Liquid water will thus first accumulates around the coldest regions of the planet. Interestingly, on a synchronous planet, these are located on the night side where no photons are available for photosynthesis.
If the spin is non-synchronous, the volatiles would likely first concentrate toward the poles. This seems to be the case on Titan where methane lakes are mostly seen at high latitudes \citep{SEL07}.


The range of water inventory for which this configuration (liquid water on nightside) may subsist not only depends on the water inventory but also on the topography. Note that in the case of a tidally-locked planet, the topography features may not be randomly distributed, because tidal locking could tend to favor the alignment of large scale gravitational anomaly  (correlated with topography anomaly) with the star-planet axis. For instance there is a clear difference between the near side and far side of the Moon \citep{Zube:94}, and the deep Sputnik Planum basin on Pluto is located near the anti-Charon point \citep{Moor:16}. Thus it is conceivable that Proxima~b may have its largest topographic basin either near the substellar point at the anti-stellar point.

\begin{figure}
\begin{center}
\includegraphics[scale=0.15]{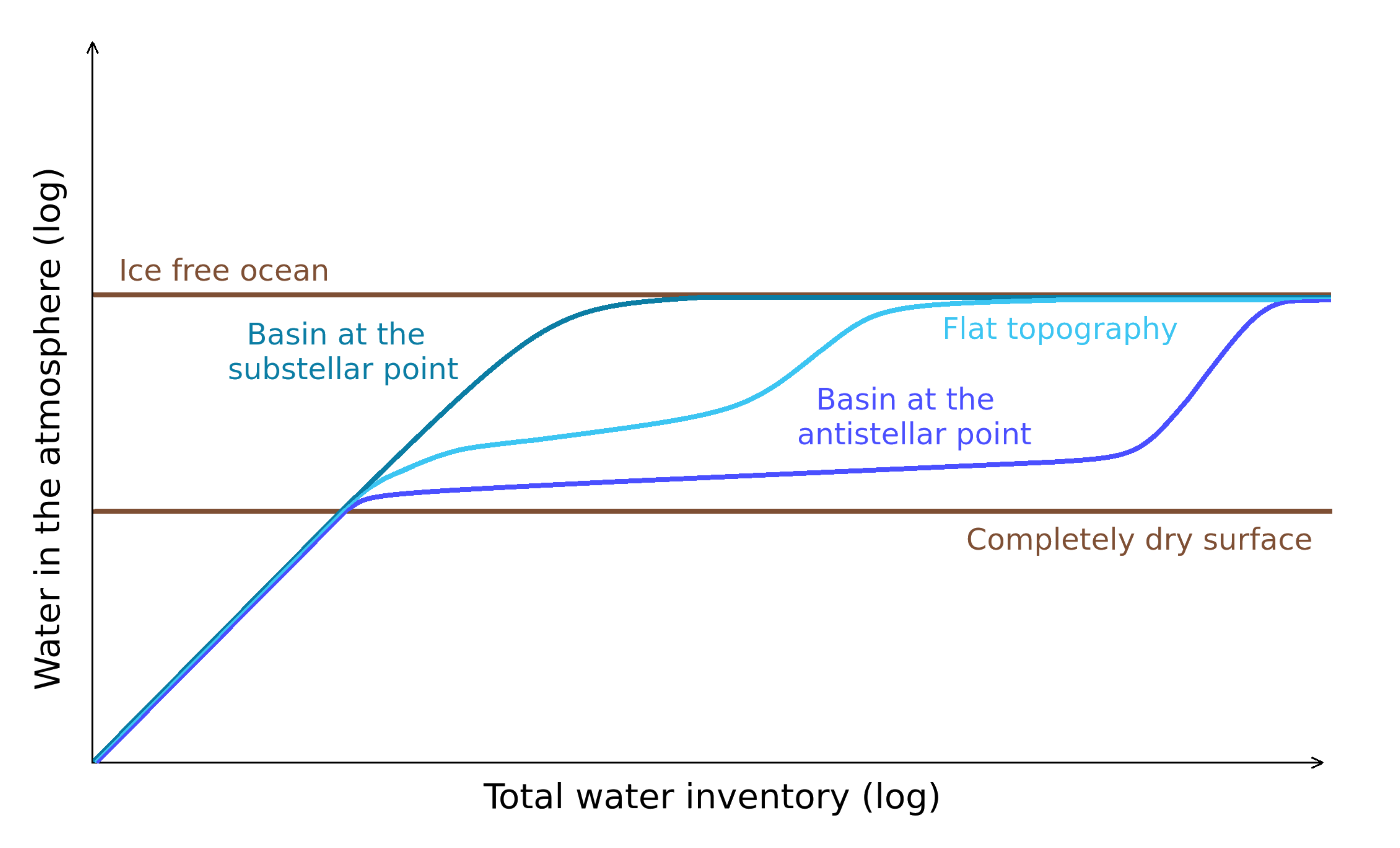}
\caption{Qualitative evolution of the amount of water vapor in the atmosphere as a function of the total water inventory, assuming a synchronous rotation and a liquid water runoff activated, for three different scenarios: a quasi-flat topography, a basin at the substellar point and a basin at the anti-stellar point. We mention that, for high pCO$_2$, the range of water inventory for which the three scenarios diverge would flatten on a logarithmic scale, because potentially high amount of water could be vaporized.}
\label{topo_vapor}
\end{center}
\end{figure}

Interestingly, as the water inventory grows, the response of the climate (amount of water vapor, atmospheric temperatures, ...) might be significantly different depending on the topography. 
Figure~\ref{topo_vapor} illustrates qualitatively the fact that, depending on the topographic setup (basin at the substellar point, basin at the anti-stellar point, or no basin at all -- quasi-flat configuration), the distribution of water between the surface and the atmosphere might significantly differ. As significant amounts of water start to condense at the surface, liquid water would likely spill toward the main topographic depression of the surface. If this depression is for example located at the substellar point, where evaporation rates are the highest, the proportion of water in the atmosphere given a fixed total water inventory would be maximum and definitely much higher than in the extremely opposite case (a basin at the antistellar point).

Therefore, it is important to mention that assessing the proportion of water vapor in the atmosphere of Proxima~b might not be sufficient to get information on the stability/location of surface liquid water, and vice versa, knowing the exact water inventory available on Proxima~b would not be totally relevant to deduce its possible climatic regime.

\section{Large Water reservoir}
\label{large_water_reservoir}

Despite the large amount of hydrogen that could have escaped within the lifetime of Proxima~b, the quantity of water now available on the planet also depends significantly on its initial water inventory. 
As argued by \citet{Ribas2016}, if Proxima Centauri~b formed beyond the ice-line in a similar fashion to Solar System icy satellites, it could still possess enough water to be an \textit{aquaplanet}\footnote{Here, the term aquaplanet refers to a planet where water is abundant enough to flow efficiently on a planetary scale. In practice, in our simulations, this means that the surface acts as an infinite source of water.}.

In this section, we will thus make an attempt at quantifying how much is \textit{enough.} Then, we will discuss the various climate regimes available to an aquaplanet and point out the main differences with the dry case.

\subsection{Transition from small to large water inventory}
\label{transit-water-rich-world}

The exact water inventory for which Proxima~b would transition from a landplanet (previous section) to an aquaplanet (this section) is hard defined. 
In particular, the nature of the transition depends of the amount of greenhouse gas in the atmosphere.

\begin{enumerate}
\item For a low atmospheric greenhouse effect (meaning that ice is stable somewhere on the planet), there is a bistability between two possible climate regimes around the transition.

To understand this bistability, let us make a thought experiment where we consider a planet with two very different initial conditions

\begin{itemize}

\item[$\bullet$] Case 1: Warm- and/or water-rich- start. Let us assume that the planet starts with a global ocean, possibly covered by sea-ice wherever cold enough. This experiment has been done by \citet{YLH14}\footnote{This paper only looked at a synchronous planet, but the argument remains qualitatively valid for a non-synchronous one, even though numerical values will surely change significantly.} with a 325\,m-deep ocean. They showed that in such a configuration, winds carry sea-ice toward hot regions and the ocean carries heat toward cold ones, so that an equilibrium can be found with less than 10\,m of sea-ice in the coldest regions\footnote{They however acknowledge that the presence of continents could significantly alter this value.}. Therefore, a water content as small as $\sim$~10$^2$\,m GEL (and maybe even lower) is sufficient to maintain aquaplanet conditions. If water were to be removed or the temperature decreased, at some point, the oceanic transport would shut down and the planet would transition to a dry, cold regime with glaciers.
\item[$\bullet$] Case 2: cold- and/or water-poor-start. As discussed in the previous section, if the planet started cold or dry enough, all the water would be trapped in ice caps and glaciers. Now, we demonstrated earlier that in this state, more than $\sim\,10^3$\,m~GEL of ice could be stored this way.
If the water inventory and/or the temperature were to be increased, glaciers would progressively spill toward hotter regions. This state could resist until the oceanic transport becomes more efficient than the atmospheric one. After that, an ocean would accumulate and, in turn, warm the cold regions, speeding up the transition to the aquaplanet regime.

\end{itemize}

Therefore, between roughly $\sim\,10^{2}$ and $10^{3}$\,m~GEL\footnote{On a non-synchronous planet, the upper limit would decrease as the cold traps are less efficient. At the same time, a smaller global ocean might be needed to efficiently warm the poles.}, two distinct climate regimes coexist, depending on the history of the planet. As both the water inventory and the amount of greenhouse gases play a role, it is even possible for the planet to undergo hysteresis cycles between the two states. But note that the mechanism is rather different from the one involved in snowball climates (which will be discussed later on). Indeed, here, the albedo feedback does not play a major role. What controls the transition between a water-rich $\&$ warm world toward a dry $\&$ cold one is now the oceanic transport. 
\item For warmer climates (in the sense that ice cannot form at the surface), the transition would happen whenever water would be abundant enough to flow. Topography would thus be the key parameter (see \sect{intermediate_co2_content}). 
\item A third situation can occur when the amount of greenhouse gases in the atmosphere is such that the partial pressure of water vapor at the surface can exceed that of the critical point. Then we expect no transition at all as there is no phase transition between liquid and gaseous phase above the critical point.  
\end{enumerate}


Finally, let us note that in any case, another transition would occur at much higher water contents ($\sim10^{5-6}$\,m GEL) when high pressure ices form.

\begin{figure*}
\centering
\includegraphics[width=\linewidth]{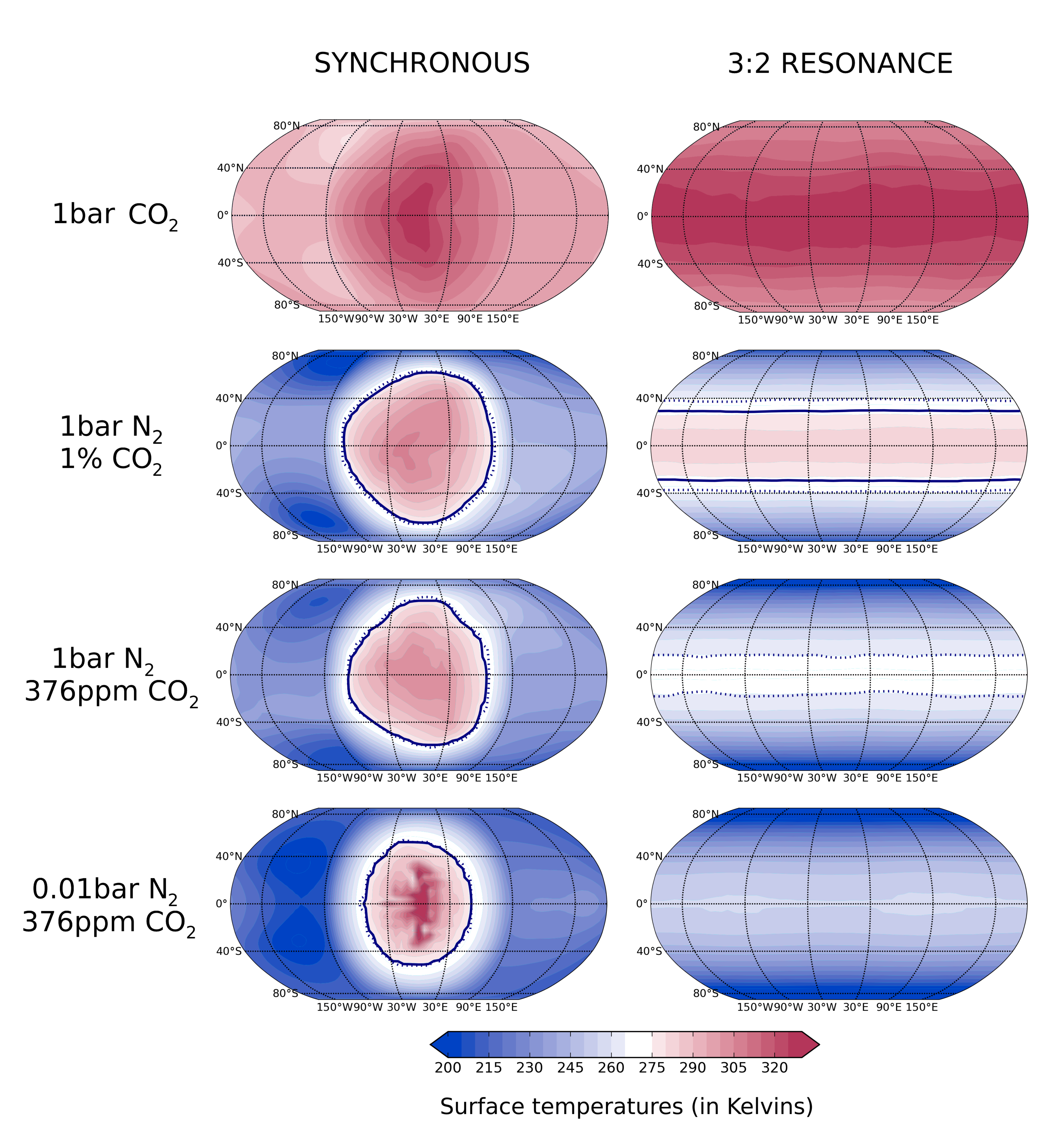}
\caption{Biennal mean surface temperatures of atmospheres made of 4 various cocktails of N$_2$/CO$_2$ for 2 orbital configurations (synchronous and resonance 3:2), assuming planets initially completely covered by liquid water. Solid lines/dashed lines contours correspond respectively to the location of permanent/seasonal surface liquid water.}
\label{aquaplanet_case}
\end{figure*}

\subsection{Necessary conditions to have surface liquid water}
\label{thin_aquaplanet}

Assuming that Proxima~b is in the aquaplanet regime, one may wonder what is the minimal requirement to maintain liquid water stable at the surface. 
Compared to the limited reservoir case, the requirements are much less stringent. This has two main reasons:
\begin{itemize}
\item[$\bullet$] There is too much water, by definition, for complete cold-trapping to occur. As a result, the lowest temperature at the surface is rather irrelevant. Instead, only the highest temperature matters. And it is much easier to have a planet with one region above freezing than one with nowhere below.
\item[$\bullet$] Because water is readily available for evaporation, the atmosphere will, on average, be much closer to saturation compared to the dry case. Water vapor being a good greenhouse gas, this usually entails that for a given background atmosphere, the average surface temperature of the aquaplanet will be higher than its dry counterpart\footnote{This can sometimes be mitigated by the ice albedo effect, but the latter is extremely weak around red stars such as Proxima \citep{Joshi2012}, especially for synchronous spin states.}.
\end{itemize}

With that in mind, we performed a suite of aquaplanet simulations with various atmospheric compositions to assert the likelihood of surface liquid water. The results are shown in \fig{aquaplanet_case}.

\subsubsection{Synchronous rotation: ubiquity of liquid water.}

As can be seen on the left column of Figure~\ref{aquaplanet_case},
when the planet is synchronously rotating, temperatures are always high enough at the substellar point to have liquid water, whatever the atmospheric content (pure CO$_2$, Earth-like mixture, thin, 0.01\,bar atmosphere). This would hold even without a background atmosphere\footnote{In this case, the atmosphere would be composed of water vapor.}. Indeed, because the substellar point permanently receives 956~W~m$^{-2}$, it would take a surface bond albedo of 0.67 to cool this region below the freezing point. We recover the 'Eyeball Earth' regime \citep{Pierrehumbert2011} or 'Lobster Earth' regime \citep{Hu2014} (when deep ocean circulation is taken into account).

Starting from there, adding greenhouse gases to the atmosphere only (1) increases the mean surface temperature, (2) increases the size of the liquid water patch, and (3) reduces the temperature contrast.
Eventually, above some threshold ($\sim 1$\,bar in our simulations), greenhouse is intense enough to deglaciate the whole planet. 

\subsubsection{Asynchronous rotation}

In a non-synchronous spin-state, on the contrary, surface liquid water is not always possible. In fact, we recover a situation quite similar to the one on Earth: below some threshold amount of greenhouse gases, the planet falls into a frozen, snowball state. 

The reason for this difference with the synchronous state is that any point at the equator now receives on average only 1/$\pi$ of what the substellar point received at any time in the synchronous configuration. The zero-albedo equilibrium temperature corresponding to this mean flux is $\sim 270$\,K. Some greenhouse effect is thus necessary to melt the equator, especially when the albedo and circulation effects are added. 

Consequently, four different climate regimes can be reached on such an asynchronous planet, depending on the greenhouse gas content. These are depicted on the right column of \fig{aquaplanet_case}. From top to bottom, we have:
\begin{enumerate}
\item For high CO$_2$ pressures (roughly above a few tenths of bars\footnote{We did not perform our simulations on a fine enough grid to be more precise but the limit should be lower than 1~bar (see \fig{aquaplanet_case}).}), the planet is covered by an ice-free ocean. 
\item At lower levels of CO$_2$ (down to $\sim$0.01\,bar with 1\,bar N$_2$) the planet can keep a permanently unfrozen equatorial belt.
\item For lower greenhouse gas contents, diurnal patches of liquid water lagging behind the substellar point subsist. 
\item For thin enough atmospheres, a completely frozen snowball state ensues. 
\end{enumerate}

The 0.01~bar atmosphere (on Figure~\ref{aquaplanet_case}, right column) is colder 
than the 1~bar because 1. the pressure broadening of CO$_2$ absorption lines by N$_2$ is drastically reduced; 
and 2. the absolute amount of CO$_2$ in the atmosphere is 100 times lower. 
Both effects are responsible for the appearance of an equatorial band of relatively warmer surface temperatures for the 3:2 resonance~/~1 bar of N$_2$ (+~376ppm of CO$_2$) case. 

Let us note that the well known "snowball hysteresis" could potentially exist between our states 2, 3 and 4, although the weak ice-albedo feedback around M-dwarfs certainly makes it less likely than on Earth \citep{SBM14}. A confirmation of this would necessitate numerous additional simulations.

\subsection{Subsurface oceans?}

Whether or not surface liquid water is possible, we now try to assess the possibility of the presence of a subsurface ocean. 
Indeed, if Proxima~b has been able to keep a large enough water inventory, the steady release of geothermal heat entails a rise in temperature with depth through the ice cap.

To assess a lower limit, we consider our coldest, and only fully glaciated case (0.01~bar of N2~/~376~ppm of CO$_2$, 3:2 resonance; see ~Figure~\ref{aquaplanet_case}, right bottom corner). Following a similar approach to section~\ref{thin_co2}, we estimate that a subsurface ocean could exist for water Global Equivalent Layers greater than 600~meters (assuming a geothermal heat flux of 180~mW~m$^{-2}$). Any effect warming the surface would tend to lower this threshold. 

Above such water inventories, Proxima b could thus be at least considered a class III or IV (if there is enough water to form high pressure ices) habitable planet \citep{Lammer2009,Forget2013drake}.

\subsection{Thin atmospheres - Implication for water loss}

When considering water loss around planets in the Habitable Zone of small stars, it is tempting to disregard water losses occurring after the end of the initial runaway greenhouse phase \citep[which can be important, e.g.][]{Ribas2016}. 
The reason for this is that the tropopause usually acts as an efficient cold trap. 
The amount of water vapor available for escape in the upper atmosphere is thus limited by diffusion \citep{Kasting1993,Wordsworthpierrehumbert2013}. 

However, this conclusion is often based on calculation including a relatively massive, Earth-like background atmosphere. 
Because escape may have been very important for Proxima~b in the past, early atmospheric escape may have removed an important fraction of the background atmosphere \citep{Ribas2016}. 
It is thus primordial to infer whether a less massive background atmosphere is still able to shield water vapor from escape once the planet has cooled down. 

\begin{figure}[htbp] 
 \centering
\subfigure{ \includegraphics[scale=.65,trim = 0cm .cm 0.2cm 0.cm, clip]{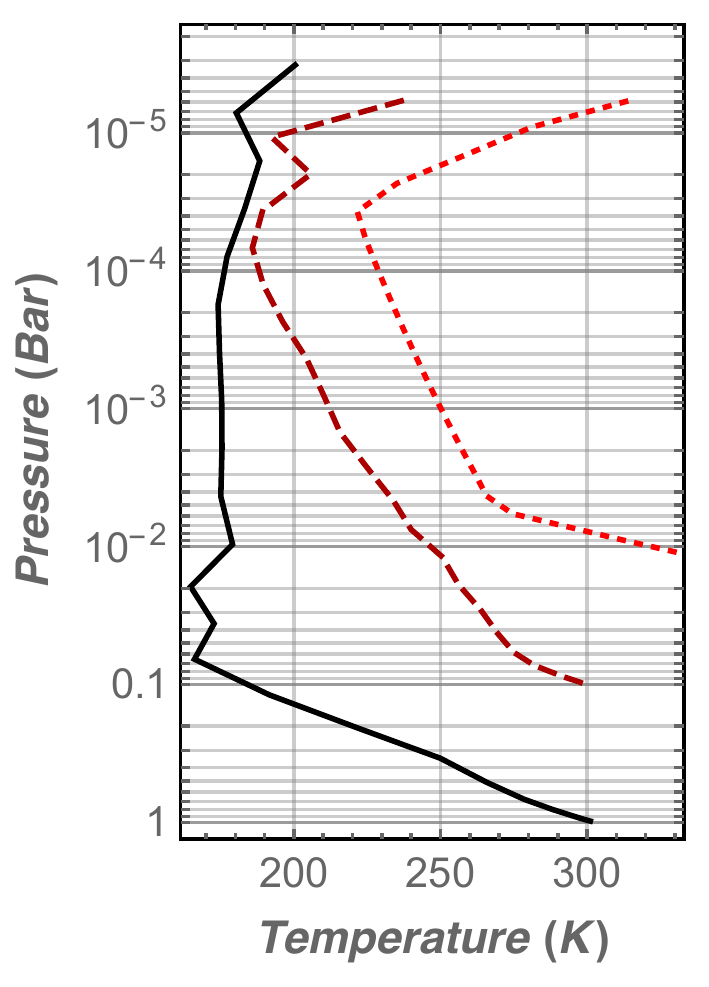} }
\subfigure{ \includegraphics[scale=.65,trim = 1.75cm .cm 0.cm 0.cm, clip]{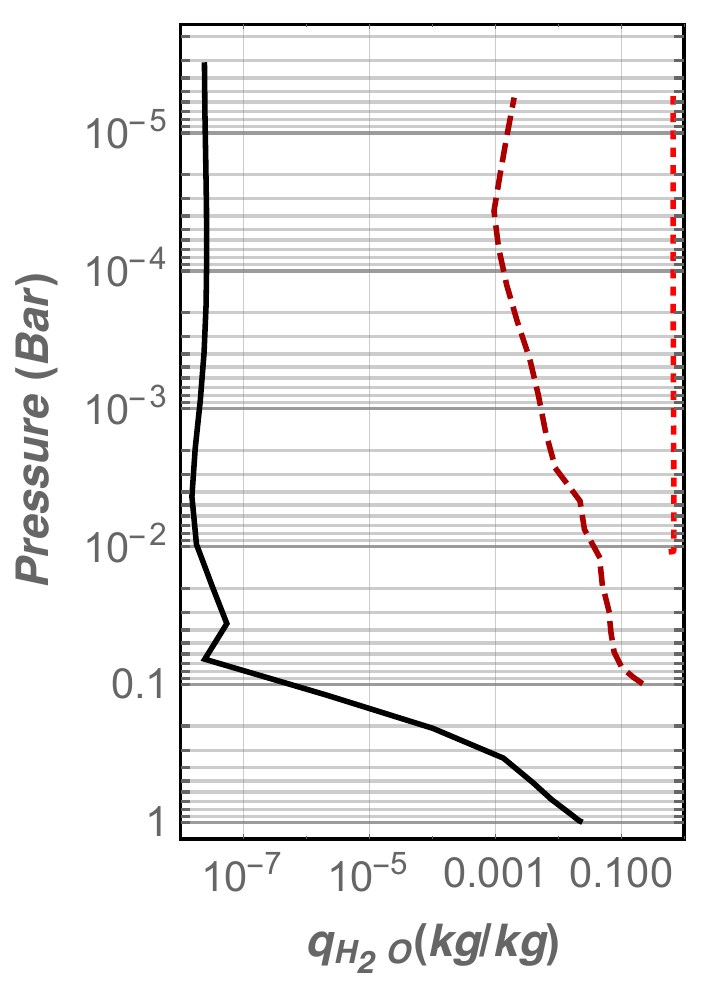} }
\caption{Annual mean atmospheric temperature (left) and water vapor (right) 
vertical profiles of atmospheres with different total surface pressure (solid black: 1\,bar; dashed, dark red: 0.1\,bar; dotted red: 0.01\,bar). The composition of all three atmospheres is N$_2$ with 376\,ppm of 
CO$_2$ and variable amount of water vapor. The profiles are shown at the substellar point, but horizontal variations are fairly small above the 0.1\,mbar level due to an efficient transport. Decreasing the surface pressure increases drastically the water vapor ratio in the upper atmosphere.
}
 \label{waterloss}
\end{figure}

To that purpose, we performed simulations with an Earth-like atmospheric composition, but a lower background surface pressure (namely 0.1 and 0.01\,bar; bottom panels of Figure~\ref{aquaplanet_case}). 
The substellar temperature and vapor mixing ratio profiles are shown in Figure~\ref{waterloss}, along with the reference 1\,bar case. 

The water vapor mixing ratio increases drastically in the atmosphere when the background pressure decreases, even at a given pressure level. 
This results from the fact that:
\begin{enumerate}
\item When the surface pressure decreases, the surface temperature cannot change drastically to remain in radiative equilibrium.
\item Throughout the troposphere, the temperature follows a (moist) adiabat and is thus determined by the ratio of the local to the surface pressure. At a given pressure level, temperatures in the troposphere thus increase when the surface pressure decreases. 
\item Because of the Clausius-Clapeyron law, this increases the mixing ratio of water vapor throughout the troposphere \item Finally, the strong absorption bands in the near-infrared---the peak of the stellar spectrum---provides a positive feedback that tends to humidify the tropopause even more.
\end{enumerate}

Moreover, the water vapor mixing ratio increases globally because it is advected by the largescale circulation.
As a consequence, hydrogen escape is not limited by the diffusion of water vapor anymore, even after the runaway phase. Low atmospheric background gas contents may thus lead to increased rates of hydrogen (and thus water) loss to space.

\section{Observability}
\label{observability}
At the time of this study we do not know whether the planet transits or not and flares complicate the search. There is only 1.3\% of chance that the inclination of the orbit of Proxima~b produces transits. So we do not consider characterization by transit spectroscopy in this work. \\
\subsection{Prospects for direct imaging}
Proxima~b may be the Habitable-Zone terrestrial exoplanet offering the best
combination of angular separation and contrast for imaging. The angular
separation between the planet and its star varies from 0 (for a $90^{\circ}$
inclination) to 38~mas (0.05~AU at 1.29~pc). The planet/star contrast of a
1.1~R$_\oplus$ purely Lambertian sphere (surface albedo = 1) at 0.05~AU from
Proxima is $2\times10^{-7}$ when seen with a 90$^{\circ}$ phase angle, and
approaches $6\times10^{-7}$ as the phase angle approaches 0$^{\circ}$. Current
instrumentation using adaptive optics and coronography on 10~m class
telescopes (like SPHERE/VLT, GPI/Gemini) aims at achieving a contrast of
$10^{-6}-10^{-7}$ but with an inner working angle of a few $\lambda/D$, not
smaller than 100-200~mas depending on the band \citep{2012Lawson}. 
\citet{Lovis2016} suggest that the detection can actually be achieved with VLT by
coupling SPHERE with the future high-resolution spectrometer ESPRESSO (first light
expected in 2017). The idea is to first use SPHERE to reduce the stellar light by a
factor $10^3-10^4$ at 37~mas from Proxima ($\sim$ 2~$\lambda/D$ at 700~nm) and then
to search with ESPRESSO for the Doppler shifted planet signature on the $\sim
37$~mas-radius annulus around the star when RV ephemeris predict maximal angular
separation. The planet and stellar signals could indeed be separated thanks to
cross-correlations with molecular high-resolution fingerprints (that could be
specific to the planet, like O$_2$, or reflected, and Doppler shifted in both cases
by the planetary orbital motion). Such a planet-star disentangling was already
achieved for non-transiting unresolved hot Jupiters with a contrast of $\sim 10^{-4}$
\citep{Brogi2014}. \citet{Lovis2016} expect a similar efficiency, which on top of the
stellar extinction provided by SPHERE would allow them to reach the $\sim 10^{-7}$
contrast of Proxima~b. \\
The combination of contrast and separation that is required to image Proxima~b
should be achieved with future larger telescopes like the E-ELT
(39~m) or the TMT (30~m). For the E-ELT, 37~mas corresponds to 7$\lambda/D$ at
$1~\mu$m. At such angular separation, E-ELT instrumentation such as PCS
\citep{PCS2013} is planned to achieve contrasts of $10^{-7}-10^{-8}$, a 
performance that is sufficient to aim at directly characterizing Proxima~b. 

\begin{figure}
   \centering
   \includegraphics[width=\columnwidth]{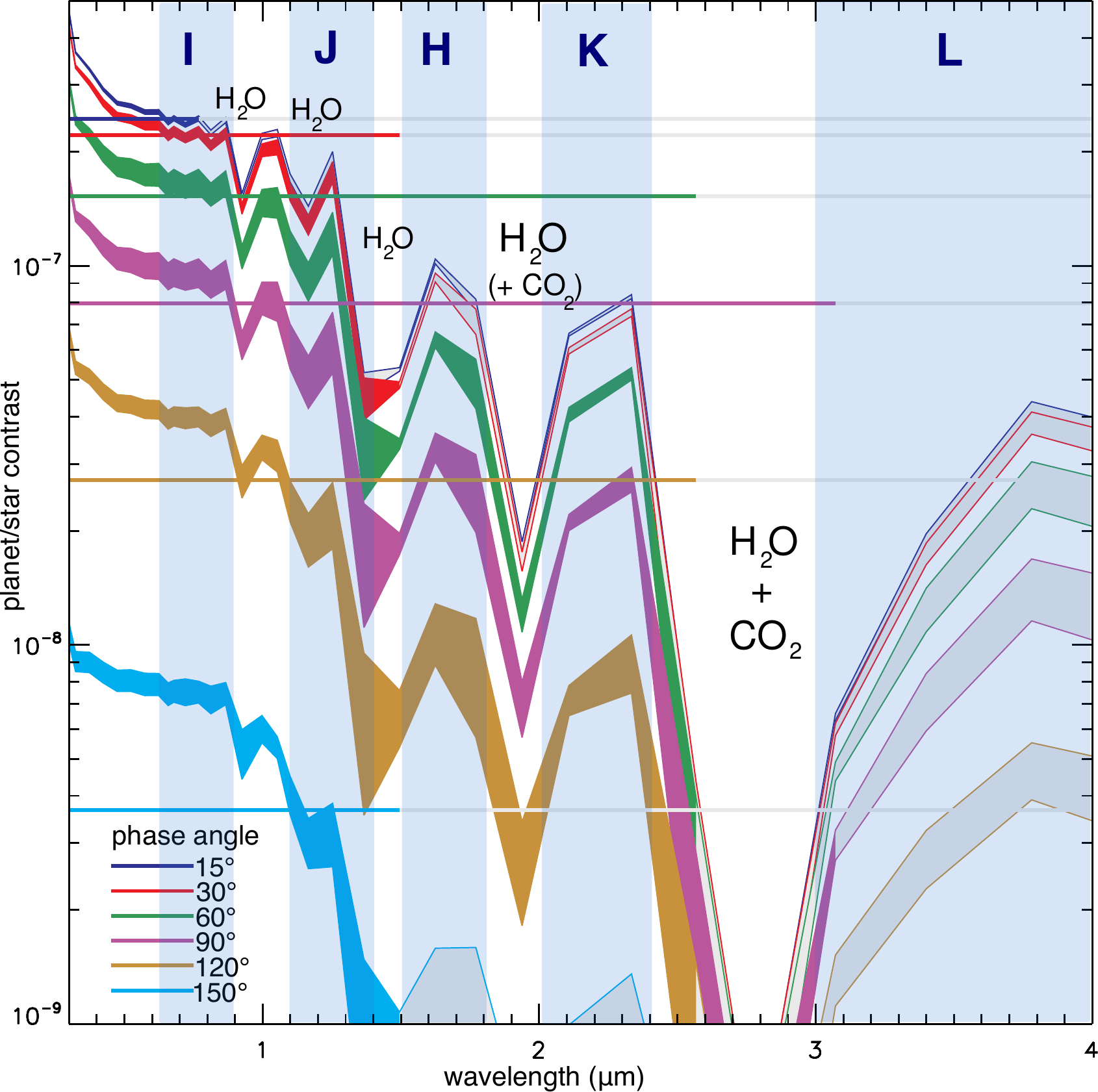}
      \caption{Reflection spectra computed for the synchronous case with an Earth-like atmosphere. Each color corresponds to a phase angle 
($0^{\circ}$
meaning that the observer looks at the substellar point and $90^{\circ}$ at a
point on the terminator). The thickness of the curve indicates the range of
possible values depending on the actual observing geometry (see text and Fig.~\ref{fig:phase}). 
Straight lines are calculated for a constant surface albedo of 0.4. Curves are
plotted in grey when the angular separation falls below twice the diffraction 
limit of the E-ELT ($2\times1.2 \lambda/D$). These plots are obtained with a 
fixed planetary radius of 1.1~$R_\oplus$. Because these plots do not include 
the contribution from the thermal emission, the contrast at 3.5--4~$\mu$m is
underestimated by a factor of $\sim 2$. }
         \label{fig:contrast}
   \end{figure}

   \begin{figure}
   \centering
   \includegraphics[width=\columnwidth]{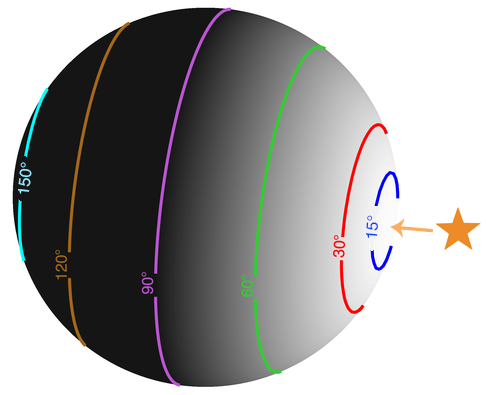}
    \caption{Observing geometries for spectra computation. Emission and reflection spectra are presented in this article for phase angles of 15, 30, 60, 90 120 and 150$^{\circ}$. }
\label{fig:phase}
\end{figure}

   \begin{figure*}
   \centering
   \includegraphics[width=\linewidth]{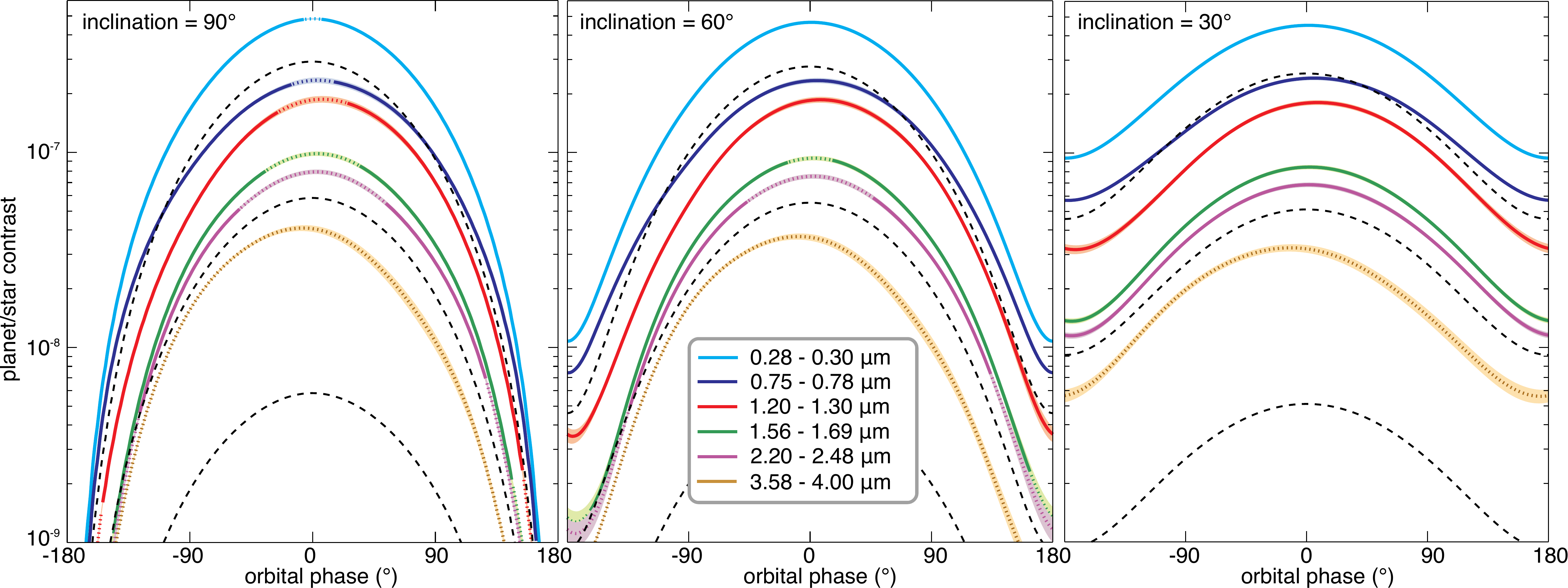}
      \caption{Reflection phase curves computed for the synchronous case with an Earth-like atmosphere for 4 spectral bands falling into the $I$, $J$, $H$
$K$ and $L$ windows. Shadowed areas indicate the 1-sigma variability due to 
meteorology
(mainly changing cloudiness). The dashed curves are calculated with a constant
surface albedo of 0.5, 0.1 and 0.01. Curves are dotted when the planet is inside an
inner working angle of twice the diffraction limit of the E-ELT ($2\times1.22
\lambda/D$). Contrary to Fig~\ref{fig:contrast}, these plots include a
dependency of the planetary radius on the inclination: $R \propto 
(M_{min}/\sin i)^{0.27}$. Note that $60^{\circ}$ is the median value for a 
random distribution of inclinations. }

\label{fig:phasecurve}
   \end{figure*}
  \begin{figure}
   \centering
   \includegraphics[width=\columnwidth]{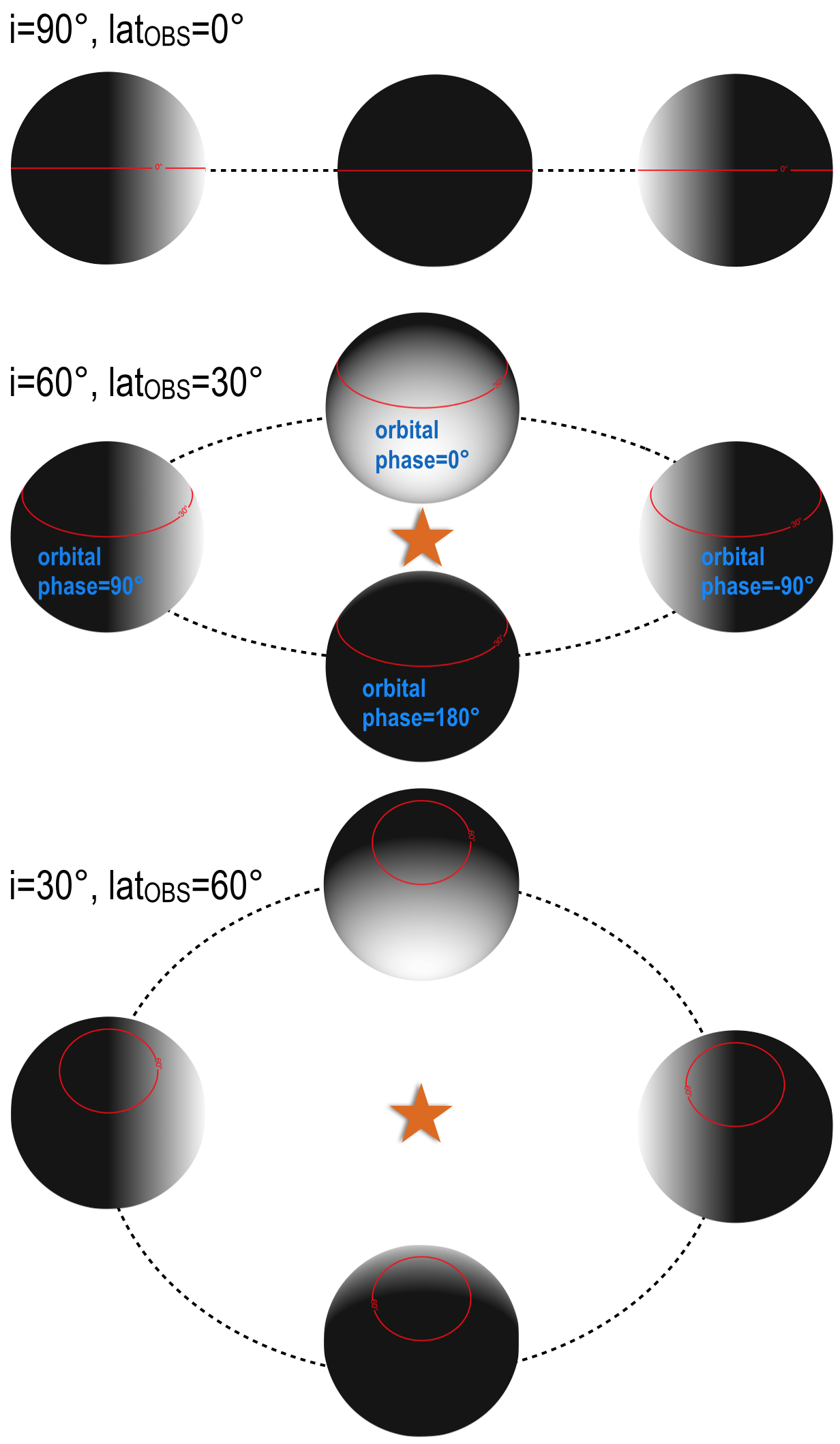}
    \caption{Phase curves and observing geometries. We computed reflected and emitted light curves for 3 inclination : 90, 60 and $30^{\circ}$ corresponding to sub-observer latitude of 0, 30 and $60^{\circ}$, respectively (as the we assumed a null obliquity). The sub-observer latitude is indicated with a red line. In the Appendix~\ref{appendix:obs}, phase curves are shown only for an inclination of $60^{\circ}$ (the median value for randomly oriented orbits).}
\label{fig:inclination}
\end{figure}

Following \citet{selsis2011}, we can use GCM simulations to compute
disk-integrated fluxes in the spectral bands of the GCM and for any observing
geometry. Figure~\ref{fig:contrast} and \ref{fig:phasecurve} present synthetic
observables at visible-NIR wavelengths obtained with a simulation for an aquaplanet with an Earth-like atmosphere (1 bar
of N$_2$, 376~ppm of CO$_2$, variable H$_2$O) and a synchronous rotation.
Figure~\ref{fig:contrast} shows
reflection spectra at the spectral resolution of the GCM for different phase
angles as well as the main bands observable through Earth's atmosphere. The
thickness of the curves indicates the range of contrast values that is 
obtained
for a given phase angle, depending on the inclination of the system. 
For instance, polar and equatorial observers both see the planet with a
$90^{\circ}$ phase angle but they do not receive the same spectral irradiance.
In Figure~\ref{fig:contrast} the radius of the planet is kept constant so the
contrast variation at a given wavelength is only due to the observing
geometry. However, the actual mass of the planet depends on the inclination of
the orbit.  In Figure~\ref{fig:phasecurve}, which shows reflection phase curves
for three different inclinations and four spectral bands, we assumed the
following relationship between radius and inclination: $R_{\text{p}} \propto (M/\sin
i)^{0.27}$. 

We can see on these phase curves that low inclinations have the
advantage of keeping the full orbit outside an inner working angle of twice 
the diffraction limit without losing contrast due to the increased planetary 
radius.
Of course, for very low inclinations, the planet could no longer be considered
"Earth-like". For instance, if $i<10^{\circ}$ ($1.5~\%$ of randomly oriented 
orbits) then the
planet would be at least eight times more massive than the Earth. Small phase
angles produce the highest contrast but implies small angular separations and
short wavelengths. For phase angle smaller than $30^{\circ}$, imaging with a
39~m telescope is doable only in bands I and J, but with a contrast larger 
than
$10^{-7}$. For phase angle between 60 and 90$^{\circ}$, the $H$ and $K$ bands
can also be considered but with contrasts below $10^{-7}$. Imaging seems out 
of
reach in the $L$ band ($3-4$~$\mu$m) with a 39-m aperture. Considering the
wavelength-dependency of the contrast, the diffraction limit and the fact that
adaptive optics is challenging in the visible, the $J$ band seems to
represent a promising opportunity.


Sensitivity should not be an issue to directly detect Proxima~b with the 
E-ELT. We calculated the exposure duration required to achieve SNR=10 per 
spectral channel with a spectral resolution of 100, assuming that the angular 
separation
would be sufficient for the noise to be dominated by the sky background
(continuum + emission lines\footnote{We used the ESO documentation for the
background :
\url{http://www.eso.org/sci/facilities/eelt/science/drm/tech_data/background/}})
and not Speckle noise. We assumed an overall throughput of 10\%, integrated
the background over an Airy disk and used the planetary signal derived from 
the
GCM simulations for an inclination of $60^{\circ}$ and a phase angle of
$60^{\circ}$. We obtained integration times of 5~min at $0.76$, $1.25$ and
2.3~$\mu$m, and 30~min at 1.6~$\mu$m.

Assuming that adaptive optics would provide the sufficient efficiency to reach
the required contrast and angular resolution at such short wavelength, an 
O$_2$ signature could be searched at 0.76~$\mu$m. This would require a high
resolution in order to separate the planet lines from the telluric ones taking
advantage of the Doppler shift of the two components. 
\citet{Snellen2015} suggest that a resolution of 100,000 may be 
necessary
for that, which matches that of the planned HIRES instrument\footnote{Documentation for E-ELT instruments HIRES and IFU can be found at \url{https://www.eso.org/sci/facilities/eelt/docs/}}. This would imply
tens of observing nights to reach a SNR of 10 per spectral channel but 
previous
observations \citep[e.g.][]{Snellen2010,Brogi2012,Snellen2014} have shown that a several-$\sigma$ detection of a
high-resolution signature by cross-correlation can be achieved with a much
lower SNR per channel ($<1$) and hence a much shorter integration time. 
\citet{Snellen2015} estimated that it could be doable in 10~hours with the 
instrument IFU/E-ELT.

\begin{figure}
   \centering
   \includegraphics[width=\linewidth]{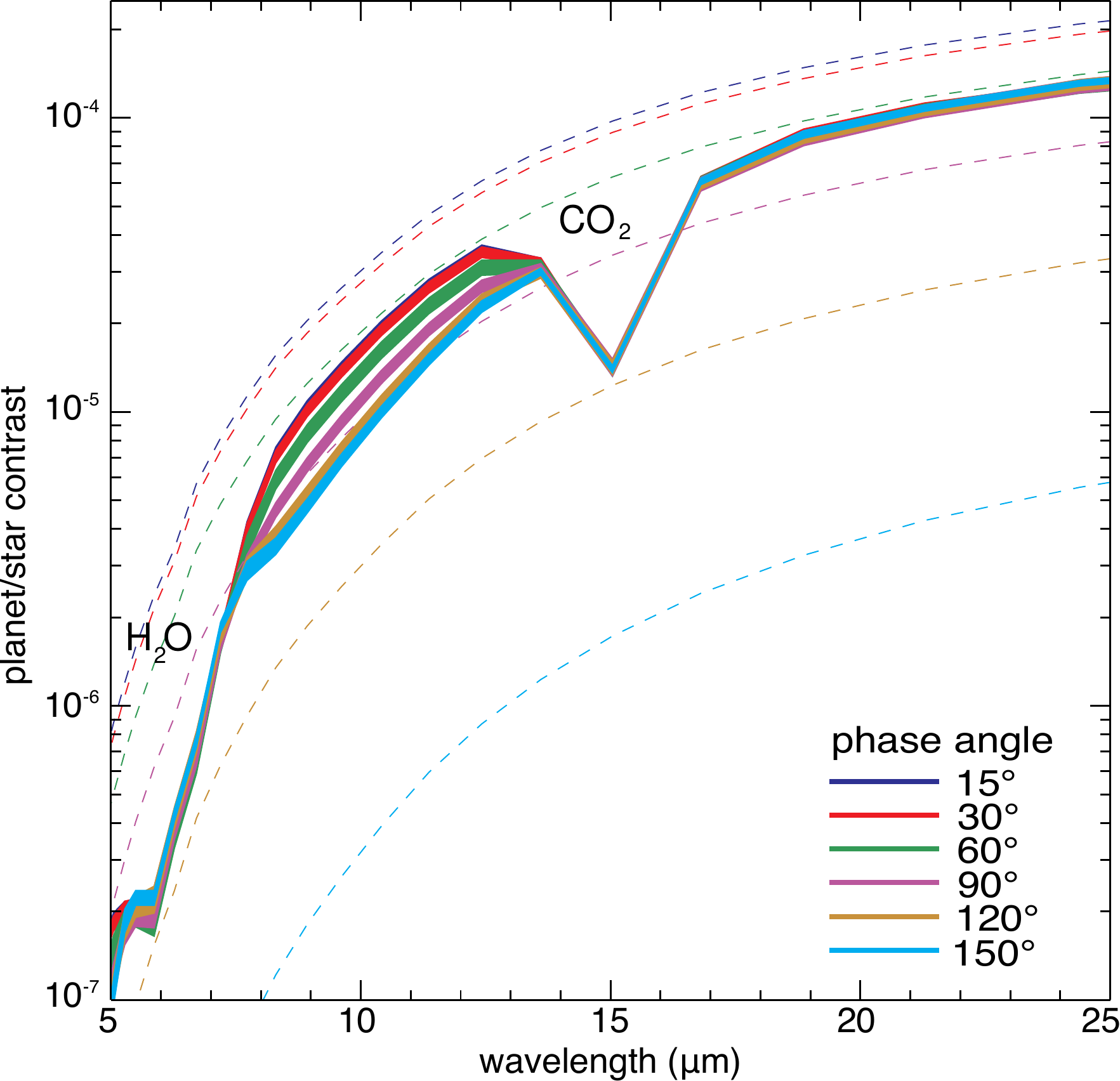}
      \caption{Emission spectra computed for the synchronous case with an Earth-like atmosphere. Each color corresponds to a phase angle. The thickness of the curves indicate the variability associated with inclination. Dashed lines are calculated for a planet with no atmosphere with a constant surface albedo of 0.2. These plots are obtained with a fixed planetary radius of 1.1~R$_\oplus$, wathever the inclination of the orbit.}
         \label{fig:contrast_IR}
   \end{figure}
\begin{figure}
   \centering
   \includegraphics[width=\linewidth]{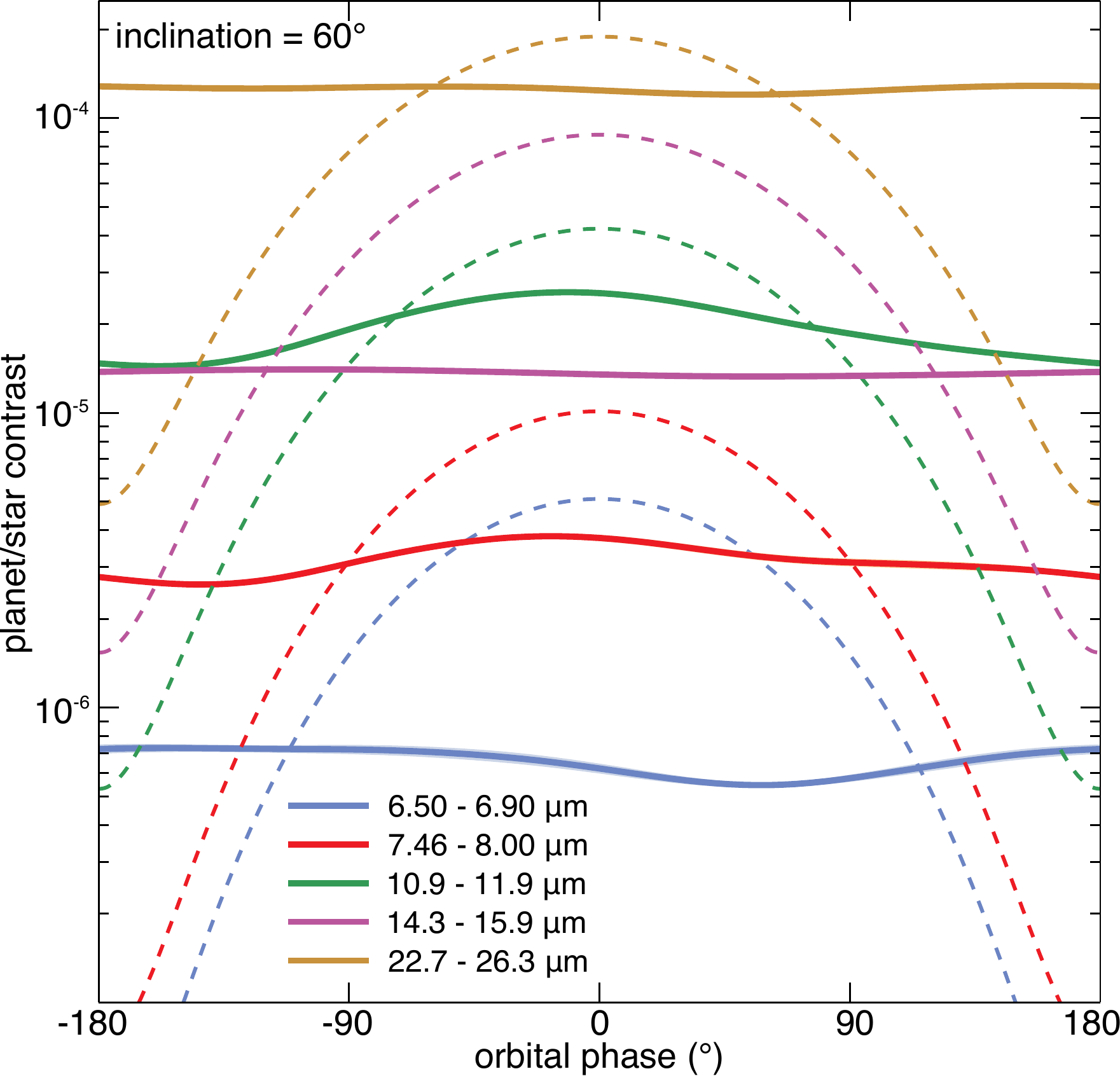}
      \caption{Emission phase curves computed for the synchronous case with an Earth-like atmosphere, for 4 spectral bands used in the GCM. Grey areas indicate the 1-sigma variability due to meteorology (mainly changing cloudiness). Dashed curves are calculated for a planet with no atmosphere with a constant surface albedo of 0.2. Contrary to fig~\ref{fig:contrast_IR}, these plots include a dependency of the planetary radius on the inclination : $R \propto (M/\sin i)^{0.27}$.}
         \label{fig:phase_curves_IR}
   \end{figure}

At 3~$\mu$m, the maximum angular separation of 37~mas corresponds to twice the
diffraction limit of a 39~m aperture. Above this wavelength, imaging Proxima~b
requires larger apertures and the Earth atmosphere (absorption and
emission) becomes a major obstacle. The planet/star contrast at thermal
wavelengths can, however, be orders of magnitude higher than the contrast
produced by reflected wavelengths. The Earth/Sun contrast reaches $\sim 5\times10^{-7}$ at 10-12~$\mu$m but Proxima is a star $\sim 1000$ times dimmer
than the Sun while its planet b could emit about the same as Earth. Contrast
values of up to $\sim 5\cdot10^{-4}$ could thus be expected. In addition,
thermal wavelengths provide a unique way to constrain atmospheric properties
(temperature mapping at different pressure levels, day-night heat
redistribution, greenhouse effect, detection of IR absorbers like H$_2$O,
CO$_2$, O$_3$, CH$_4$). For this reason, space telescopes using IR nulling
interferometry have been considered in the past (Darwin, TPF-I) and
will certainly have to be re-assessed in the future as one of the main ways to
characterize the atmosphere and climate of terrestial planets in nearby
systems. In this context, we computed mid-IR spectra
(Figure~\ref{fig:contrast_IR})  and  thermal phase curves
(Figure~\ref{fig:phase_curves_IR}) obtained with the same simulation used to
produce short wavelength observables. 

The phase curves we obtain are rather flat except in the 8--12~$\mu$m
atmospheric window where the day side emits significantly more than the night
side. Because the flux received by the planet is rather low, the updraft of
clouds and humidity on the day side remains moderate and restricted within a
small region around the substellar point. For this reason we do not find that
most of the cooling occurs on the night side as found by \citet{Yang2013},
\citet{GomezLeal2012} or \citet{Bolmont2016} for planets with an Earth-like
irradiation.



  Considering the history of the planet, and in particular its hot past, an Earth-like atmosphere may not be the most relevant case to address observation prospects. In the Appendix~\ref{appendix:obs} we present spectra and phase curves obtained with different compositions and rotations. As the planet could have experienced a fate similar to Venus \citep{Ingersoll1969}, we also present reflection spectra and phase curves for a Venus-like atmosphere \citep{LEL15}, including sulphur-bearing aerosols known to produce a high albedo at visible wavelength, and then exposed to the irradiance of Proxima at the orbital distance of Proxima~b. 
\subsection{Prospects with JWST}
\begin{figure}
   \centering
   \includegraphics[width=\columnwidth]{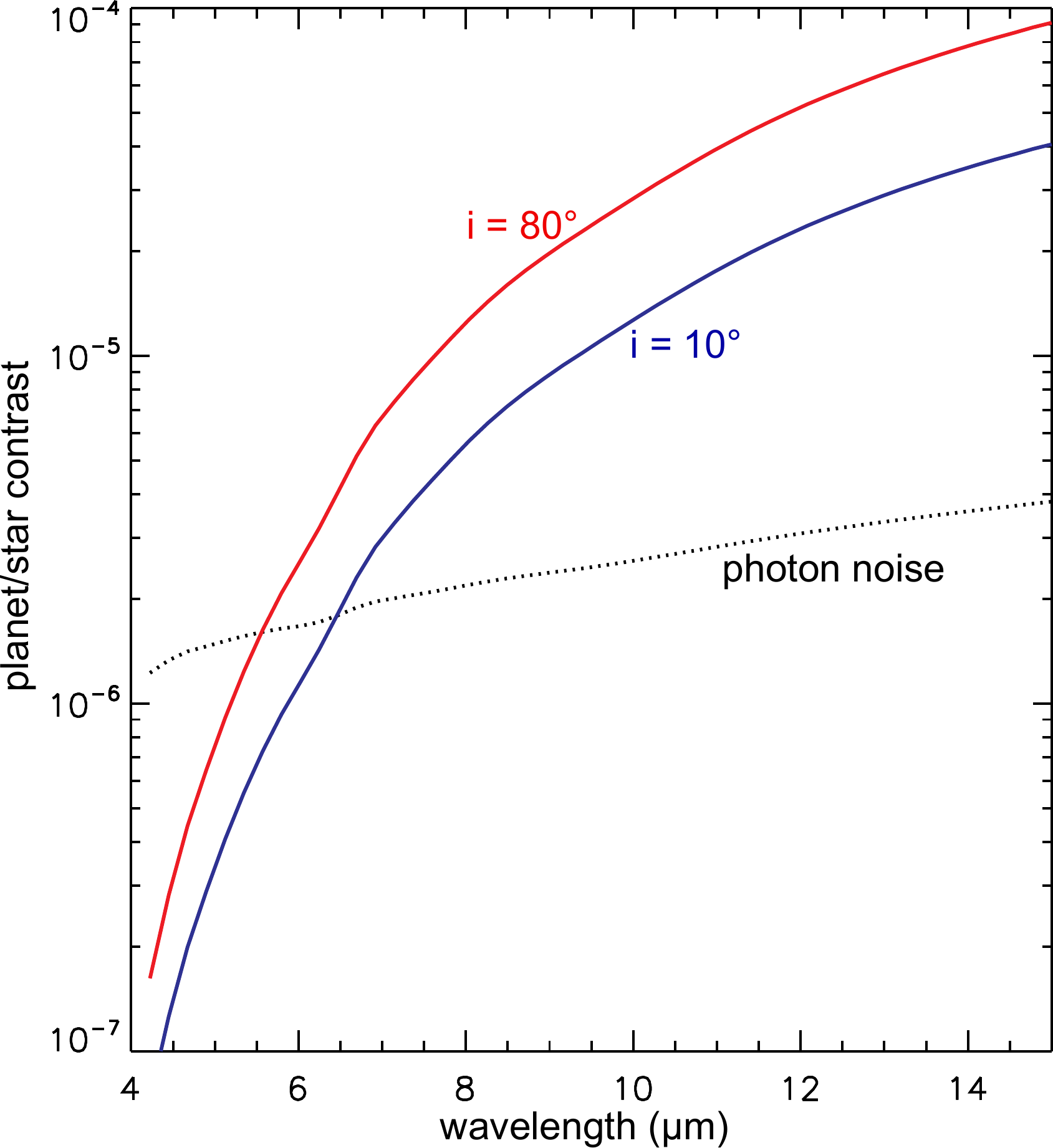}
    \caption{Observability of the photometric modulation due to thermal phase curves with JWST. The amplitude of the modulation is calculated for a planet with no atmosphere and a Bond albedo of 0.2 and with a radius that scales with the inclination as : $R \propto (M/\sin i)^{0.27}$. The noise is computed for a 1~hr exposure, a spectral resolution of 10, and the collecting area of the JWST (25~m$^2$).}
\label{fig:jwst}
\end{figure}

Observing the modulation due to thermal phase curves does not require a transit and has been achieved by photometry in the case of both transiting and non-transiting hot Jupiters \citep{Crossfield2010}. Although very challenging, in particular due to the star variability, this observation can be attempted with JWST (James Webb Space Telescope). The planet-star contrast in the mid-IR can reach $10^{-4}$ but only the amplitude of the modulation can be detected. This amplitude depends strongly on the thickness of the atmosphere as shown by the phase curves presented in the Appendix~\ref{appendix:obs}. Dense atmospheres lower the day-night temperature contrast and therefore produce rather flat lightcurves and modulation below $10^{-5}$ in contrast. On the other hand, planets with no or a tenuous atmosphere ($<10$~mbar) produce contrast amplitudes of $10^{-5}-10^{-4}$ in the wavelength range $8-15~\mu$m. Fig~\ref{fig:jwst} shows the amplitude of the contrast modulation for a planet with no atmosphere, in radiative equilibrium, and for two inclinations. The modulation decreases with the inclination but this decrease is compensated by the increase of the mass and thus the radius. One can see that 1~hr exposure with the JWST at R=10 allows to beat the stellar photon noise above $6~\mu$m. According to \citet{Belu2011}, the total noise is usually within 2-3 times the stellar photon noise for wavelengths below $15~\mu$m (above this limit the thermal emission from the telescope dramatically degrades the observations). 
Detecting these modulations with JWST would be extremely challenging due to stellar variability and flares. But flux variations are smaller in the infrared and the orbital period and ephemeris of the planet are known, which considerably helps planning short exposure over several orbits, in particular near the peak at superior conjunction. Measuring a modulation would point to planet with no dense atmosphere like Mercury or Mars. In theory, measurement at different wavelengths could be used to find atmospheric signatures \citep{selsis2011}, constrain the radius, the albedo and the inclination of the planet \citep{Maurin2012}, as well as its rotation \citep{Selsis2013}.

\section{Discussions}

The modelling work performed here to explore the possible climates and observability of Proxima b remains speculative. Major suprises in the composition of the atmosphere or the nature of the planet cannot be discarded. Within the known uncertainties, we can list the following points:

\begin{enumerate}
\item \textbf{The luminosity of Proxima Centauri is not perfectly known:} Using  interferometry with two VLT telescopes, \citet{Demory2009} measured the
radius of Proxima~b as well as its effective temperature and found $R$=$0.141 \pm
0.007\,R_{\odot}$ and $T_{eff}=3098 \pm 56$~K, which yields a bolometric luminosity
of $0.00165 \pm 0.00012\,L_{\odot}$. In this study we used a value of
0.0017\,$L_{\odot}$ while \citet{Anglada16} give 0.00155$\,L_{\odot}$ as a median
value, derived from \citet{Boyajian2012}. Both values are within the uncertainty of
\citet{Demory2009} and \citet{Boyajian2012}. Although changing  the actual
bolometric flux received by Proxima~b would slightly alter surface and atmosphere
temperatures found for a given atmospheric composition, this discrepancy does not
impact our results, qualitatively. Changing the actual bolometric flux received by
Proxima~b would of course alter the detailed relationship between atmospheric
pressure and temperature. Considering the importance of this system, the community
should agree on a standard and calibrated irradiance spectrum of Proxima to be used
for climate and habitability studies.

\item \textbf{The amount of background gas:} Some simulations in this work included background N$_2$ while others did not. It would require dedicated absorption coefficients to perform all the GCM simulations with a fixed N$_2$ partial pressure. Moreover, we have bad constraints on the amount of background gas available on the planet due to (1) uncertainties 
on the mass of Proxima~b and (2) possibility that the background gas was lost to space due to the high XUV 
flux of Proxima Centauri \citep{Ribas2016}.

\item \textbf{The convection scheme:} It should be kept in mind that models like \citet{Yang2013} find higher albedos in the substellar area due to a different convection scheme and possibly different cloud parametrizations. 
As Proxima~b is moderately irradiated, this discrepancy might not be as important as for the near-inner edge 
cases studied by \citet{Yang2013}.

\item \textbf{The oceanic circulation:} We remind that our results in the aquaplanet regime neglected the effect of oceanic transport. This is an important era of future improvement, although it adds several new unconstrained ingredients, such as the presence and location of continents. 

\item \textbf{Climate retroactions:} We assumed in this study that the amount of volatiles (H$_2$O, CO$_2$ and N$_2$) should be uncorrelated and therefore that each configuration of volatile inventory could be achievable by Proxima~b. In fact, climate retroactions such as the carbonate-silicate cycle \citep{Walk:81} could favor some of them by linking the amount of carbon dioxide and water.

\end{enumerate}

\section{Conclusions}

In this study, we explored the possible climates of Proxima Centauri b for a wide range of volatile inventories of water, carbon dioxide and nitrogen. It appears from our results that the habitability of the planet is possible for a very broad range of atmospheric pressures and compositions, as shown by the size of the blue regions in Figure~\ref{proxima_diagrams}. 

In a nutshell, the presence of surface liquid water requires either a large surface inventory of water (a global ocean able to resupply H$_2$O to the dayside by deep circulation) or an atmosphere with a strong enough greenhouse effect that increases surface temperatures above the freezing point of water \textit{everywhere}. 

Apart from receiving the necessary insolation, this study tells us that the key ingredient to the habitability of a planet is the retention of all volatiles---water of course, but also non-condensible gases (with a greenhouse effect or not) to warm surface cold-traps.
\citet{Ribas2016} showed that it is possible that the planet lost large amounts of water.
However, this work shows that even with extremely low amounts of water (a few $10^{-3}$ Earth ocean content for the synchronous case and less than $10^{-5}$ for the asynchronous case), there are CO$_2$ pressures which allow surface liquid water.

More generally, these conclusions are not restricted to the case of Proxima-b. In fact, any low-obliquity planet within the classical habitable zone of its star should be in one of the climate regimes discussed here, although the limits between the various regimes would shift quantitatively with the planet parameters (e.g. the insolation).

Prospects for direct imaging with E-ELT are extremely promising: the star-planet separation reaches $9.5~\lambda/D$ à 760~nm (wavelength of the O$_2$ band) with a contrast of $0.9-5\times 10^{-7}$ (depending on atmospheric/surface composition and planetary radius) and 
$3.6~\lambda/D$ à 2~$\mu$m with a contrast of $10^{-7}-10^{-8}$. J band (1.1-1.4~$\mu$m) offers a fine trade-off in terms of separation, contrast and constraints on adaptive optics. 
The brightness of the planet should allow high resolution spectroscopy and the search for a variety of molecular signatures including O$_2$, H$_2$O, CO$_2$, CH$_4$. 

Thermal phase curve modulations are observable - in theory - with JWST, with a contrast of $\sim 10^{-5}$ at 10~$\mu$m but will be challenging due to stellar variability.
More accurate mid-IR spectroscopy would probably require space-based interferometry.

Most of our knowledge on planetary atmospheres and habitability come from the study of Venus, Mars and Earth. Proxima~b could potentially be the fourth terrestrial planet to confront all that we know on these domains.



\begin{acknowledgements}
M.~T. thanks S\'ebastien Lebonnois for his help on the modelling of a venusian atmosphere with the LMD Generic Model.
E.~B. acknowledges that this work is part of the F.R.S.-FNRS ``ExtraOrDynHa'' research project.
I.~R. acknowledges support from the Spanish Ministry of Economy and Competitiveness (MINECO) through grant ESP2014-57495-C2-2-R. 
S.~N.~R. acknowledges support from the Agence Nationale de la Recherche via grant ANR-13-BS05-0003-002 (project MOJO).
\end{acknowledgements}

%
%

   \bibliographystyle{aa} 
   \bibliography{biblio} 


\begin{appendix}

\section{Computation of maximal ice thickness before basal melting} \label{appendix:melting curve}

Because the ice thermal conductivity can vary substantially with temperature according to the relation $\lambda_{\text{ice}}(T)$=$A/T$ with A=651~W~m$^{-1}$ \citep{Petrenko2002}, the temperature profile inside an ice layer in equilibrium follows an exponential law from which we derive a maximum thickness $h_{\text{ice}}^{\text{max}}$ before melting \citep{Abbot2011}:
\begin{equation}
\label{thick_ice}
\centering
h_{\text{ice}}^{\text{max}}~=~\frac{A}{F_{\text{geo}}}~\ln~\left(\frac{T_{\text{melt}}}{T_{\text{surf}}}\right).
\end{equation}
$F_{\text{geo}}$ is the internal heat flux. $T_{\text{melt}}$ is the melting temperature of ice at the base of the glacier. For pressure lower than 100~bars (ice thickness lower than $\sim$~1~km), $T_{\text{melt}}$ is roughly constant and equal to 273~K. However, this assumption does not work for higher pressures. Thus, we use the following parametrization for the melting curve of ice \citep{Wagner1994}: 
\begin{equation}
\label{p-melt}
\begin{split} 
P_{\text{melt}}(T)~=~P_{\text{ref}}~\Bigg[1-0.626\times10^6\left(1-\left(\frac{T}{T_{\text{ref}}}\right)^{-3}\right) \\ +0.197135\times10^6\left(1-\left(\frac{T}{T_{\text{ref}}}\right)^{21.2}\right)\Bigg],
\end{split} 
\end{equation}
with $T_{\text{ref}}$/$P_{\text{ref}}$ the temperature/pressure of the triple point of water.

Using the relation $h_{\text{ice}}^{\text{max}}$~=~$\frac{P_{\text{melt}}-P_{\text{surf}}}{\rho_{\text{ice}}~g}$ and following a similar approach to \citet{Leco:13}, we can solve implicitly (and numerically) the system of equations (\ref{thick_ice}, \ref{p-melt}) and find the thickness at which melting occurs at the base of the glaciers.

\section{Spectra and phase curves} \label{appendix:obs}
We present here synthetic spectra and phase curves obtained with some of the GCM simulations. 
Reflected and emitted spectra are obtained as in Fig.~\ref{fig:contrast} and Fig.~\ref{fig:contrast_IR}, 
respectively. Reflection and thermal phase curves are computed as in Fig.~\ref{fig:phasecurve} and Fig.~\ref{fig:phase_curves_IR}, 
respectively, for a 60$^{\circ}$ inclination and a radius R$=
(M_{min}/\sin 60^{\circ})^{0.27} = 1.11~$R$_\oplus$. Color codes for phase angles and wavelengths are the same as for figures in the main text.

To describe each different type of observables (reflected spectra, reflected lightcurves, thermal spectra, thermal lightcurves), we will start from dry cases with tenuous atmosphere, which exhibit the most simple features, and progress towards dense and humid atmospheres that combine more effects, including clouds.

\subsection{Reflected spectra} 
\begin{figure*}[h]
    \centering
        \includegraphics[width=0.432\linewidth]{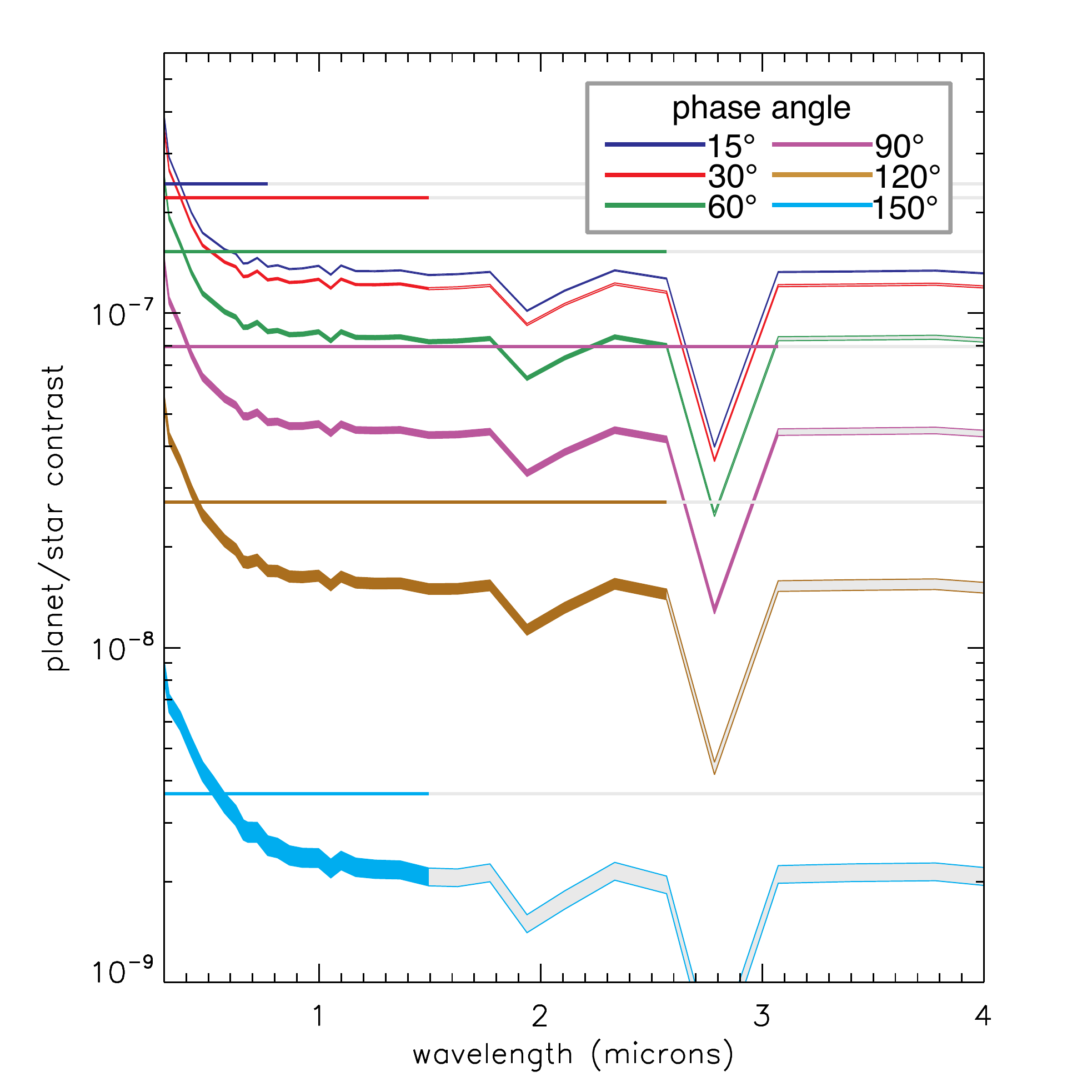}
        \includegraphics[width=0.47\linewidth]{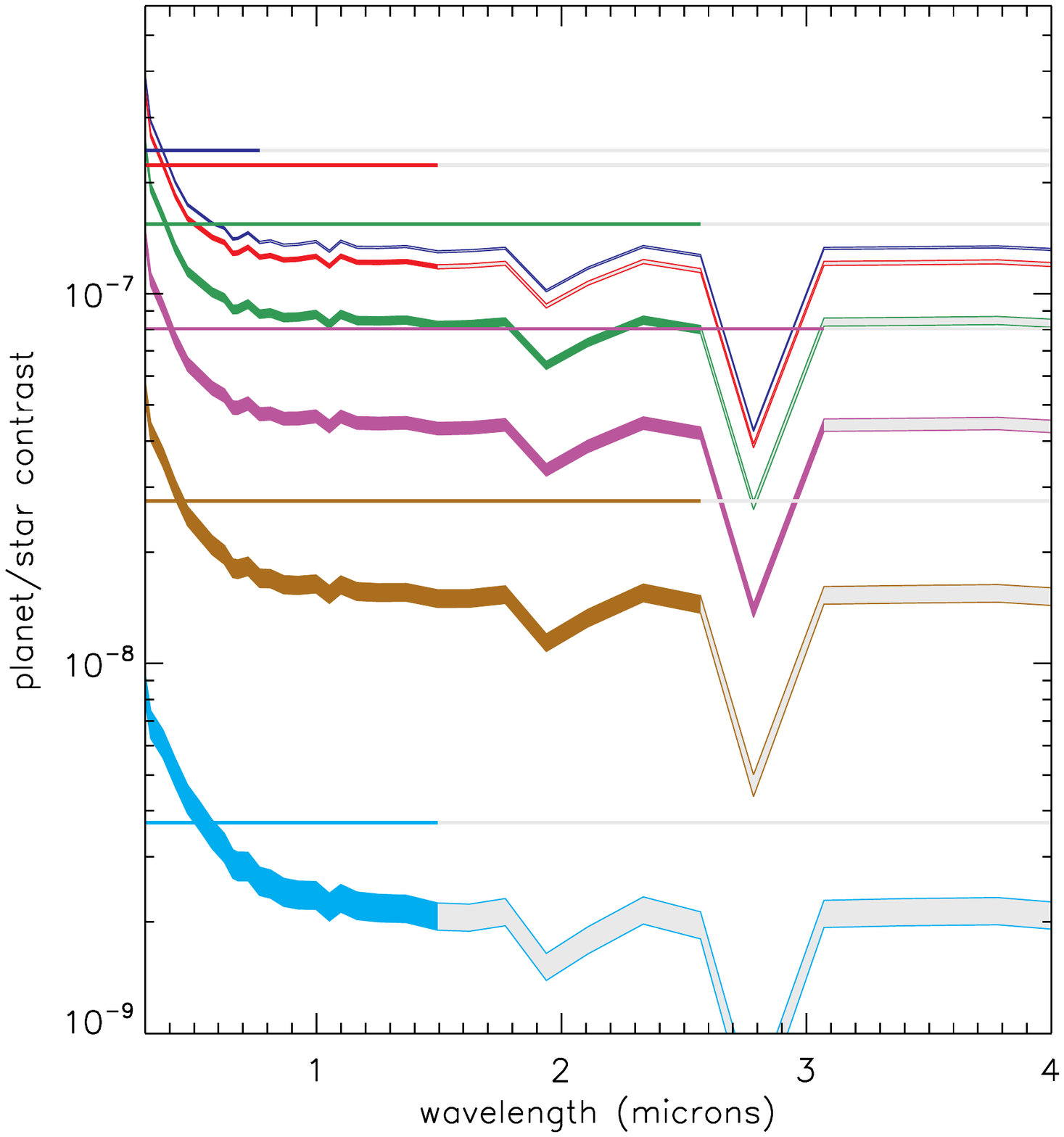}
        \caption{Reflection spectra computed for a dry planet with an 
Earth-like atmosphere in synchronous rotation (left) and in 3:2 spin-orbit resonance (right).}
\label{fig:VIspec:dryEarth}
\end{figure*}

\begin{figure*}[h]
    \centering
        \includegraphics[width=0.432\linewidth]{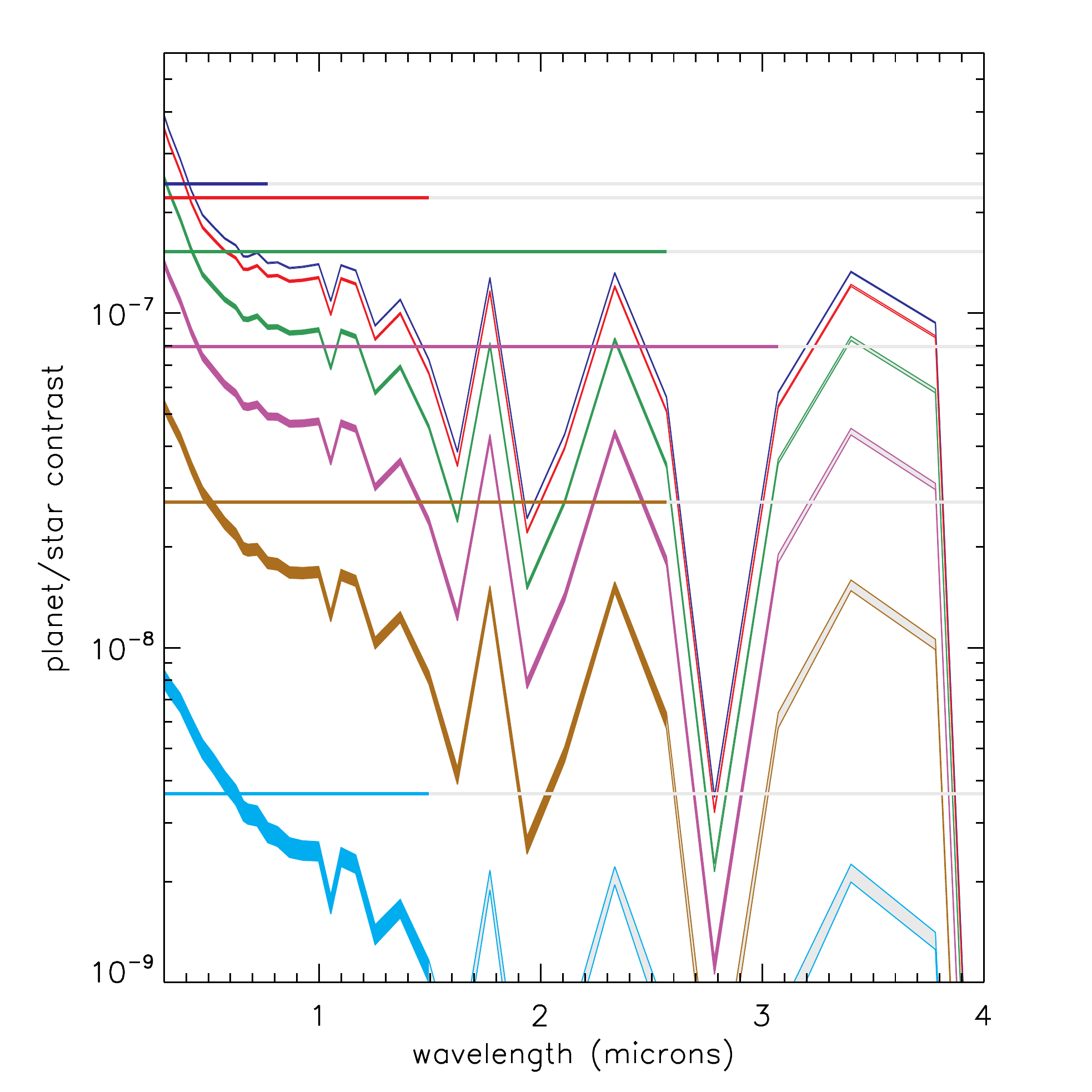}
        \includegraphics[width=0.47\linewidth]{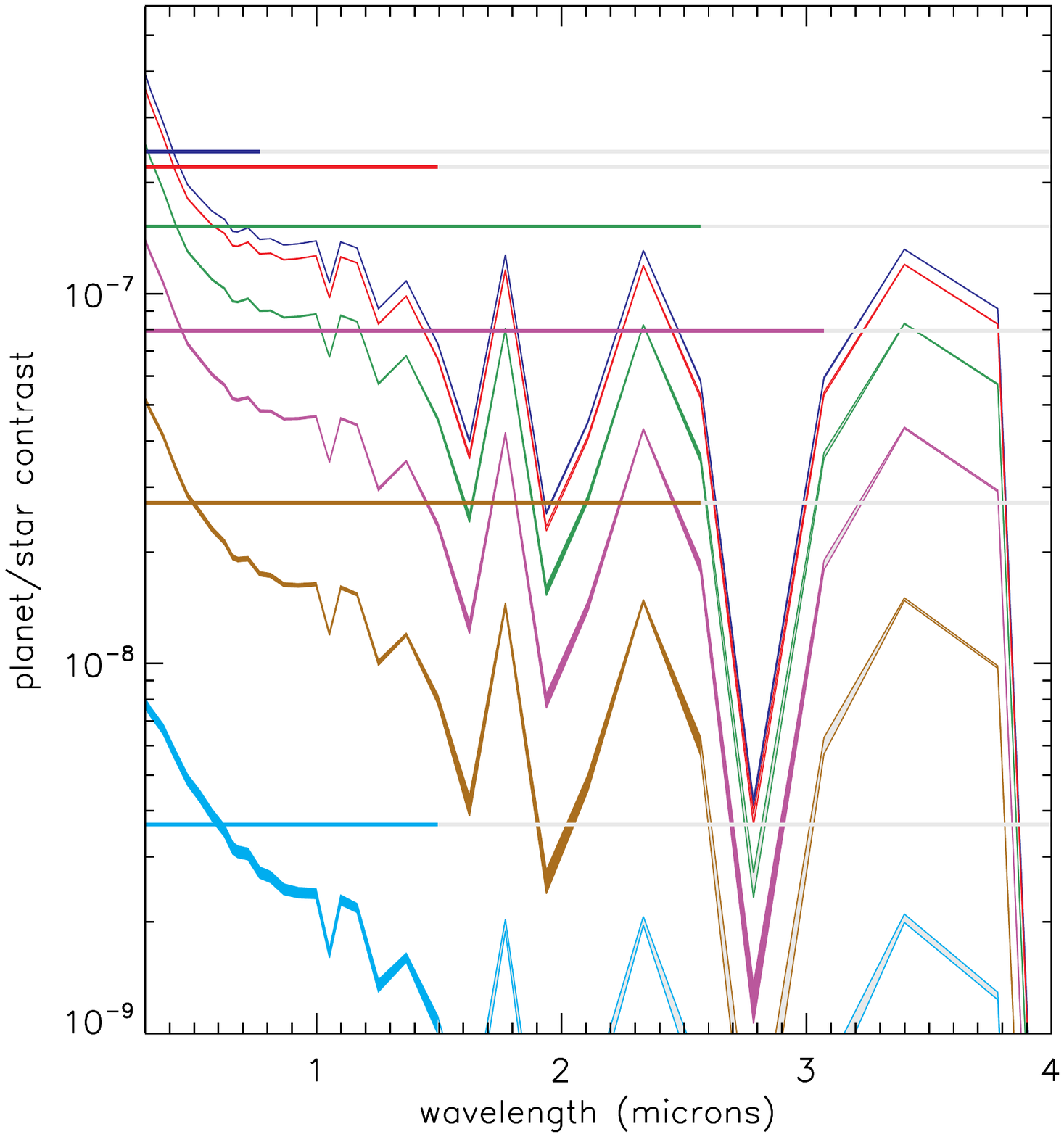}
        \caption{Reflection spectra computed for a dry planet with a 
1~bar CO$_2$-dominated atmosphere in synchronous rotation (left) and in 3:2 spin-orbit resonance (right).}
\label{fig:VIspec:dryCO2}
\end{figure*}

\begin{figure*}[h]
    \centering
        \includegraphics[width=0.47\linewidth]{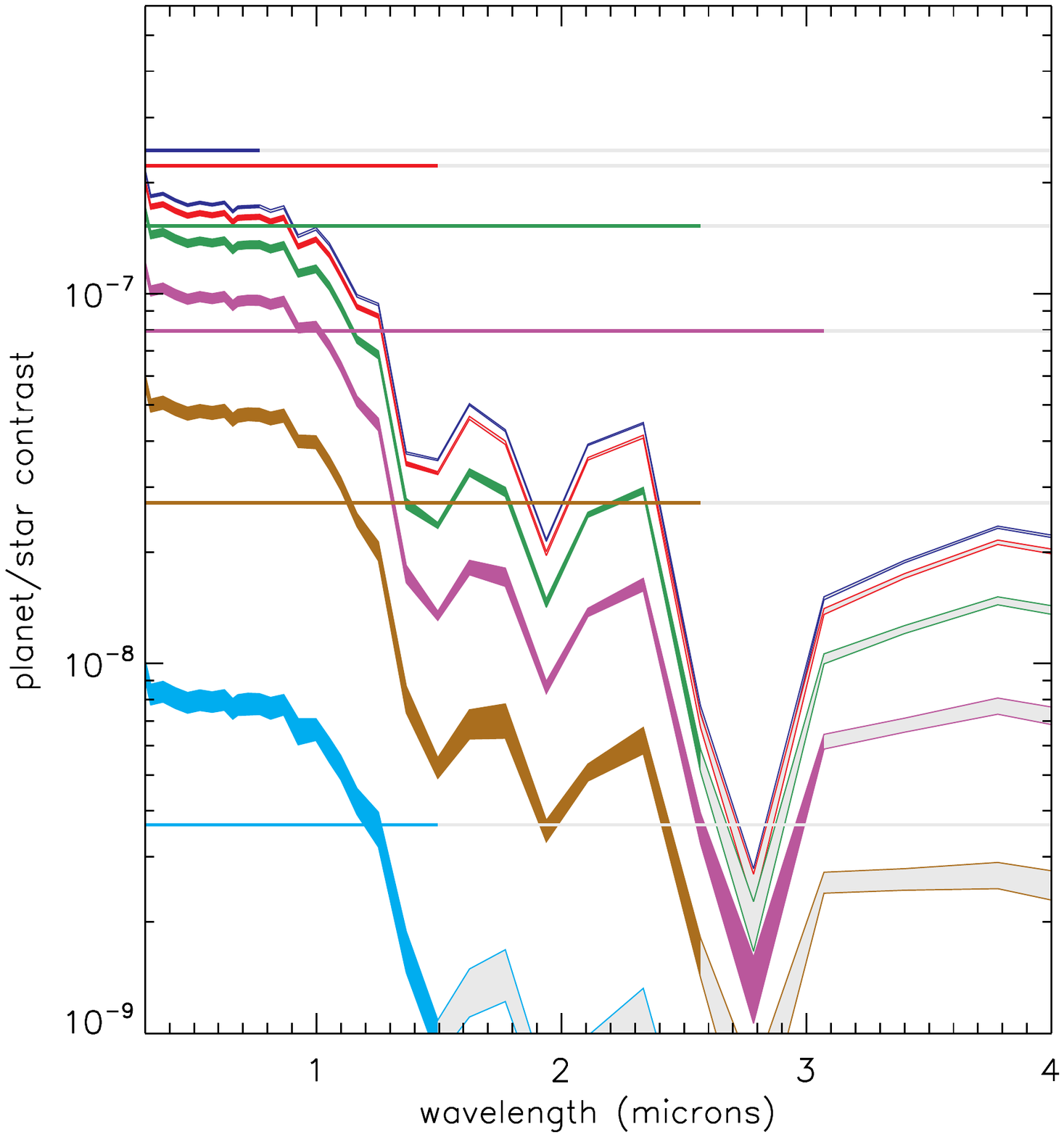}
        \includegraphics[width=0.47\linewidth]{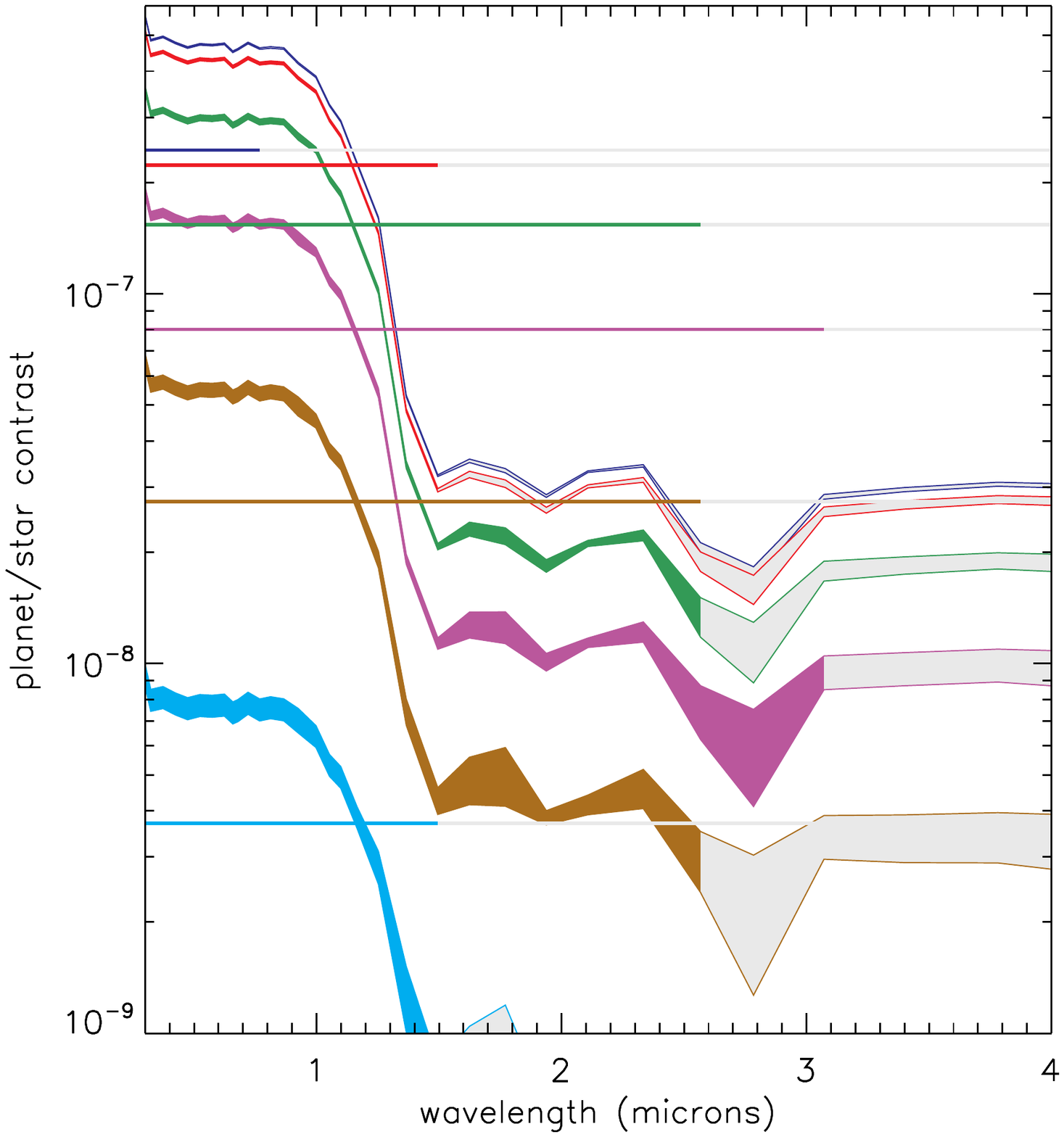}
        \caption{Reflection spectra computed for an aquaplanet with a 
10~mbar N$_2$-dominated (+~376ppm of CO$_2$) atmosphere in synchronous rotation (left) and in 3:2 spin-orbit resonance (right).}
\label{fig:VIspec:tenuousEarth}
\end{figure*}

\begin{figure*}[h]
    \centering
        \includegraphics[width=0.432\linewidth]{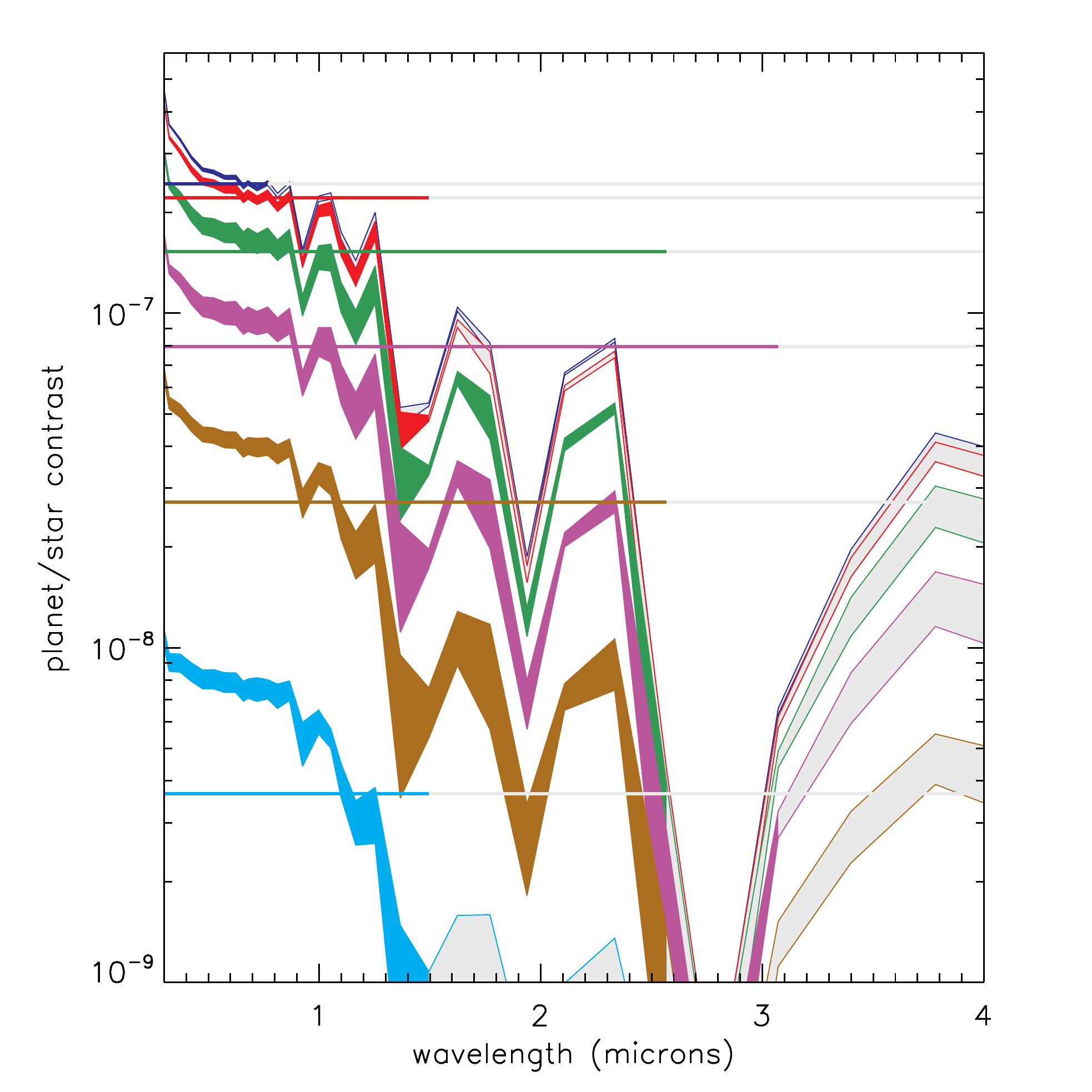}
        \includegraphics[width=0.47\linewidth]{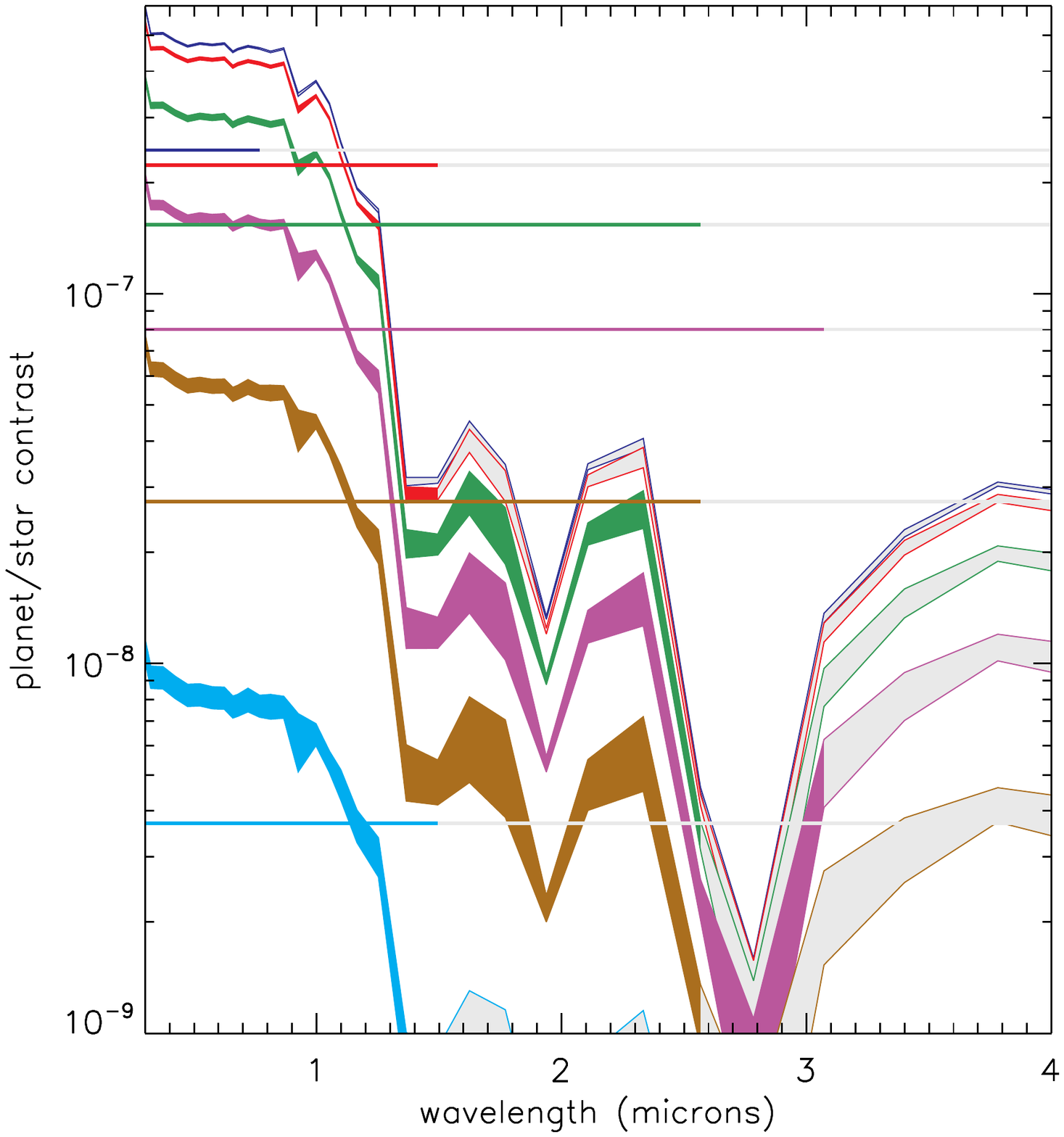}
        \caption{Reflection spectra computed for an aquaplanet with an 
Earth-like atmosphere in synchronous rotation (left) and in 3:2 spin-orbit resonance (right).}
\label{fig:VIspec:Earth}
\end{figure*}

\begin{figure*}[h]
    \centering
        \includegraphics[width=0.47\linewidth]{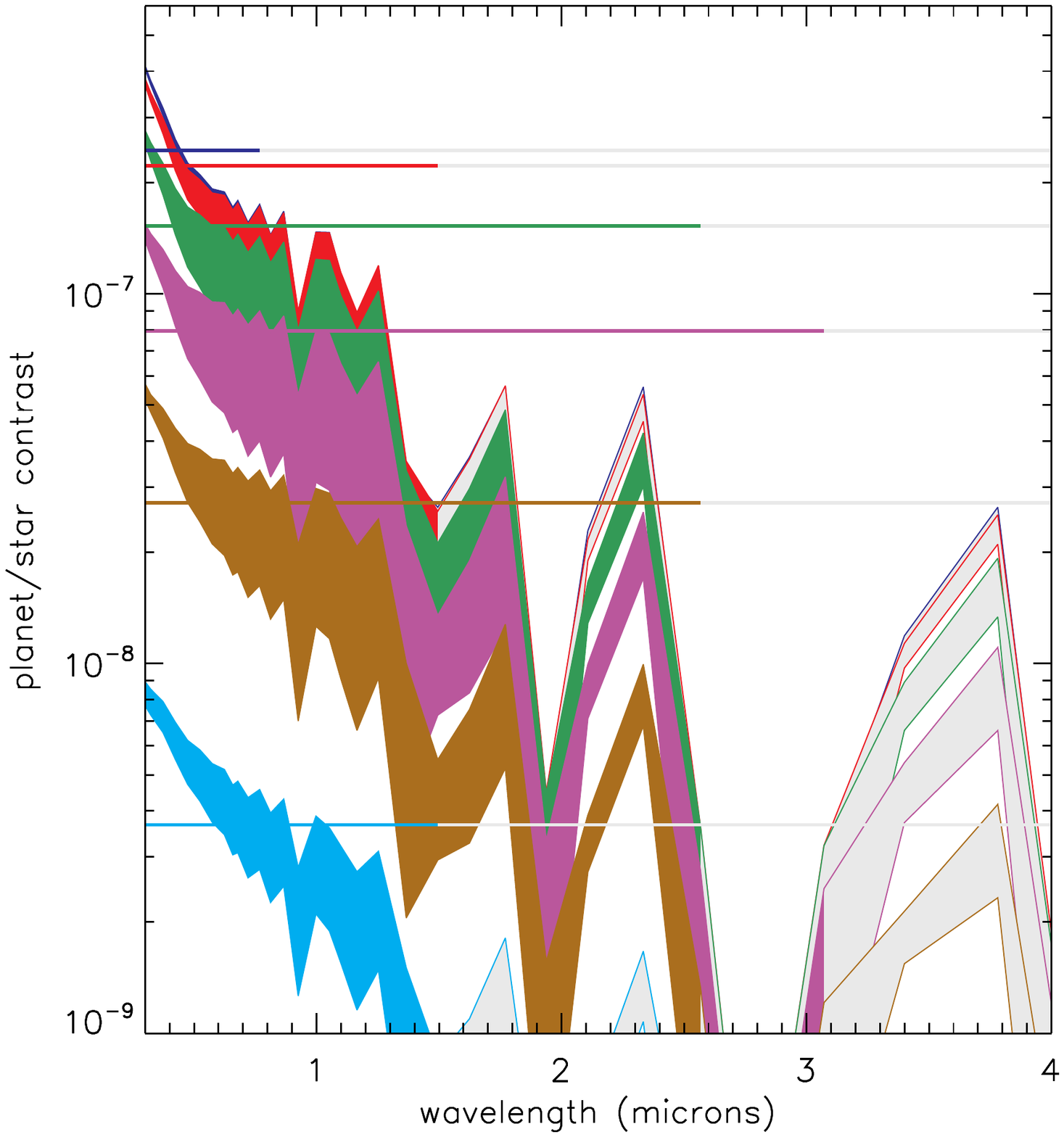}
        \includegraphics[width=0.47\linewidth]{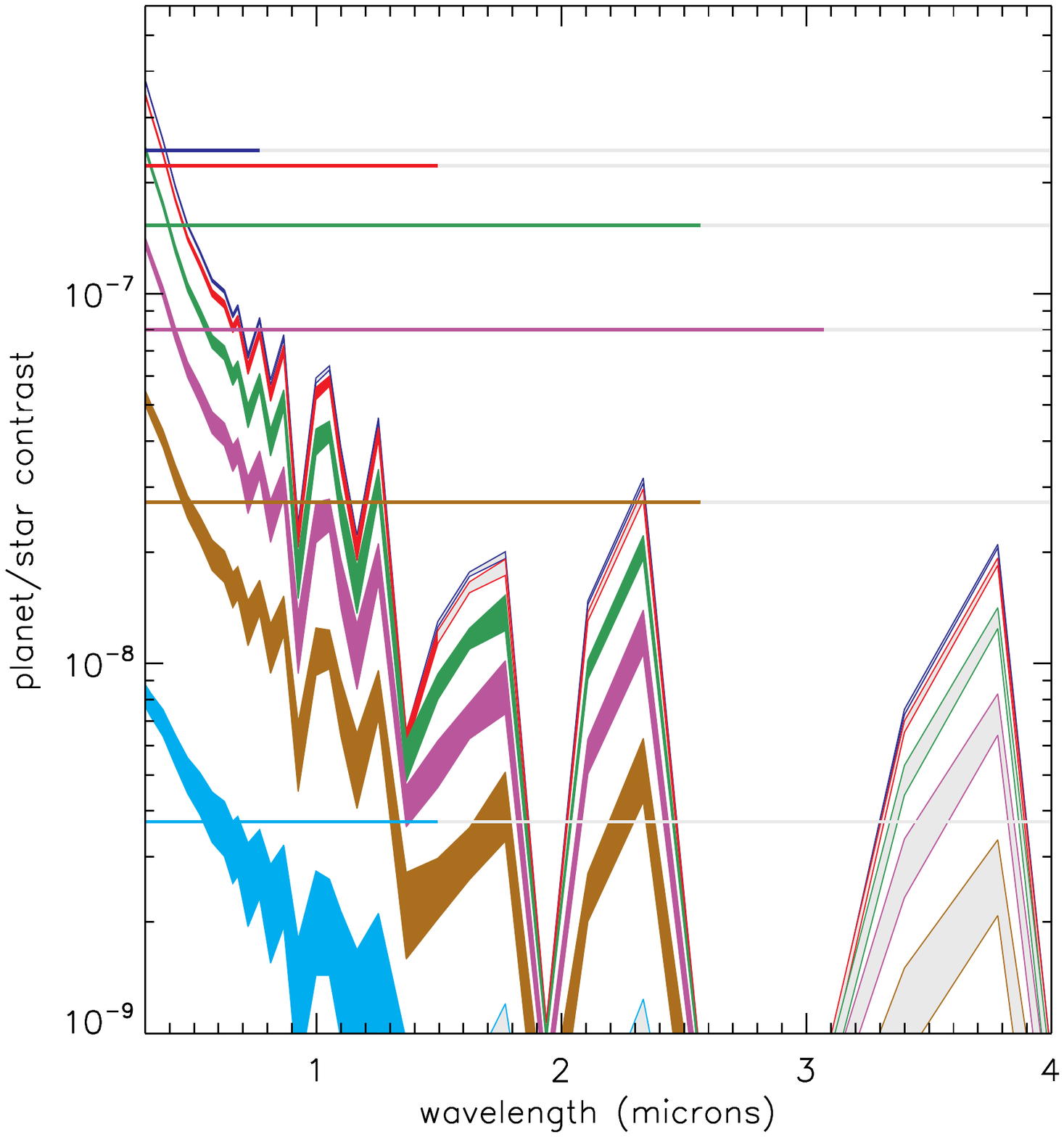}
        \caption{Reflection spectra computed for an aquaplanet with a 
1~bar CO$_2$-dominated atmosphere in synchronous rotation (left) and in 3:2 spin-orbit resonance (right).}
\label{fig:VIspec:CO2}
\end{figure*}

Reflected spectra are shaped by Rayleigh scattering from the gas, Mie scattering from clouds, molecular absorption features and surface reflectivity. \\
\noindent - Dry case, Earth-like atmosphere (Fig.~\ref{fig:VIspec:dryEarth}): we can see the decrease of the albedo with increasing wavelengths in the UV-visible domain due to the combination of Rayleigh scattering and the constant surface albedo of 0.2. CO$_2$ absorption features can be seen at 1.9 and 2.6-2.7~$\mu$m. The rotation mode of the planet does not affect the observables.\\
\noindent - Dry case, 1 bar of CO$_2$ (Fig.~\ref{fig:VIspec:dryCO2}): same as above with more and deeper CO$_2$ features.\\
\noindent - Aquaplanet, mainly frozen, 10 mbar of N$_2$; 376 ppm of CO$_2$ (Fig.~\ref{fig:VIspec:tenuousEarth}): Rayleigh scattering is negligible and the only atmospheric features are the H$_2$O bands. The drop of albedo between 1 and 1.5~$\mu$m is produced by the wavelength-dependent reflectivity of ice included in the model (Fig.~\ref{spectrum_proxima}). This drop is attenuated in the synchronized case because the dayside is partly covered by liquid water (Fig~\ref{aquaplanet_case}) whose reflectivity is constant with wavelength ($\sim$~7$\%$) and have a value close to that of ice in the infrared ($\sim$~5$\%$). For this reason, the overall albedo is also higher for the 3:2 rotation, the surface being mostly covered by ice. \\
\noindent - Aquaplanet, mainly frozen, 1 bar of N$_2$; 376 ppm of CO$_2$ (Fig.~\ref{fig:VIspec:Earth}): In addition to the features described in the previous case, this configuration also exhibits signatures of the larger water vapor content due to higher temperatures: deeper H$_2$O absorption bands and a larger spatial/temporal variability due to meteorology. In synchronous rotation, the Rayleigh slope can be seen but not in the non-synchronous case because the albedo of the icy surface dominates over the albedo due to Rayleigh scattering.\\
\noindent - Aquaplanet, 1 bar of CO$_2$ (Fig.~\ref{fig:VIspec:CO2}): this case exhibits the strongest molecular absorptions due to large columns of both CO$_2$ and water vapor. The low albedo of the liquid water surface reveals the atmospheric Rayleigh slope at wavelength lower than 600~nm. The spectrum of the synchronous case is very sensitive to the observing geometry for a given phase angle. As seen in Fig~\ref{fig:cloudmap1barCO2}, this is due to the concentration of clouds at low latitude and eastward of the substellar point in the synchronous case while the 3:2 case has very uniform cloud coverage.

\subsection{Reflected phase curves}
\begin{figure*}[h]
    \centering
        \includegraphics[width=0.47\linewidth]{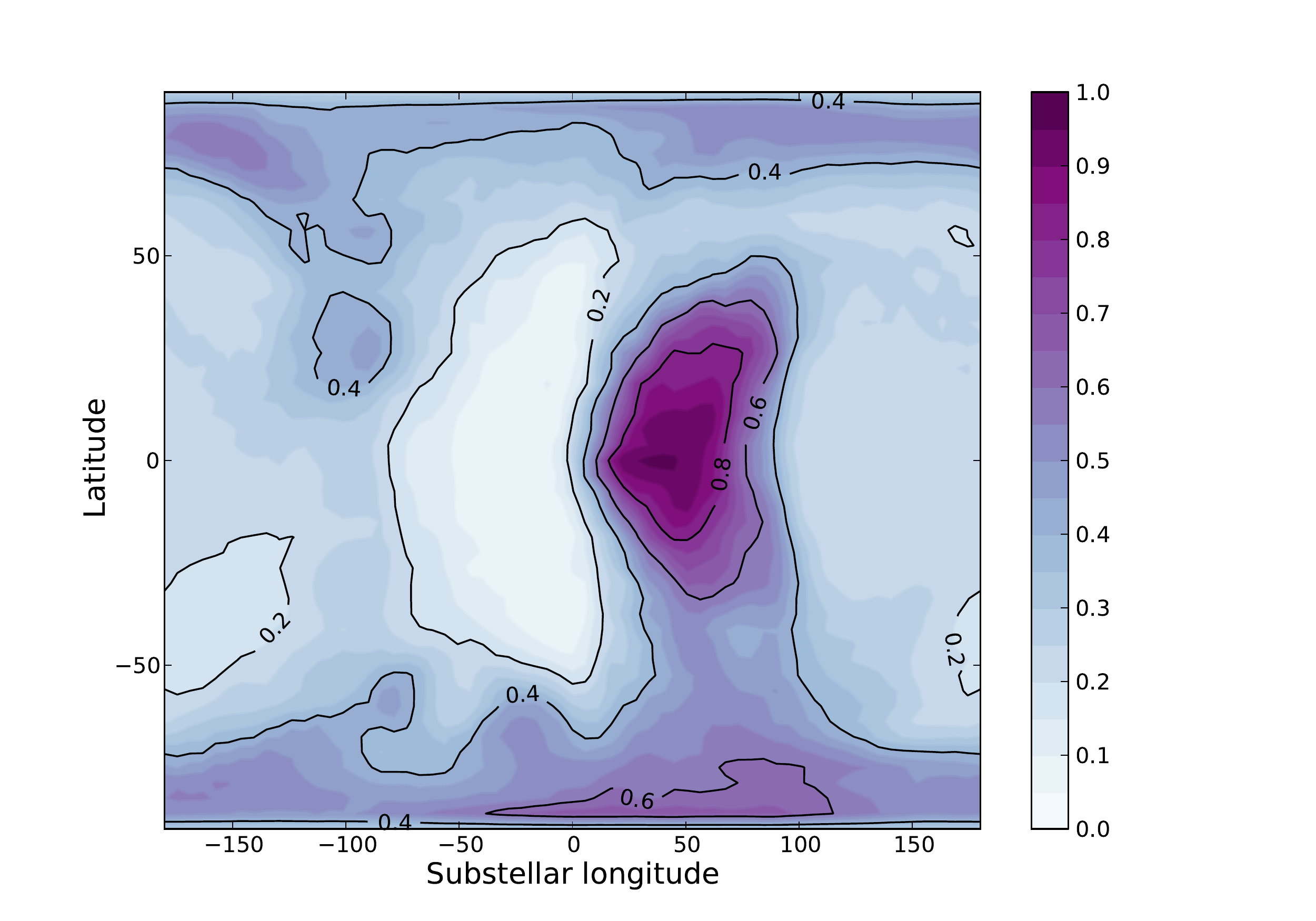}
        \includegraphics[width=0.47\linewidth]{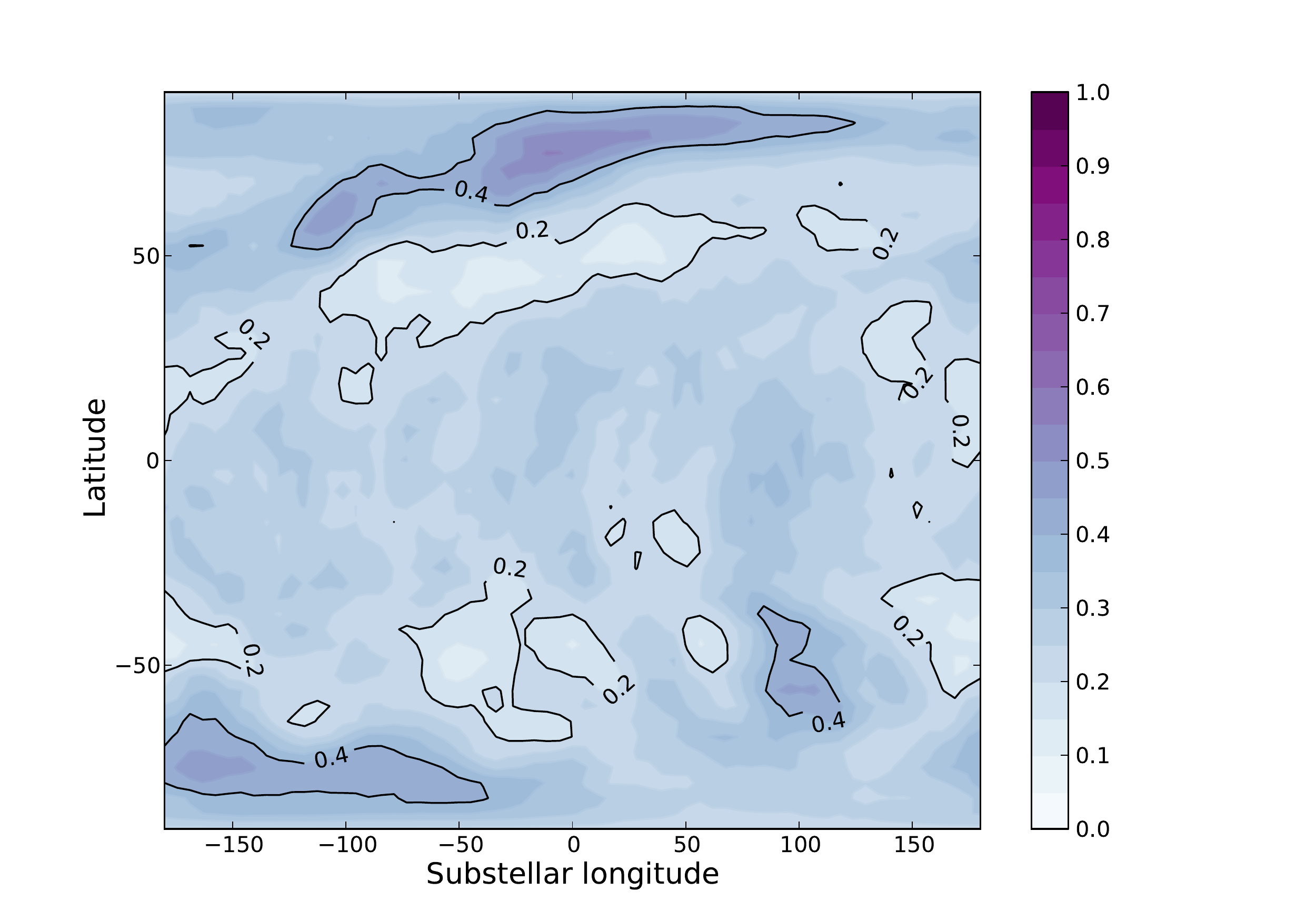}
        \caption{Cloud maps for an aquaplanet with a 1 bar CO$_2$ atmosphere. Left : synchronous rotation. Right : 3:2 spin-orbit resonance. In both cases, longitude is given relative to the substellar point. Colors indicate the fractional cloud cover, averaged over 2 orbits. }
        \label{fig:cloudmap1barCO2}
\end{figure*}

\begin{figure*}[h]
    \centering
        \includegraphics[width=0.432\linewidth]{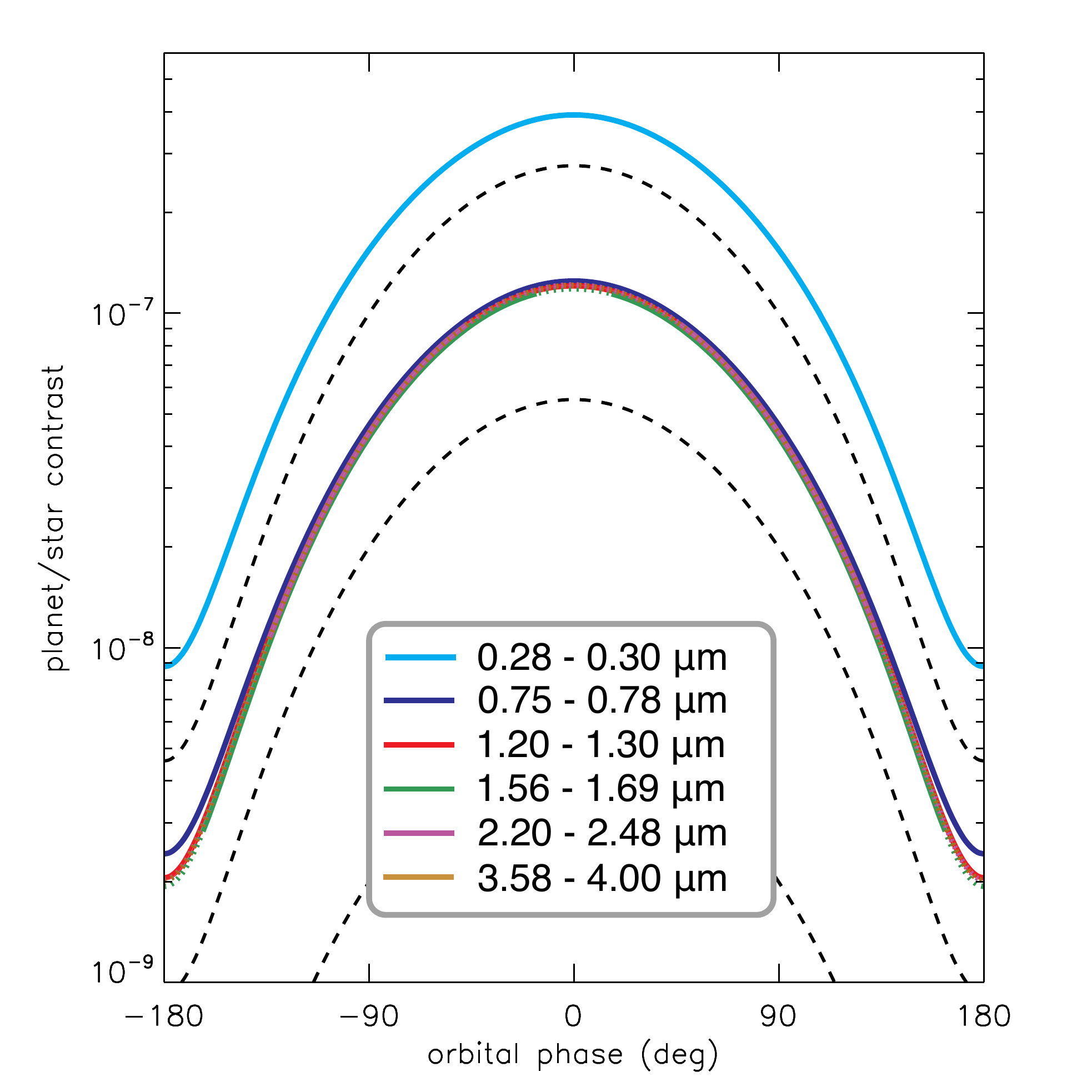}
        \includegraphics[width=0.47\linewidth]{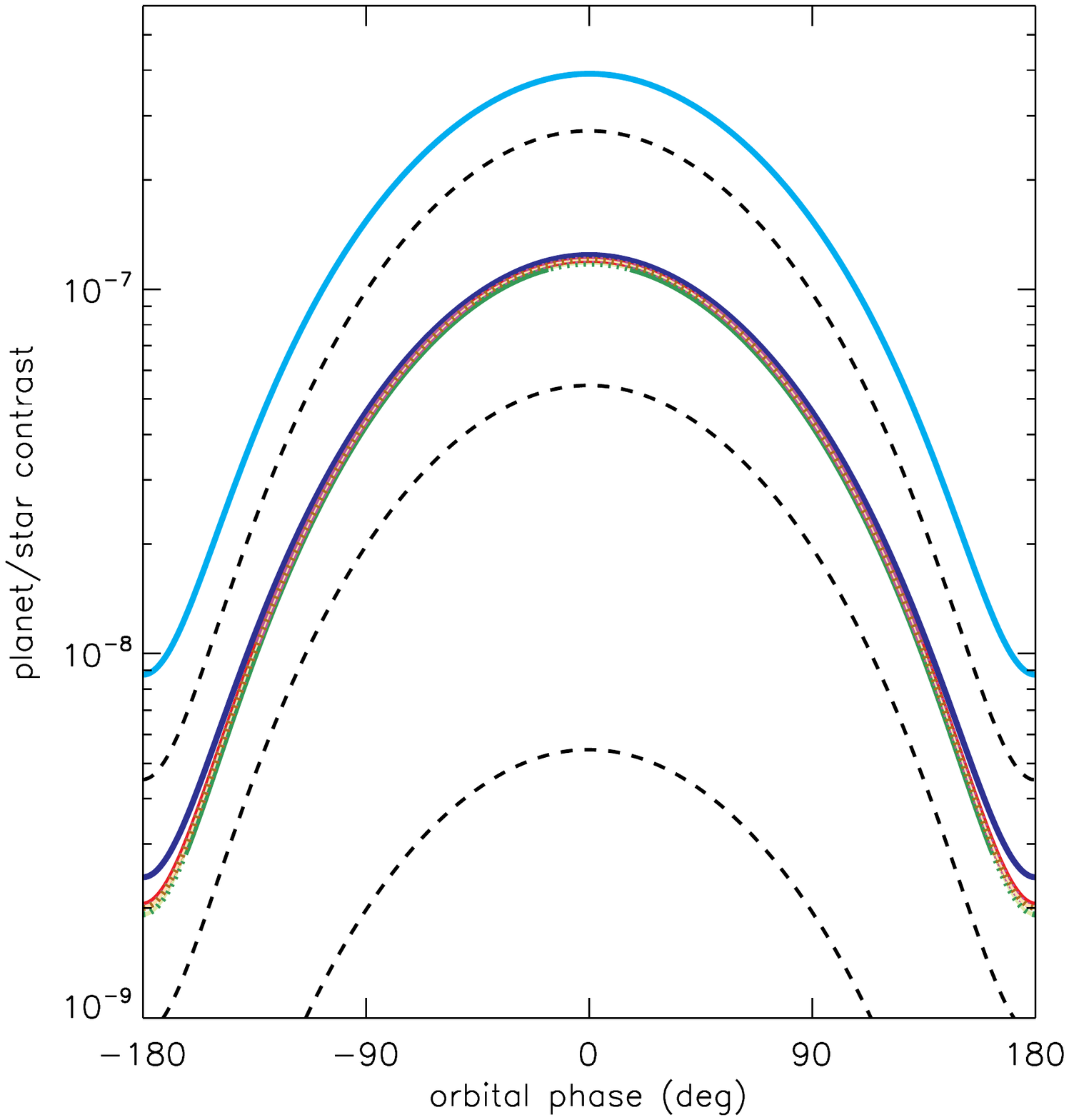}
        \caption{Reflection phase curves computed for a dry planet with an 
Earth-like atmosphere in synchronous rotation (left) and in 3:2 spin-orbit resonance (right).}
\label{fig:VIphase:dryEarth}
\end{figure*}

\begin{figure*}[h]
    \centering
        \includegraphics[width=0.432\linewidth]{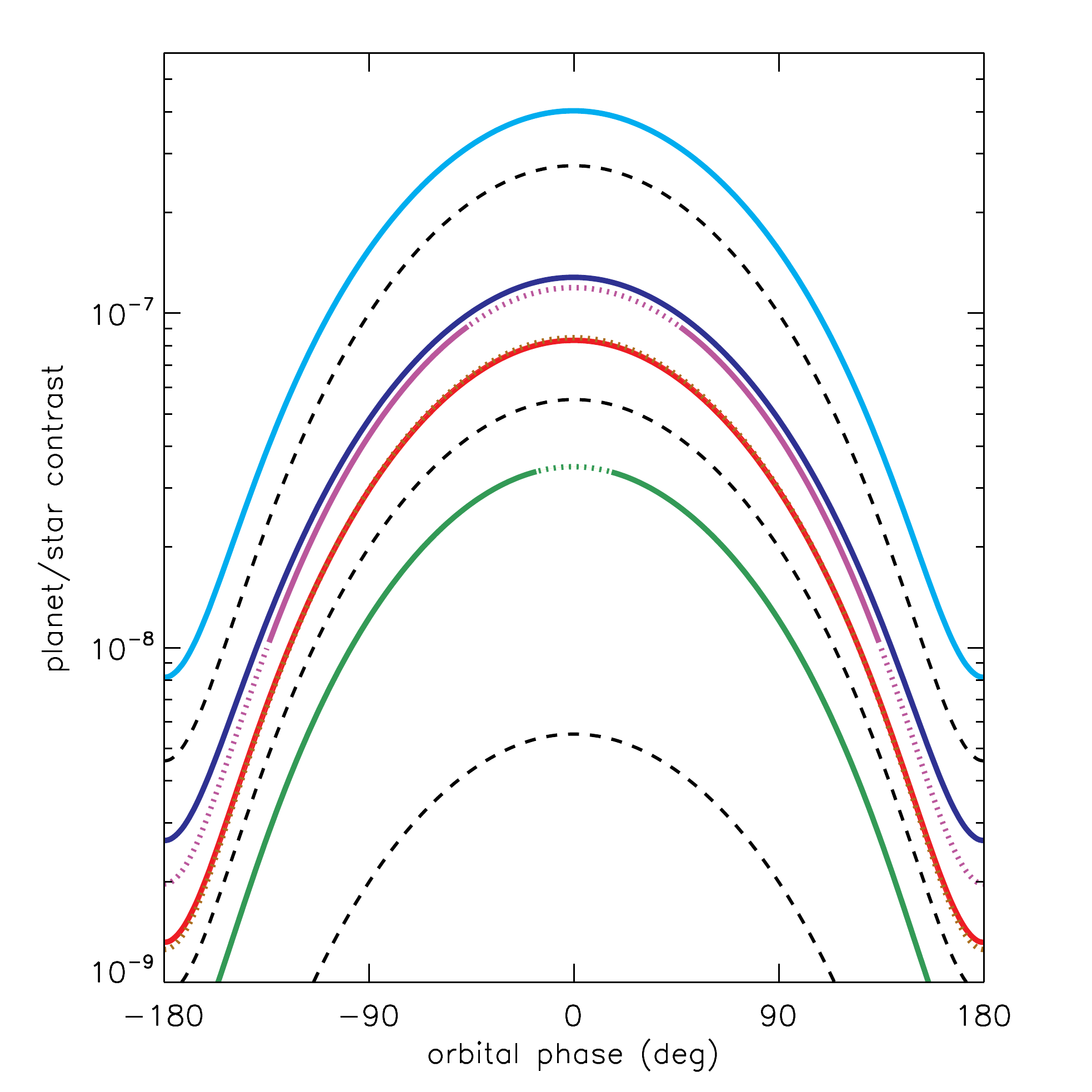}
        \includegraphics[width=0.47\linewidth]{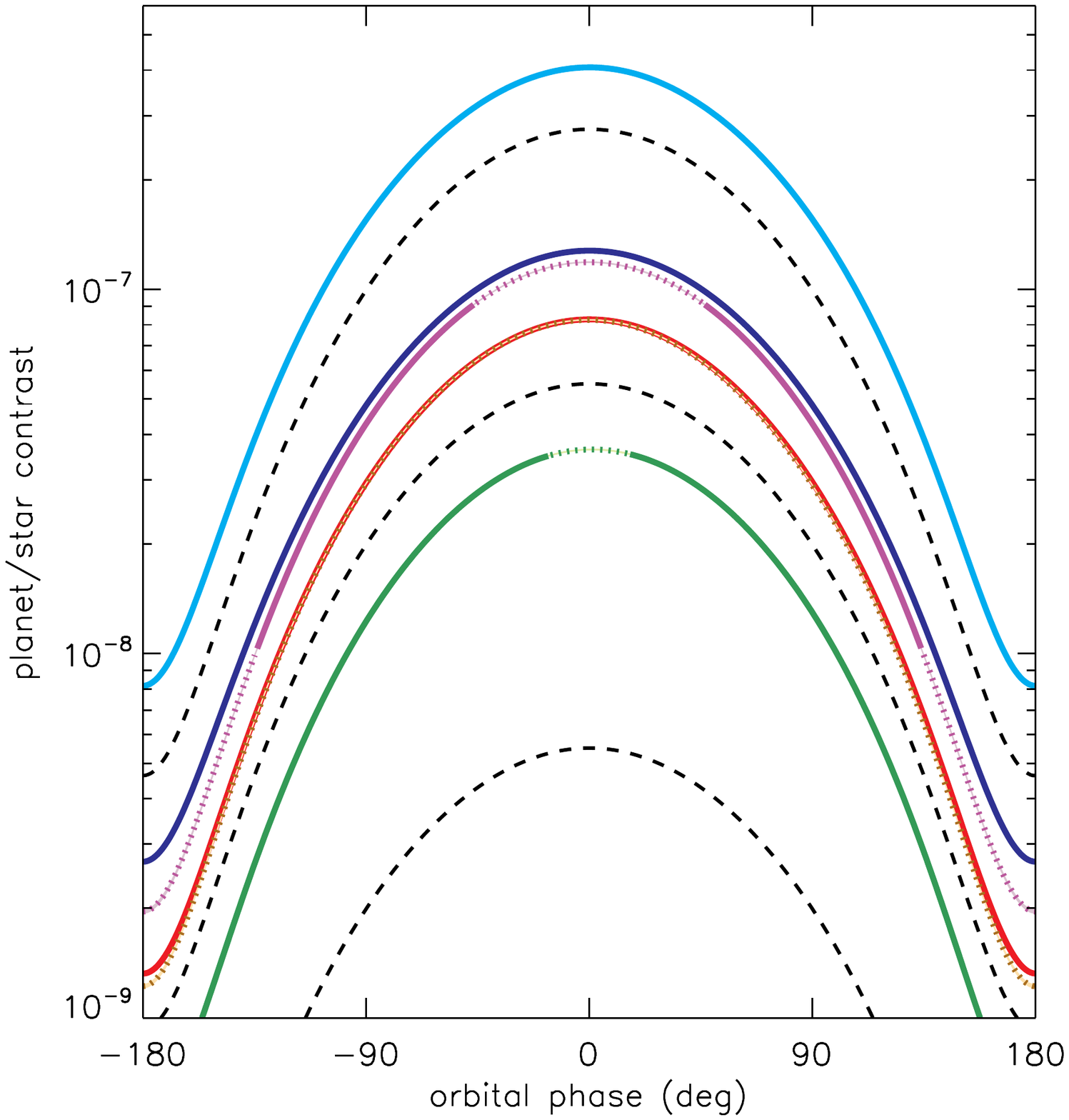}
        \caption{Reflection phase curves computed for a dry planet with a 
1~bar CO$_2$-dominated atmosphere in synchronous rotation (left) and in 3:2 spin-orbit resonance (right).}
\label{fig:VIphase:dryCO2}
\end{figure*}

\begin{figure*}[h]
    \centering
        \includegraphics[width=0.47\linewidth]{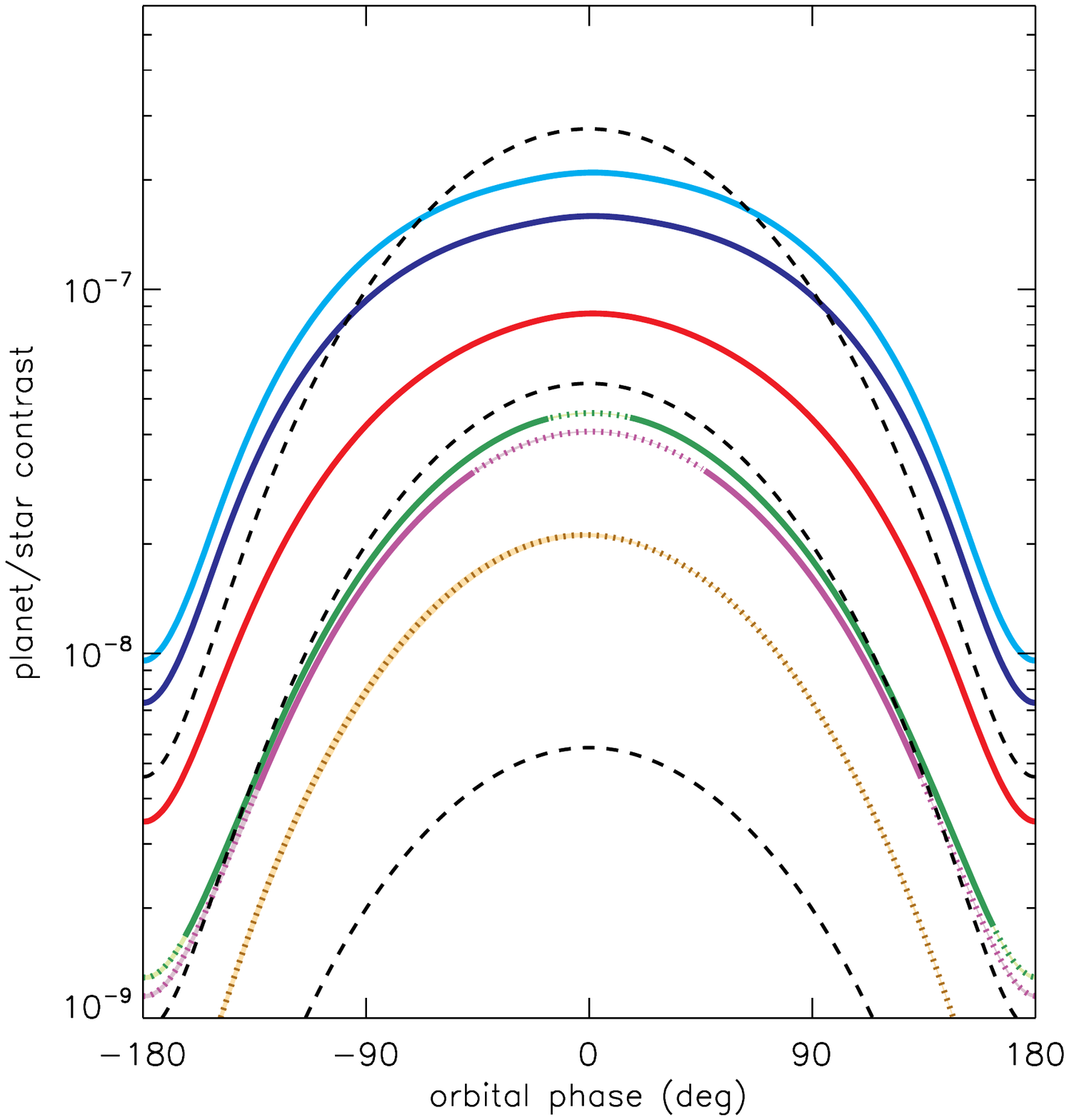}
        \includegraphics[width=0.47\linewidth]{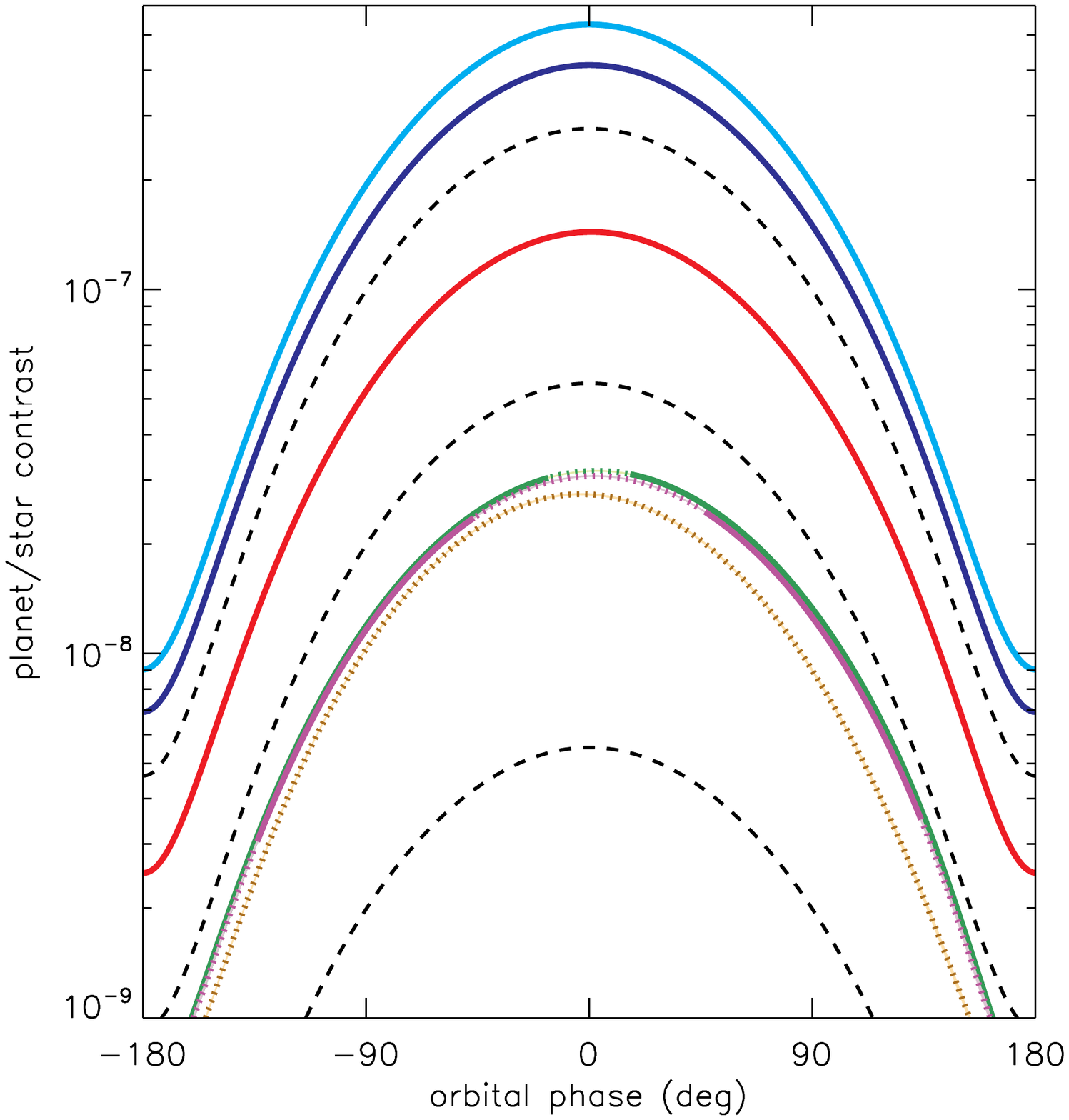}
        \caption{Reflection phase curves computed for an aquaplanet with a 
10~mbar N$_2$-dominated (+~376ppm of CO$_2$) atmosphere in synchronous rotation (left) and in 3:2 spin-orbit resonance (right).}
\label{fig:VIphase:tenuousEarth}
\end{figure*}

\begin{figure*}[h]
    \centering
        \includegraphics[width=0.432\linewidth]{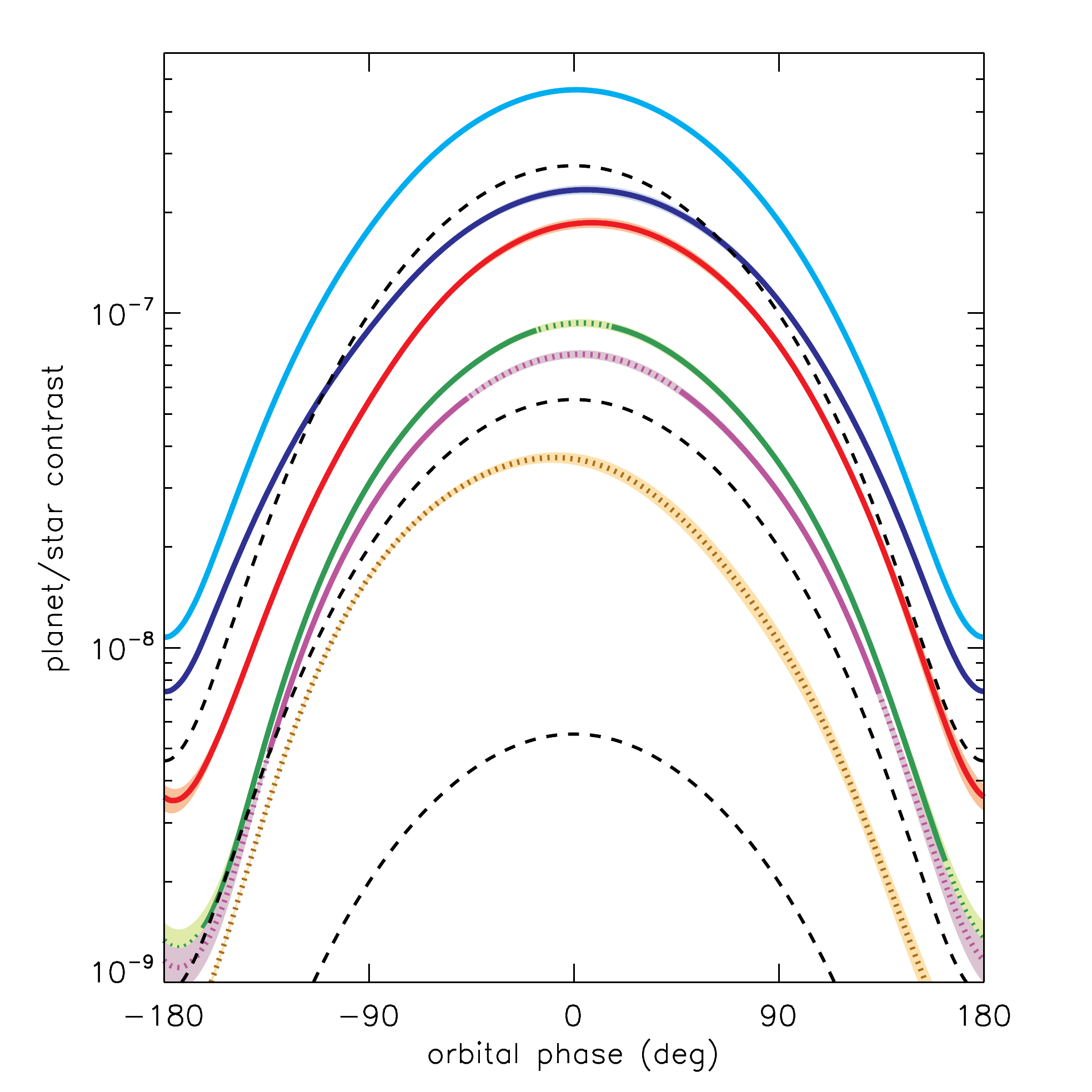}
        \includegraphics[width=0.47\linewidth]{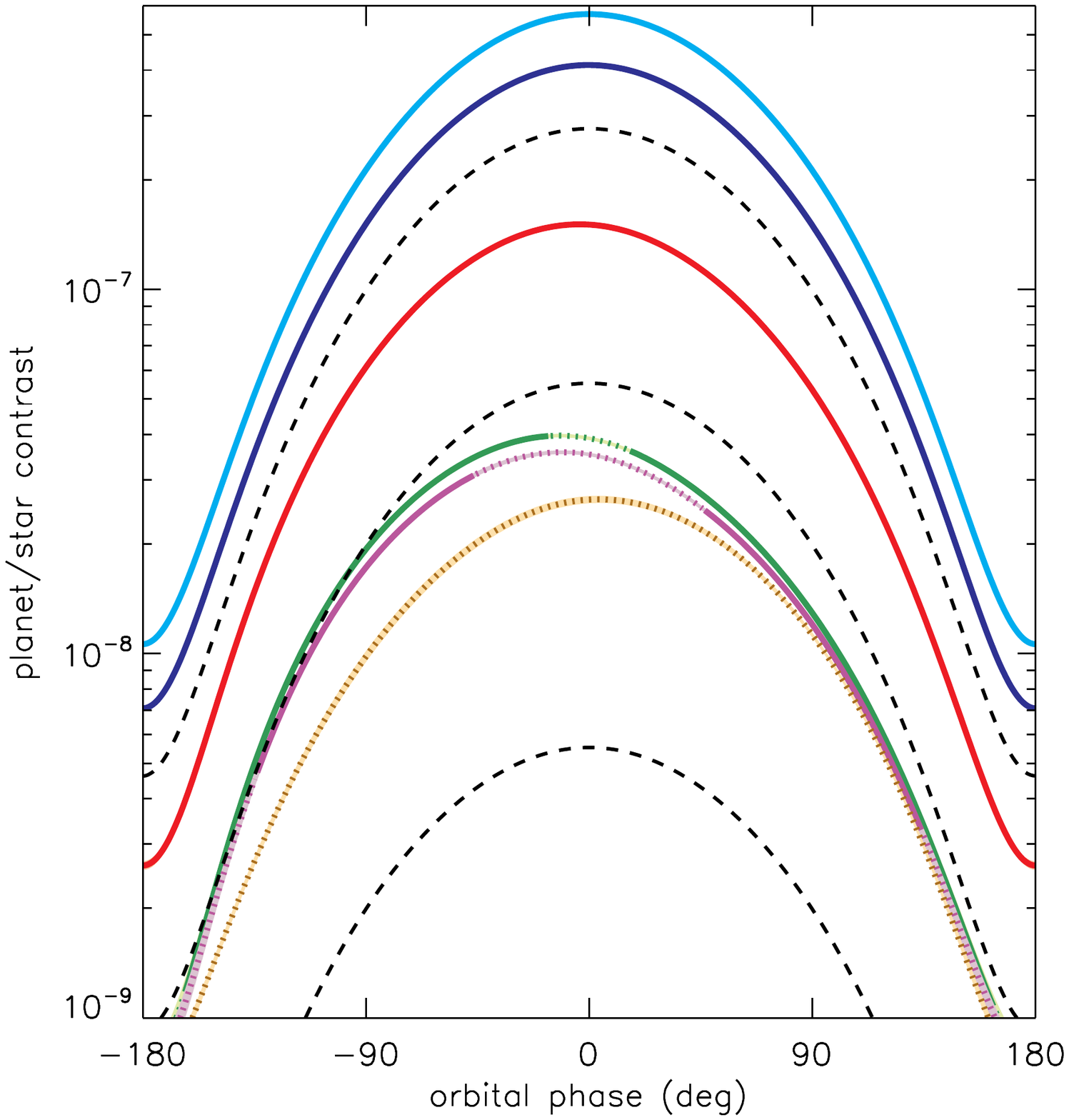}
        \caption{Reflection phase curves computed for an aquaplanet with an 
Earth-like atmosphere in synchronous rotation (left) and in 3:2 spin-orbit resonance (right).}
\label{fig:VIphase:Earth}
\end{figure*}

\begin{figure*}[h]
    \centering
        \includegraphics[width=0.47\linewidth]{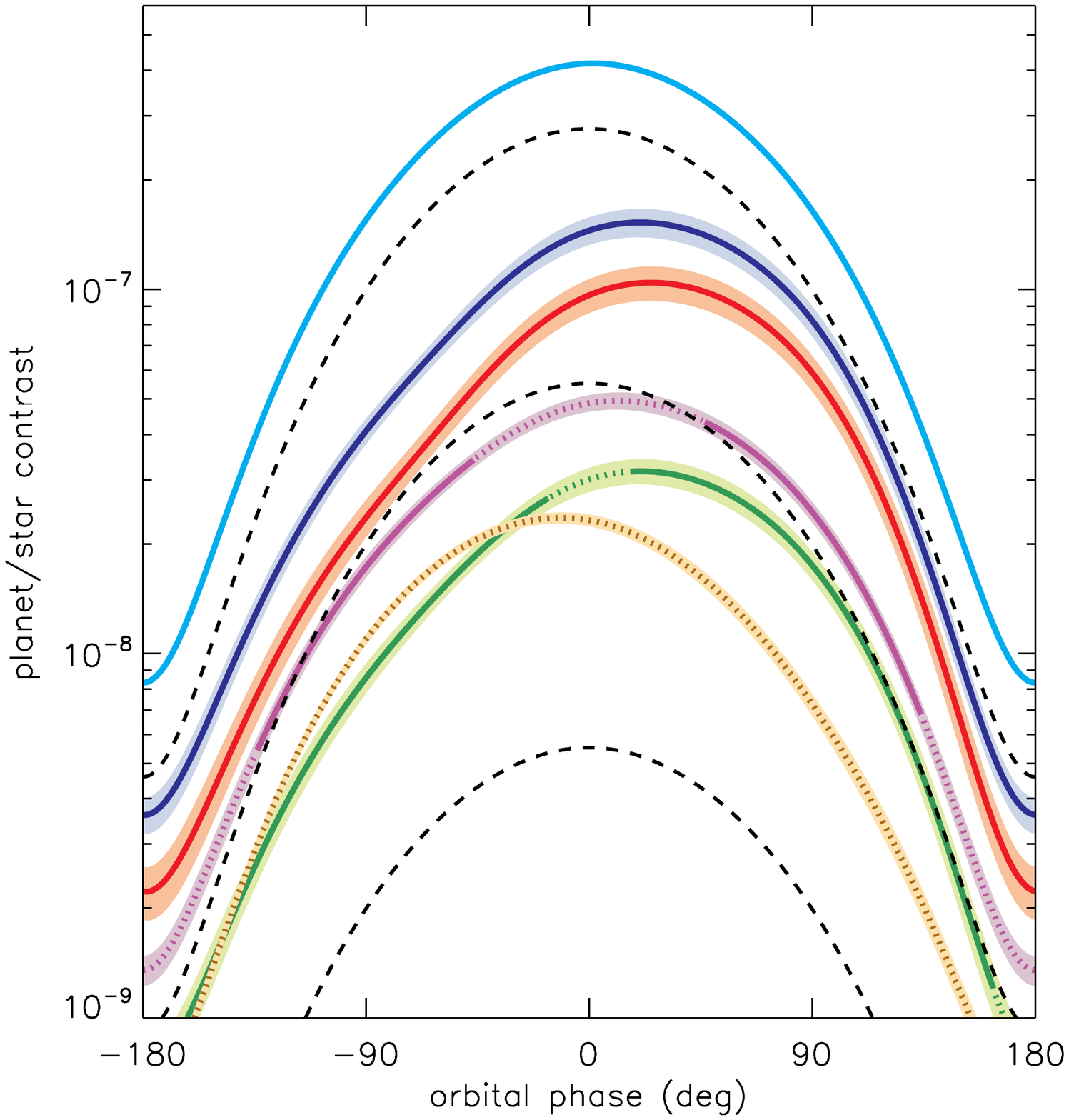}
        \includegraphics[width=0.47\linewidth]{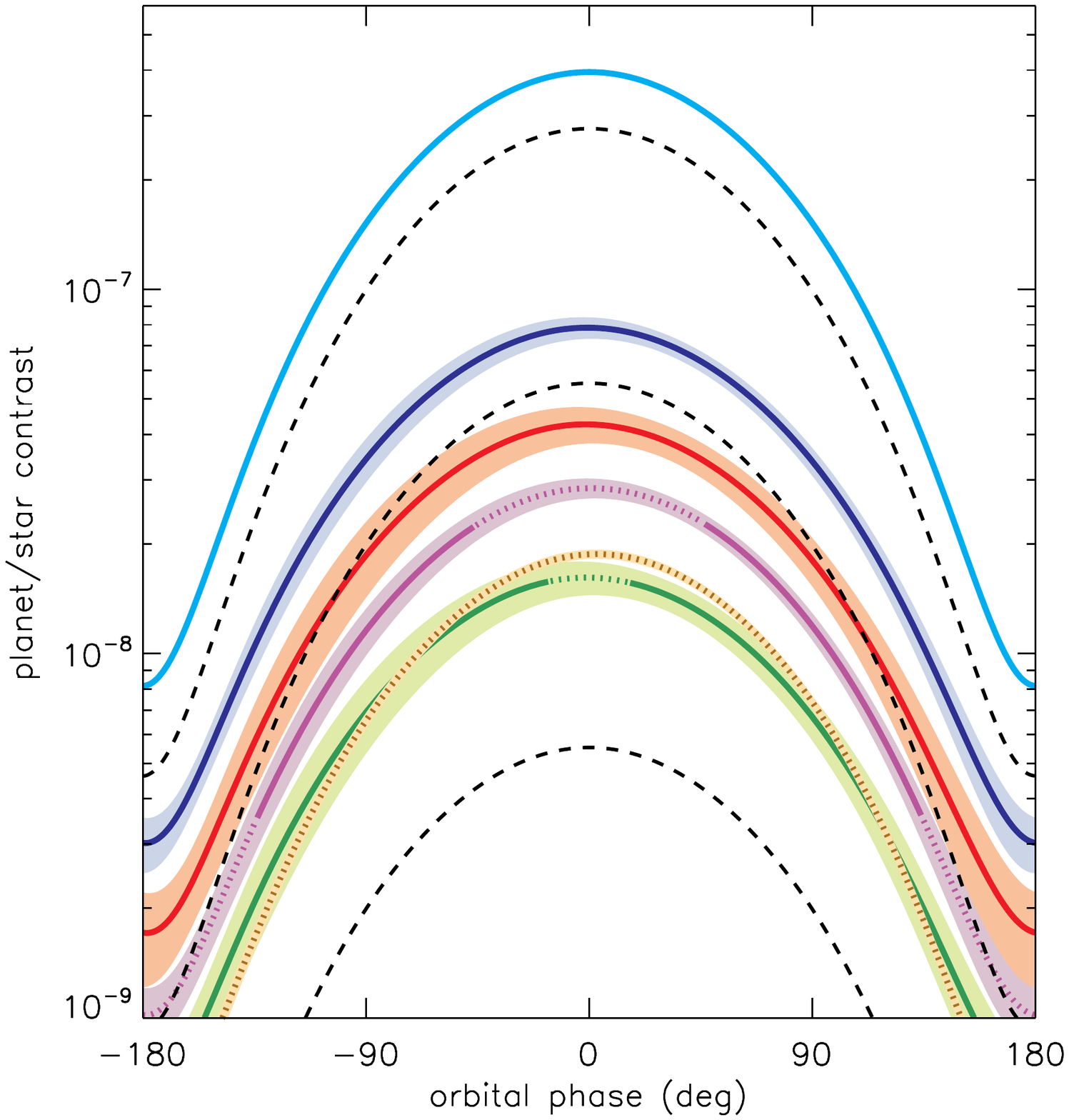}
        \caption{Reflection phase curves computed for an aquaplanet with a 
1~bar CO$_2$-dominated atmosphere in synchronous rotation (left) and in 3:2 spin-orbit resonance (right).}
\label{fig:VIphase:CO2}
\end{figure*}

Departures between the observed phase curve and the phase curve produced by a sphere with uniform surface albedo are due to longitudinal variations of reflectivity on the dayside, which - in our cases - can be due to cloud coverage or change in the nature of the surface (liquid vs icy).\\
\noindent - Dry case, Earth-like atmosphere (Fig.~\ref{fig:VIphase:dryEarth}): the lightcurves are those expected for a sphere with a uniform albedo. As the atmosphere is very transparent at visible-NIR wavelengths, the value of this albedo is 0.2  (the wavelength-independent value attributed to the dry surface in the model) for all the bands except the one in the UV (cyan) that exhibits a higher albedo due to Rayleigh scattering.\\
\noindent - Dry case, 1 bar of CO$_2$ (Fig.~\ref{fig:VIphase:dryCO2}): same as above, except for the 1.6 (green) and 1.25~$\mu$m (red) bands that are attenuated by CO$_2$ absorption.\\
\noindent - Aquaplanet, mainly frozen, 10 mbar of N$_2$; 376 ppm of CO$_2$ (Fig.~\ref{fig:VIphase:tenuousEarth}): we can note again the overall decrease of albedo with increasing wavelength (a property of ice). While the phase curves of the asynchronous case present no longitudinal change of reflectivity, the phase curves of the synchronous case show a flattening at opposition at wavelength below 1.3~$\mu$m, which is due to the different albedo of ice and liquid water. This decrease of albedo at small phase angles is therefore a signature of the "eye ball" configuration that is observable here because the atmosphere and the cloud cover are thin enough to give access to the surface reflectivity. \\
\noindent - Aquaplanet, mainly frozen, 1 bar of N$_2$; 376 ppm of CO$_2$ (Fig.~\ref{fig:VIphase:Earth}): The "eye ball" signature of the synchronous case is still noticeable at wavelengths not absorbed by water vapor and not dominated by Rayleigh scattering, so typically between 0.5 and 0.8~ $\mu$m. Clouds reduce the signature without hiding it completely as they cover only partially the dayside ocean (see Fig.~\ref{fig:cloudmapEarth}) and because their albedo is still lower than that of ice at these wavelengths. Because thick clouds tend to accumulate eastward of the substellar point, the phase curves are asymmetric except at UV wavelengths that are backscattered above the clouds. In the asynchronous case clouds are uniformely distributed in longitude and do not produce significant asymmetry. \\
\noindent - Aquaplanet, 1 bar CO$_2$ (Fig.~\ref{fig:VIphase:CO2}): Both synchronous and asynchronous case exhibit a high variability due to meteorology, which is a result of high temperatures and strong water cycle. The visible and NIR phase curves of the synchronous case are very asymmetric. The reason is the same as in the previous case but with a sharper transition in cloudiness with a clear sky west of the substellar point and a dense cloud patch east of the substellar point, as seen on Fig.~\ref{fig:cloudmap1barCO2}. \\

\subsection{Emission spectra}
\begin{figure*}[h]
    \centering
        \includegraphics[width=0.432\linewidth]{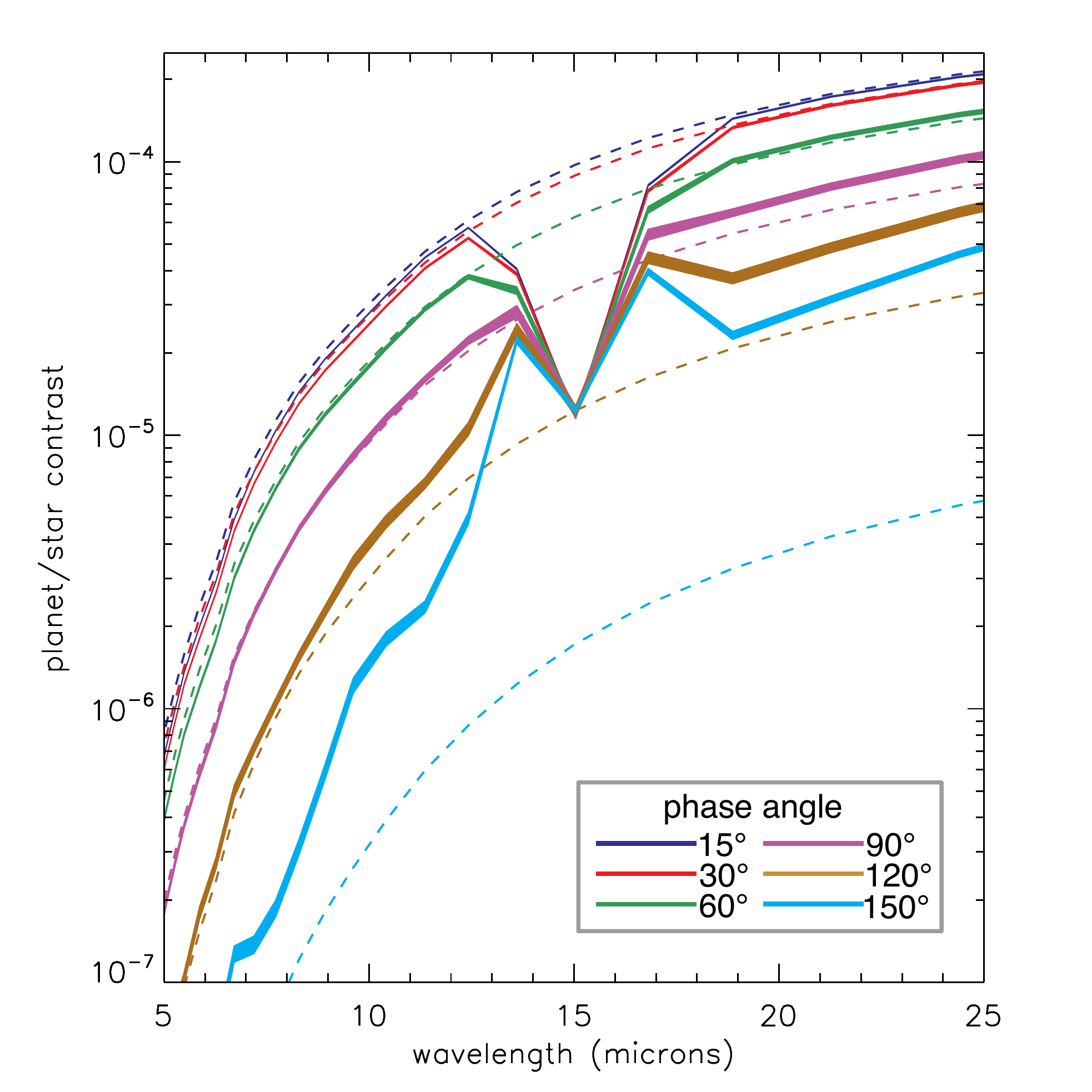}
        \includegraphics[width=0.47\linewidth]{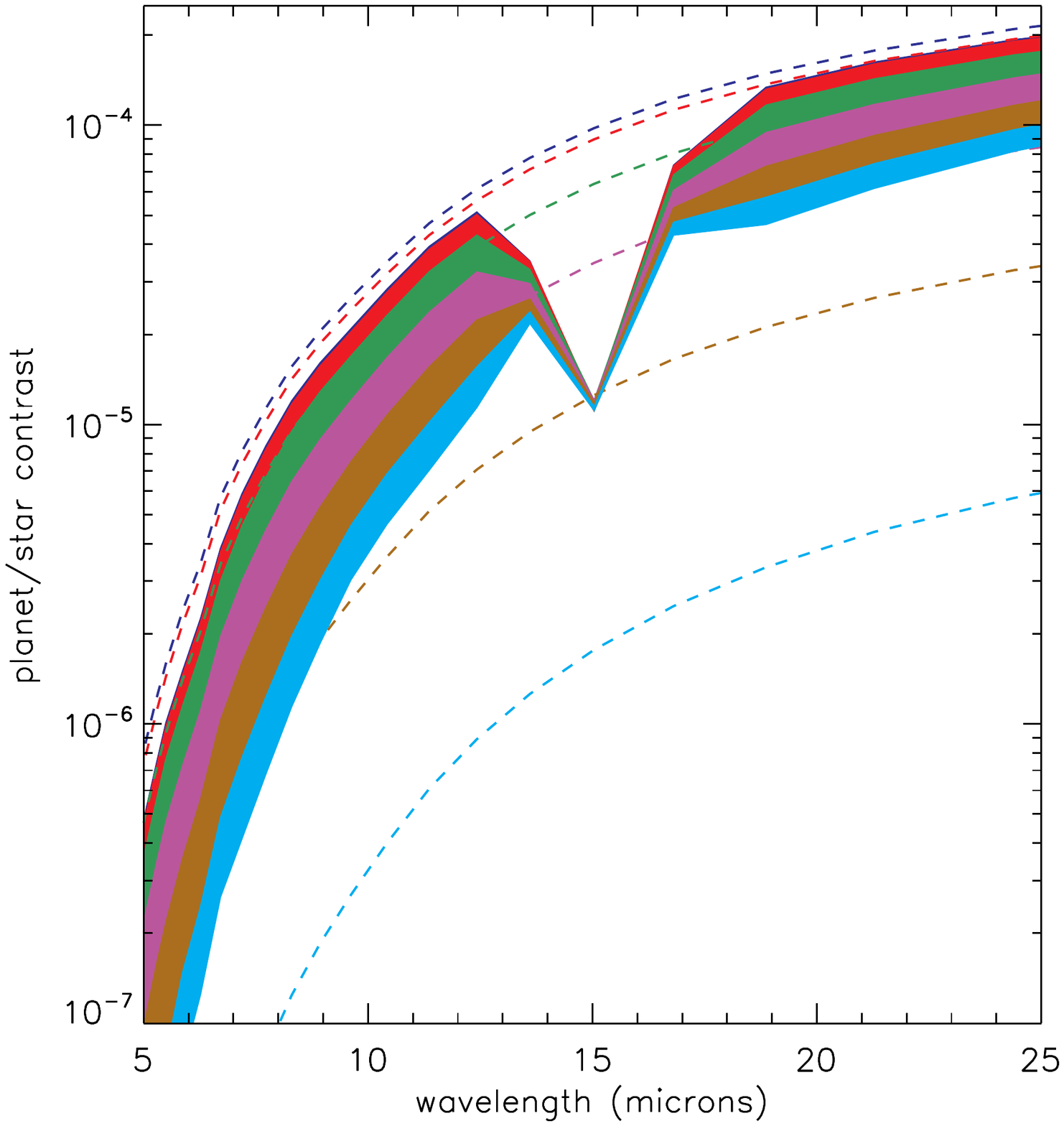}
        \caption{Emission spectra computed for a dry planet with an 
Earth-like atmosphere in synchronous rotation (left) and in 3:2 spin-orbit resonance (right).}
\label{fig:IRspec:dryEarth}
\end{figure*}

\begin{figure*}[h]
    \centering
        \includegraphics[width=0.432\linewidth]{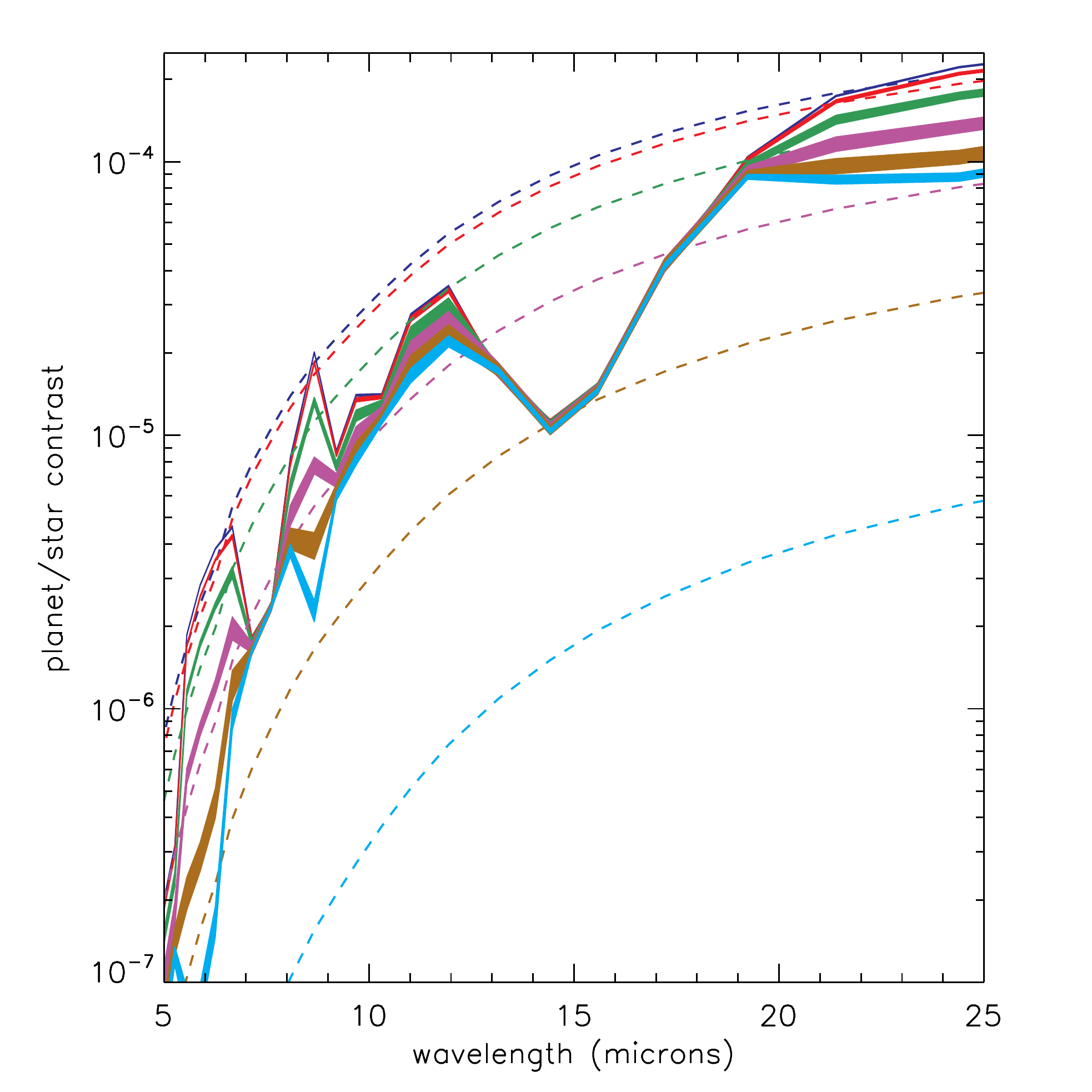}
        \includegraphics[width=0.47\linewidth]{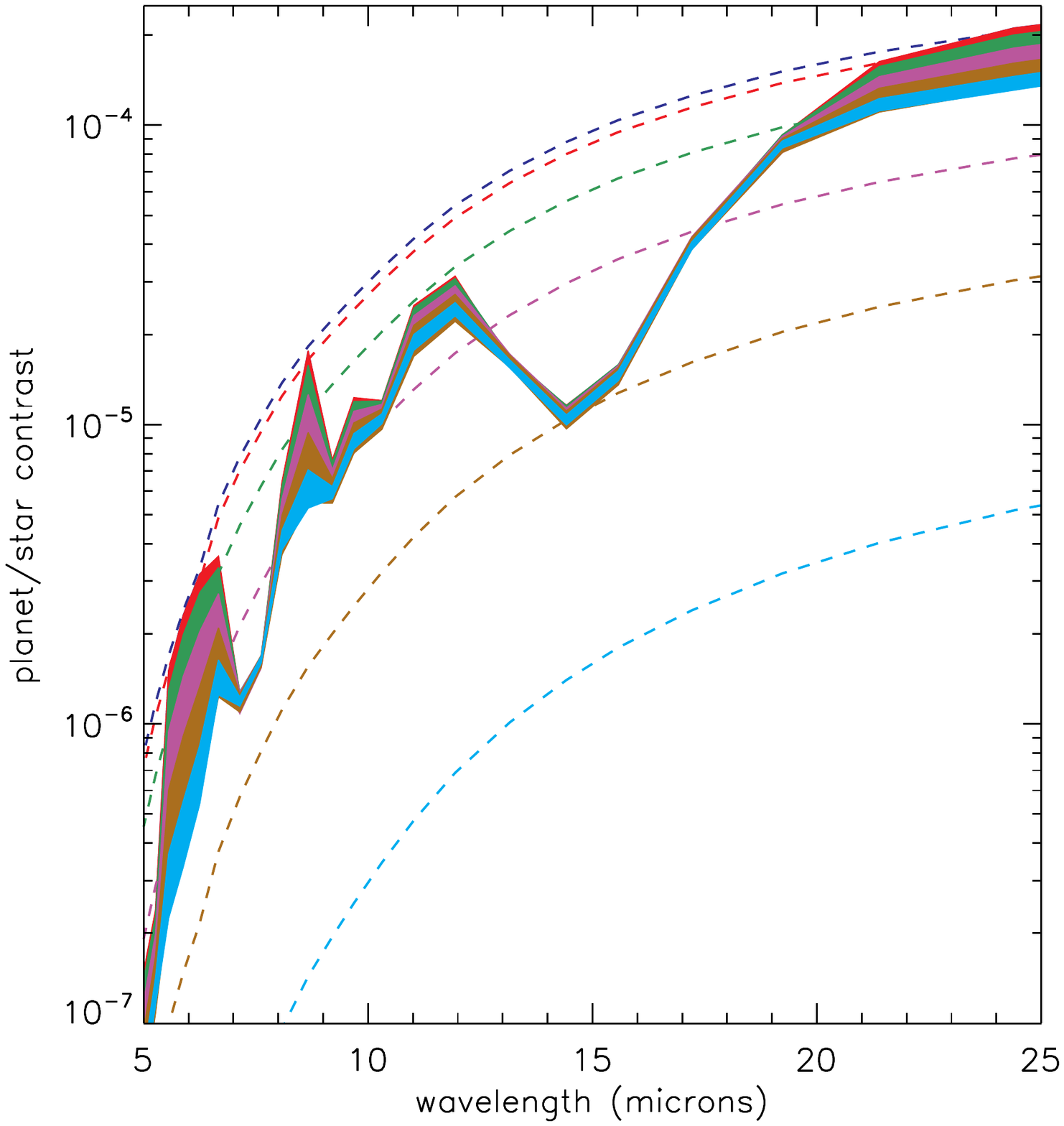}
        \caption{Emission spectra computed for a dry planet with a 
1~bar CO$_2$-dominated atmosphere in synchronous rotation (left) and in 3:2 spin-orbit resonance (right).}
\label{fig:IRspec:dryCO2}
\end{figure*}

\begin{figure*}[h]
    \centering
        \includegraphics[width=0.47\linewidth]{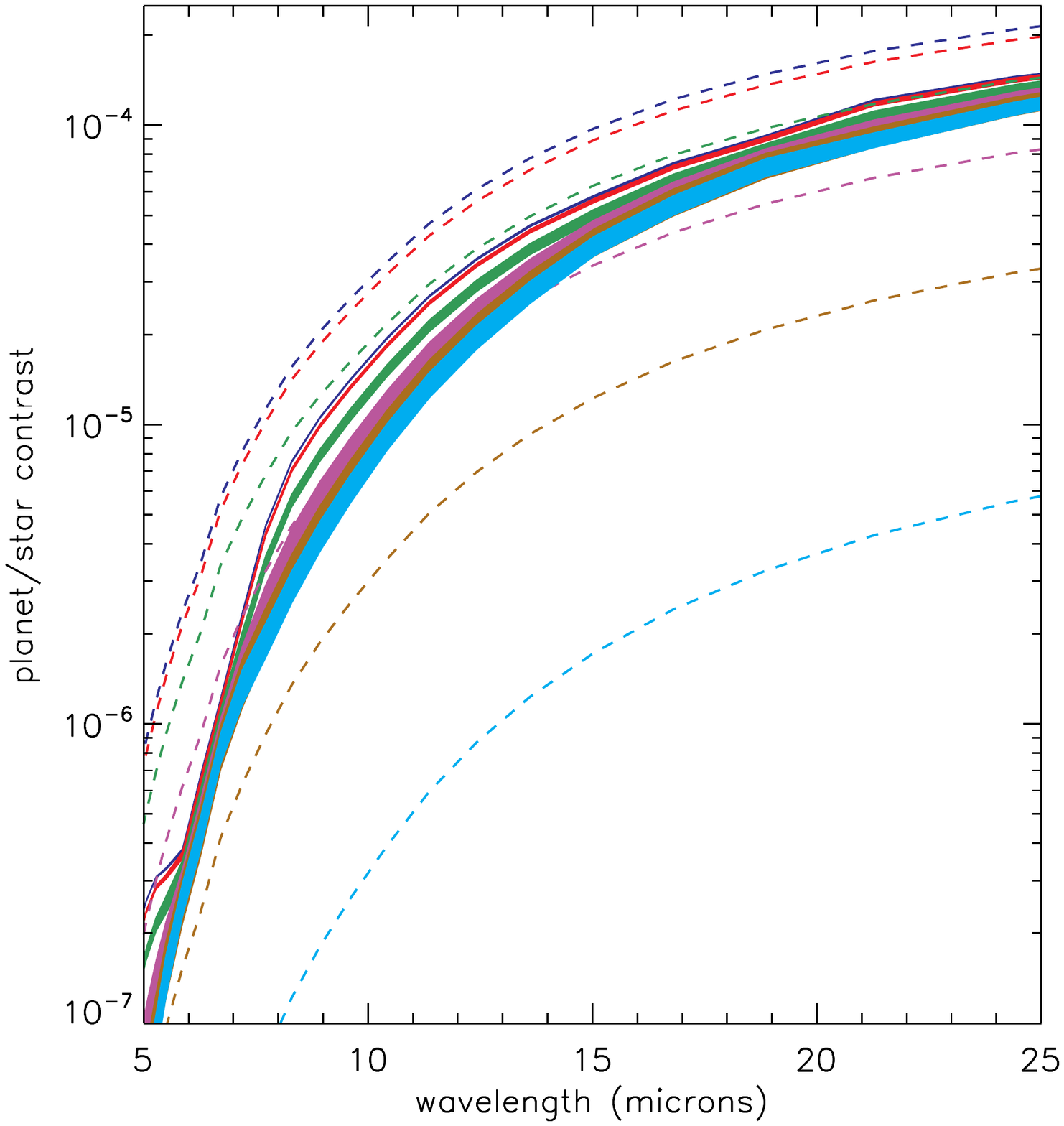}
        \includegraphics[width=0.47\linewidth]{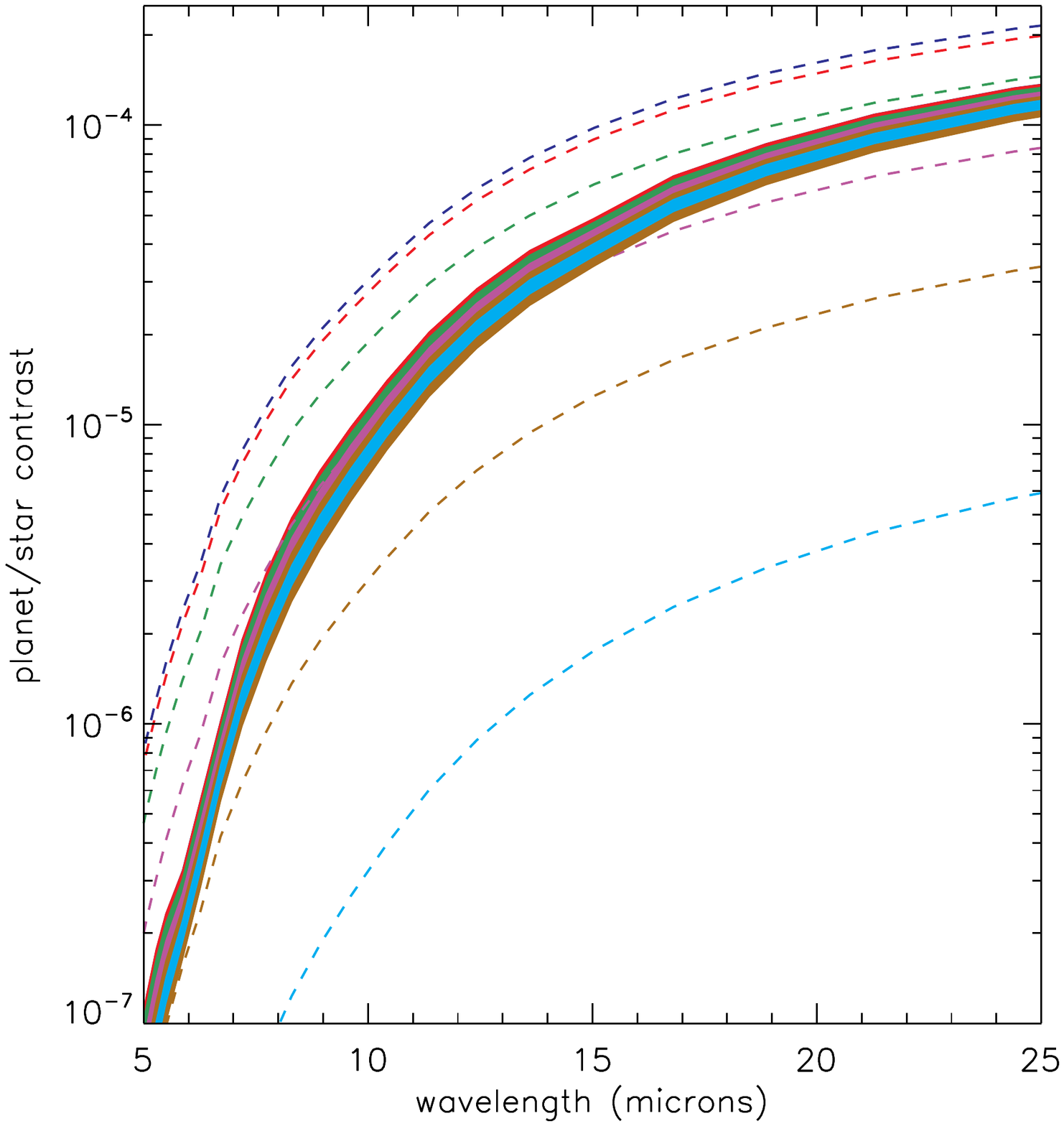}
        \caption{Emission spectra computed for an aquaplanet with a 
10~mbar N$_2$-dominated (+~376ppm of CO$_2$) atmosphere in synchronous rotation (left) and in 3:2 spin-orbit resonance (right).}
\label{fig:IRspec:tenuousEarth}
\end{figure*}

\begin{figure*}[h]
    \centering
        \includegraphics[width=0.432\linewidth]{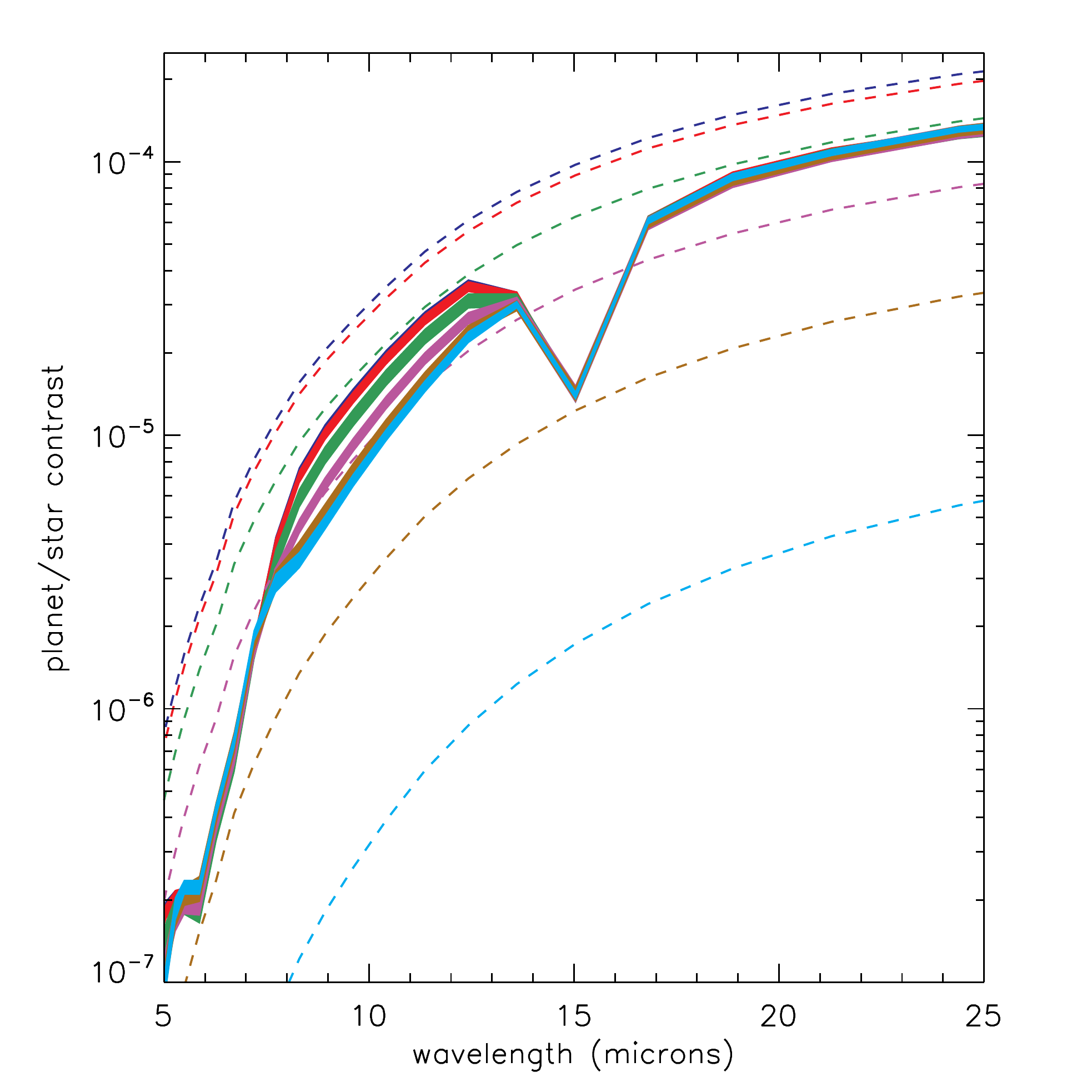}
        \includegraphics[width=0.47\linewidth]{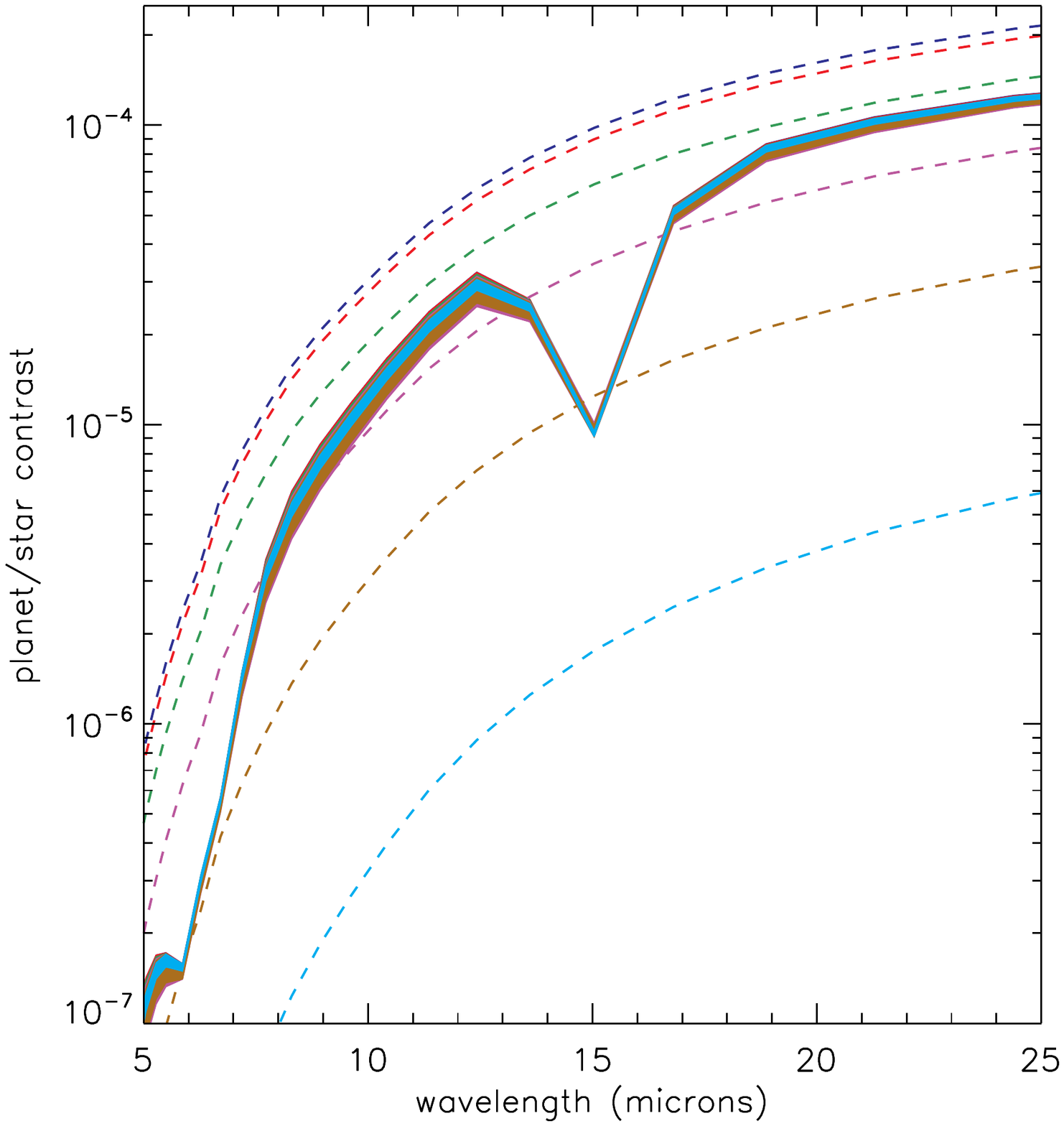}
        \caption{Emission spectra computed for an aquaplanet with an 
Earth-like atmosphere in synchronous rotation (left) and in 3:2 spin-orbit resonance (right).}
\label{fig:IRspec:Earth}
\end{figure*}

\begin{figure*}[h]
    \centering
        \includegraphics[width=0.47\linewidth]{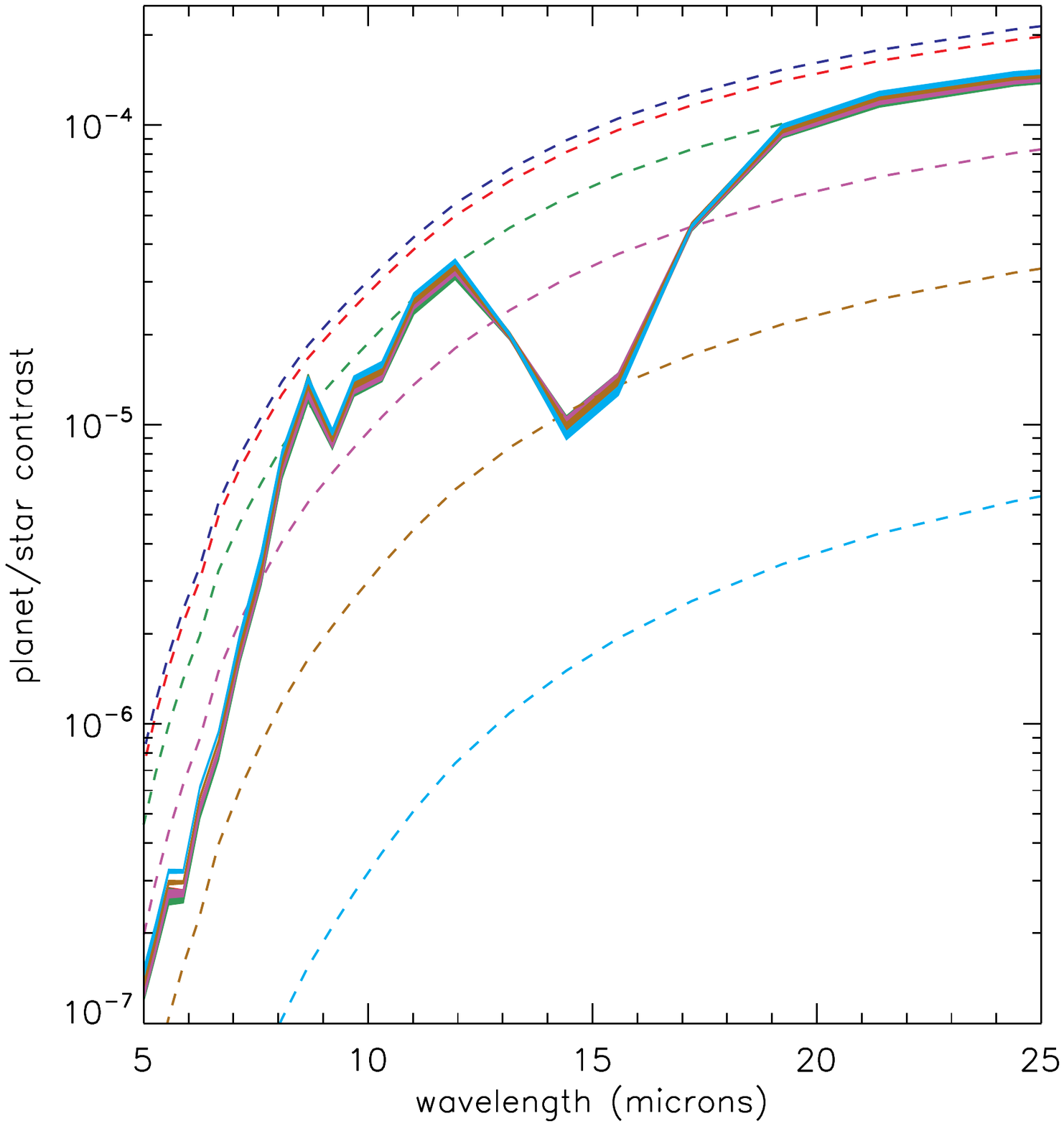}
        \includegraphics[width=0.47\linewidth]{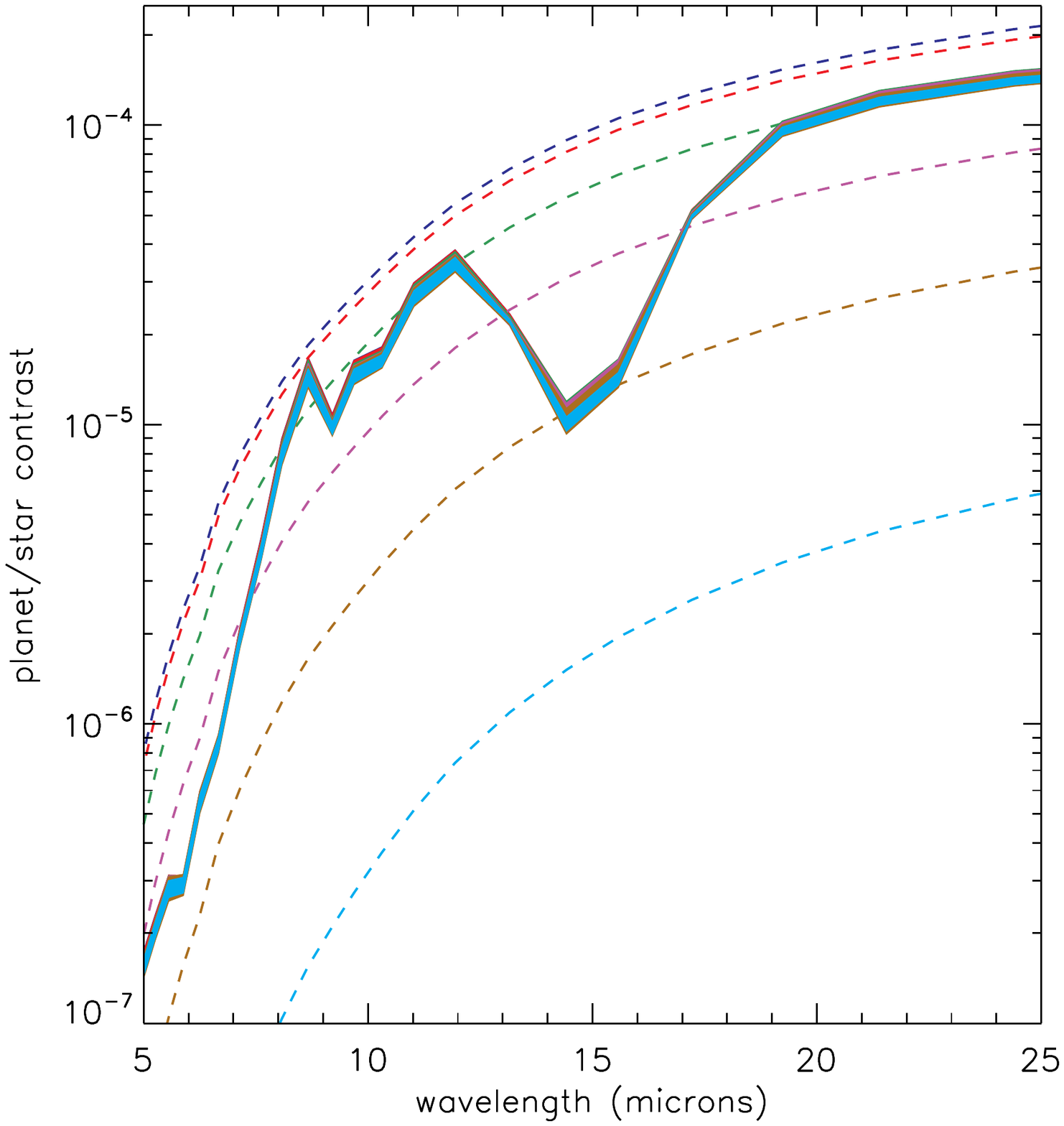}
        \caption{Emission spectra computed for an aquaplanet with a 
1~bar CO$_2$-dominated atmosphere in synchronous rotation (left) and in 3:2 spin-orbit resonance (right).}
\label{fig:IRspec:CO2}
\end{figure*}

Thermal emission spectra are influenced by the temperature distribution at the surface at wavelengths where the atmosphere is transparent and by the thermal structure of the atmosphere at other wavelengths. In theory, emission spectra are also shaped by the surface emissivity but in our model it is fixed to unity in most cases. \\
\noindent - Dry case, Earth-like atmosphere (Fig.~\ref{fig:IRspec:dryEarth}): In the synchronous case, the spectra are similar to those of sphere in radiative equilibrium except in the CO$_2$ band, which is emitted by a horizontally-uniform layer of the upper atmosphere. At high phase angle and long wavelengths, differences from the sphere in radiative equilibrium appear due to the transport of heat toward the night side that contributes to the emission. In 3:2 rotation, the spectra are similar but more sensitive to the observing geometry at a given phase angle as the surface temperature map is no longer symmetric relative to the substellar point. \\
\noindent - Dry case, 1 bar of CO$_2$ (Fig.~\ref{fig:IRspec:dryCO2}): In addition to the 15~$\mu$m band, high pressure CO$_2$ features including the hot bands at 9.5 and 11~$\mu$m and a CIA feature between 6.5 and 8~$\mu$m. These absorption features probe uniform high-altitude atmospheric layers and do not depend on the sub-observer position while windows probe the surface and exhibit a dependency on the observing geometry (synchronous) and variability (3:2). \\
\noindent - Aquaplanet, mainly frozen, 10 mbar of N$_2$; 376 ppm of CO$_2$ (Fig.~\ref{fig:IRspec:tenuousEarth}): The spectra are featureless except for a shallow H$_2$O signature at 6-7~$\mu$m in the synchronous case. They depart from the spectra of a sphere in radiative equilibrium, in particular at long wavelength, as the surface temperature does not drop below 200~K (synchronous) and 150~K (3:2) thanks to a redistribution of latent heat.\\
\noindent - Aquaplanet, mainly frozen, 1 bar of N$_2$; 376 ppm of CO$_2$ (Fig.~\ref{fig:IRspec:Earth}): The spectra are similar to a present Earth spectrum, without the O$_3$ band and with shallower water vapor absorption due to lower temperatures. \\
\noindent - Aquaplanet, 1 bar CO$_2$ (Fig.~\ref{fig:IRspec:CO2}): There are no difference between the synchronous and 3:2 cases. All atmospheric windows are closed by either CO$_2$ or H$_2$O absorption and the emerging spectrum come from different but horizontally-uniform layers. \\

\subsection{Thermal phase curves}
\begin{figure*}[h]
    \centering
        \includegraphics[width=0.47\linewidth]{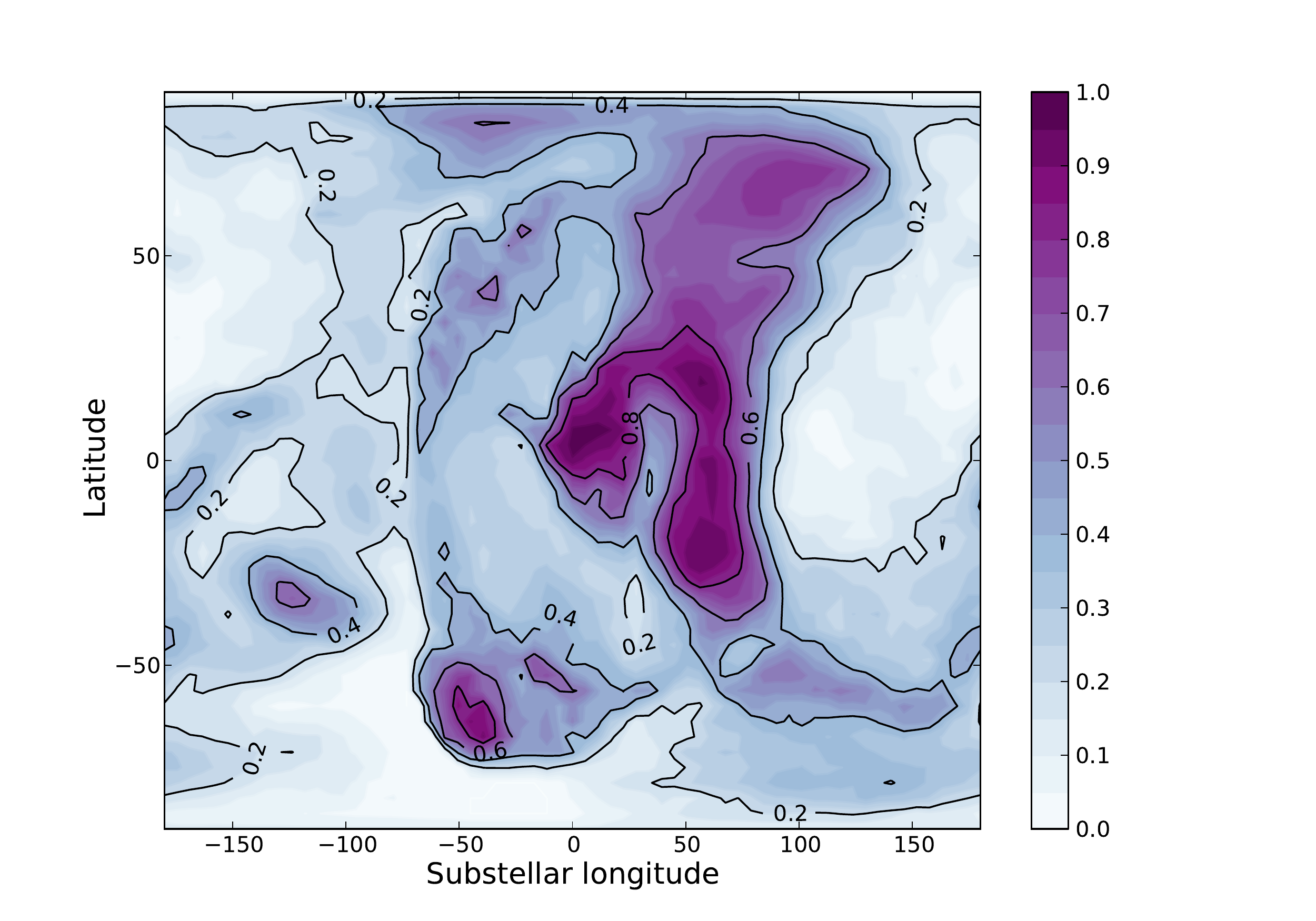}
        \includegraphics[width=0.47\linewidth]{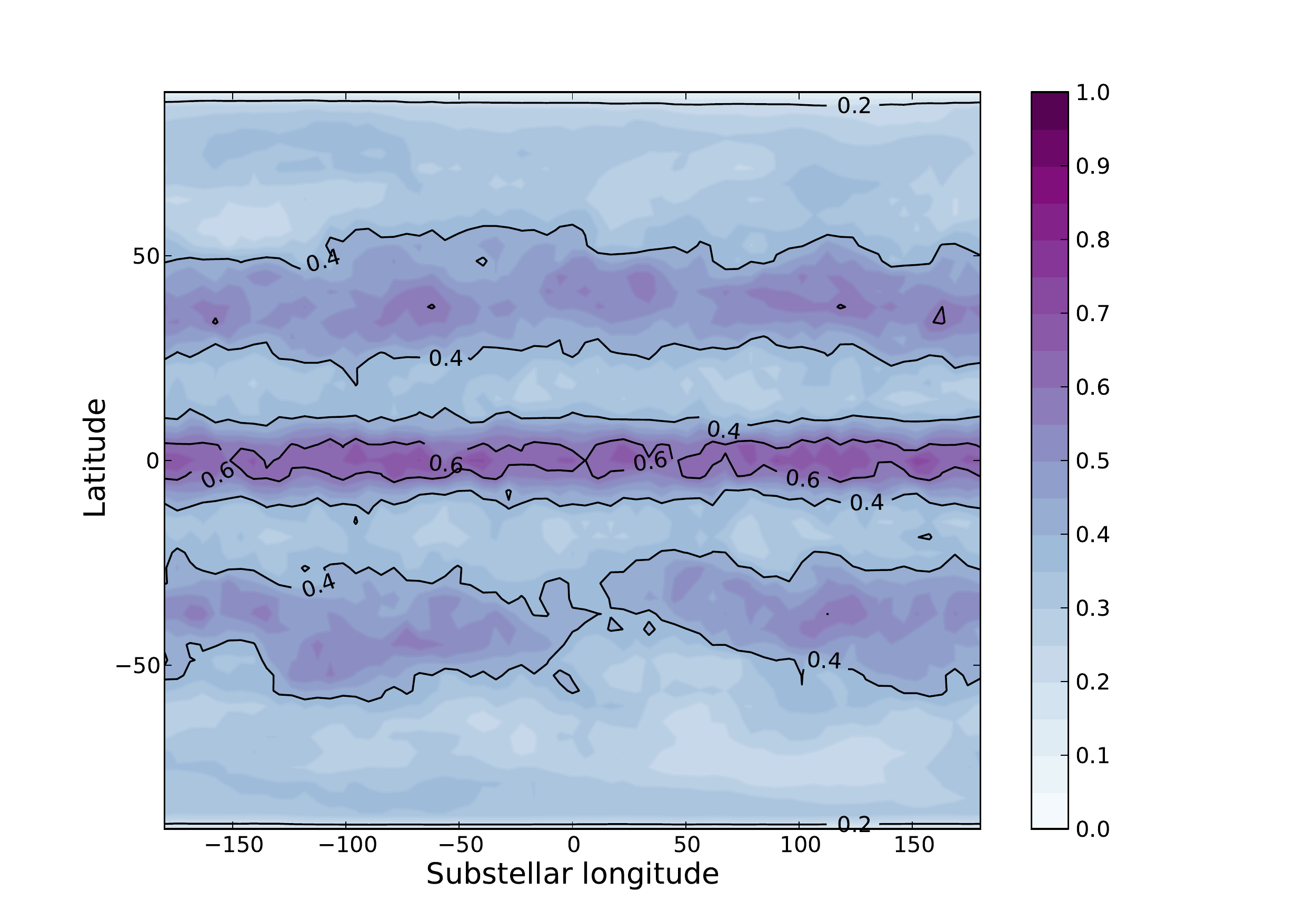}
        \caption{Cloud maps for an aquaplanet with an Earth-like atmospheric composition (1 bar of N$_2$, 376~ppm of CO$_2$). Left : synchronous rotation. Right : 3:2 spin-orbit resonance. In both cases, longitude is given relative to the substellar point. Colors indicate the fractional cloud cover, averaged over 2 orbits. }
        \label{fig:cloudmapEarth}
\end{figure*}

\begin{figure*}[h]
    \centering
        \includegraphics[width=0.432\linewidth]{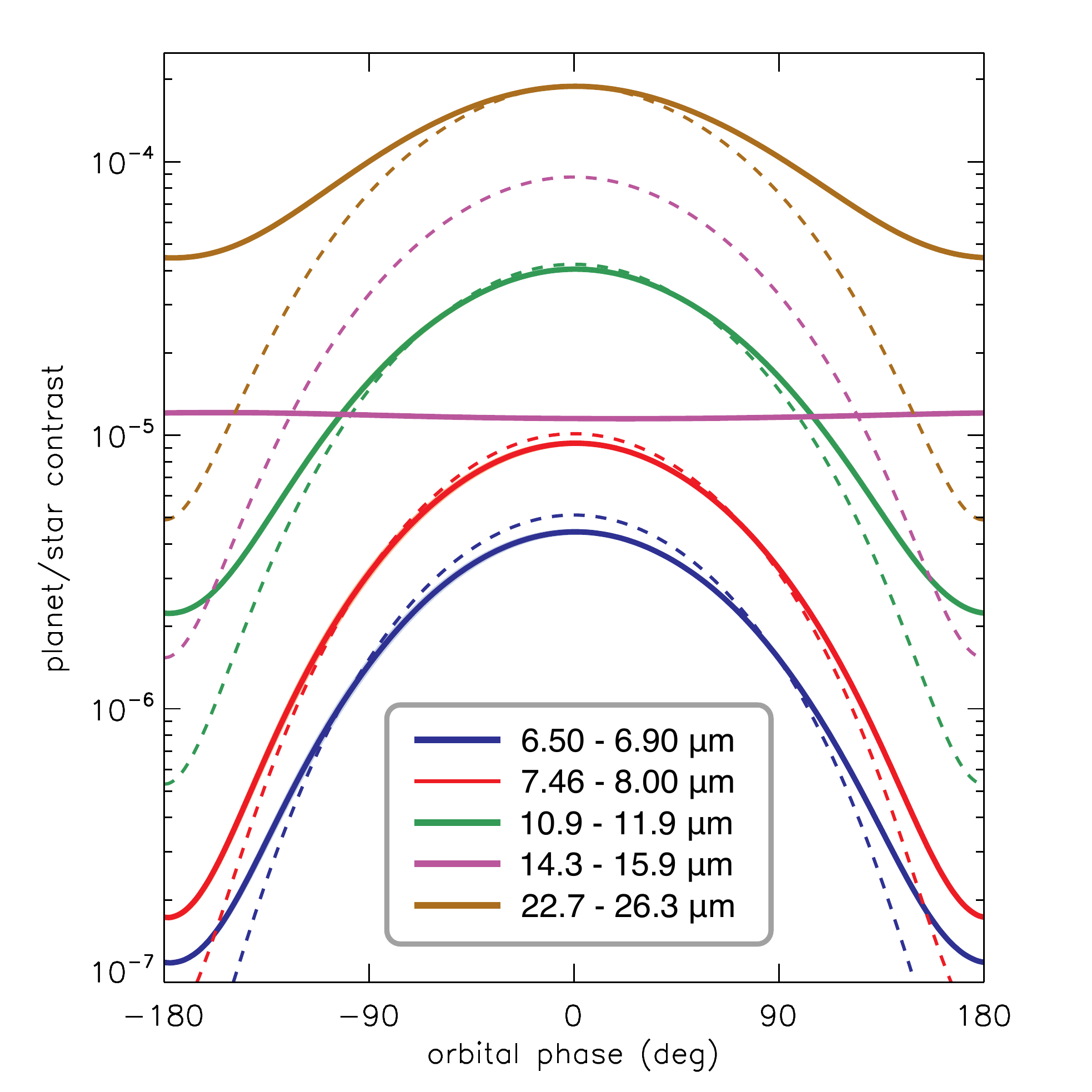}
        \includegraphics[width=0.47\linewidth]{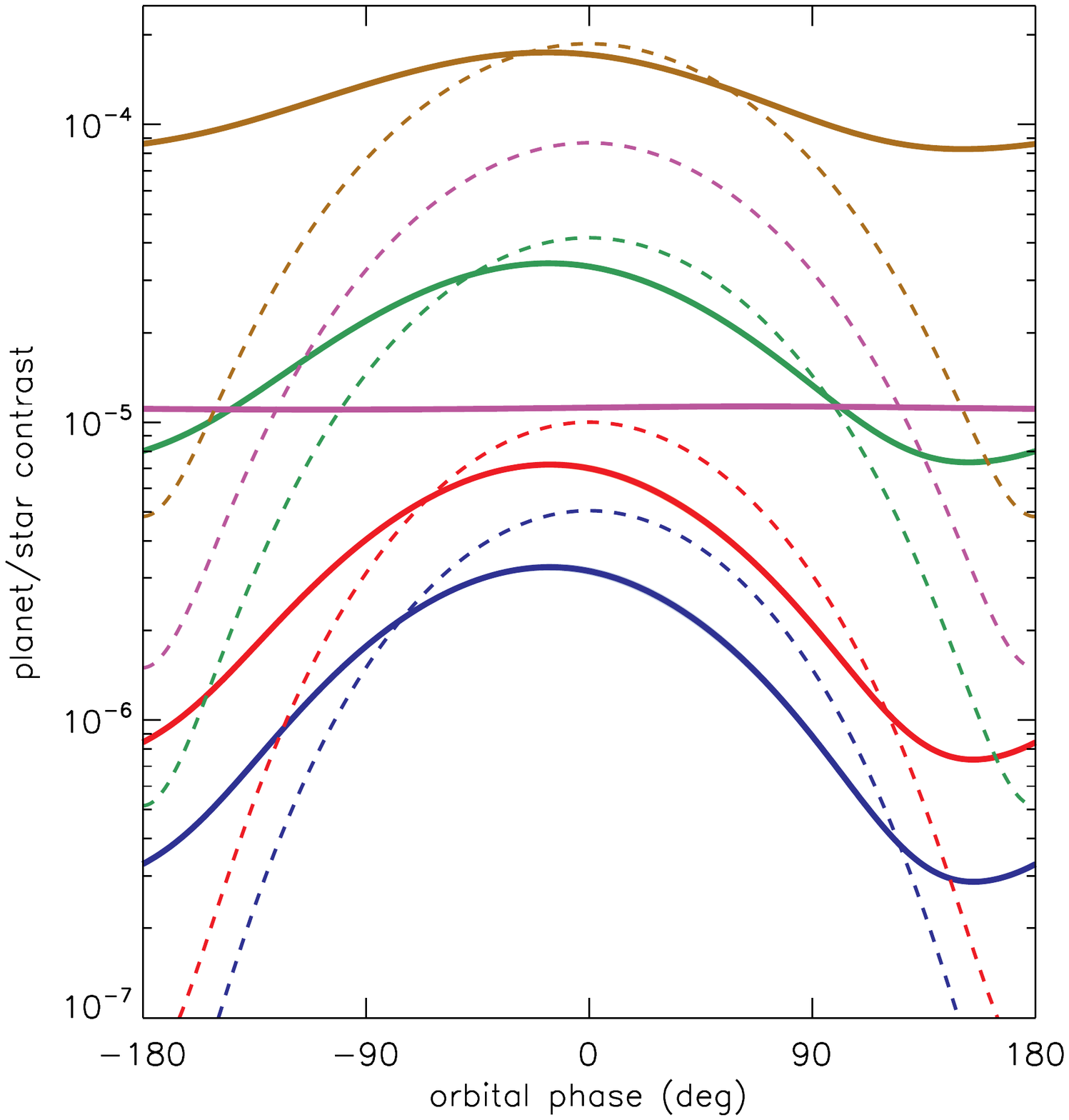}
        \caption{Emission phase curves computed for a dry planet with an 
Earth-like atmosphere in synchronous rotation (left) and in 3:2 spin-orbit resonance (right).}
\label{fig:IRphase:dryEarth}
\end{figure*}

\begin{figure*}[h]
    \centering
        \includegraphics[width=0.432\linewidth]{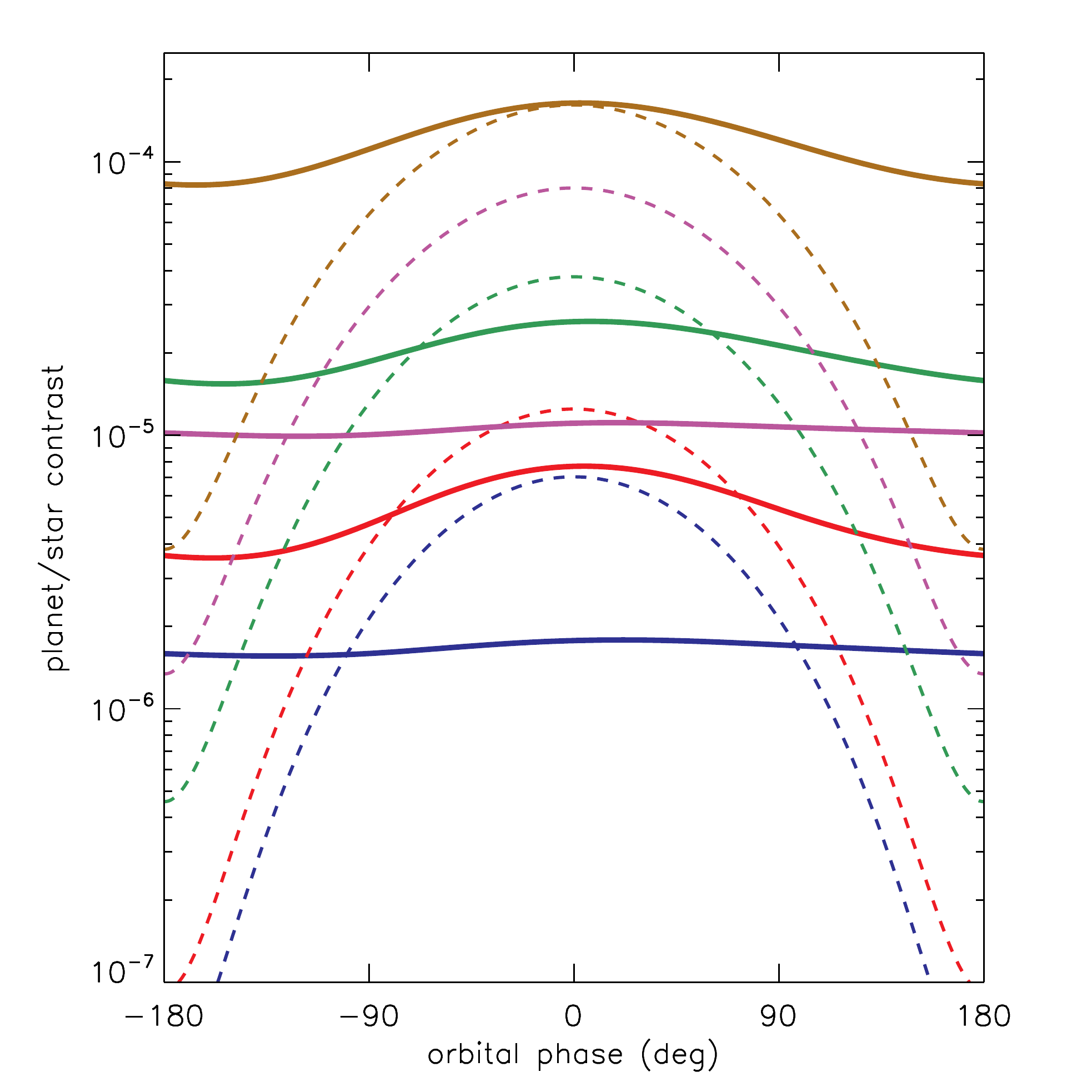}
        \includegraphics[width=0.47\linewidth]{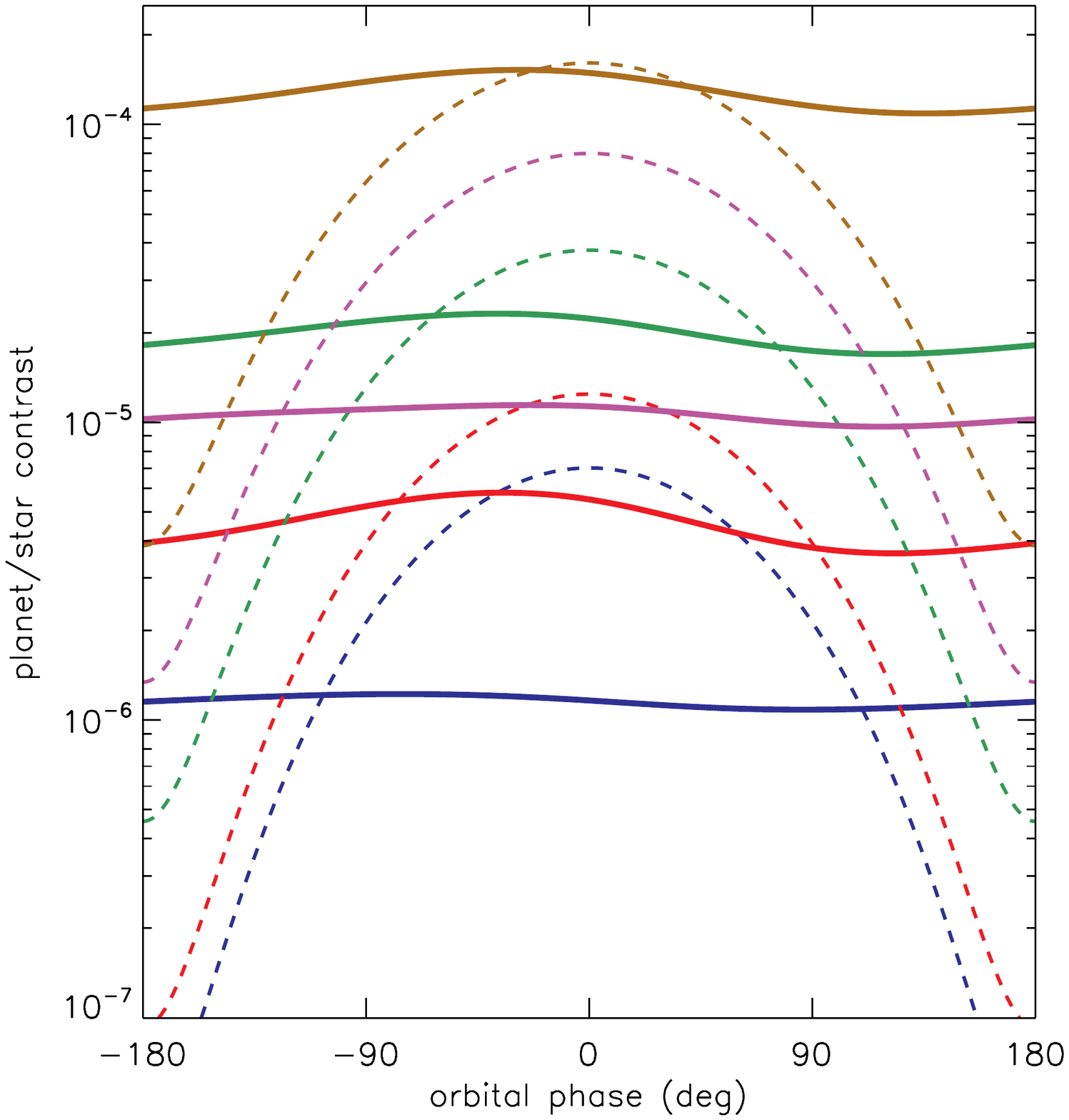}
        \caption{Emission phase curves computed for a dry planet with a 
1~bar CO$_2$-dominated atmosphere in synchronous rotation (left) and in 3:2 spin-orbit resonance (right).}
\label{fig:IRphase:dryCO2}
\end{figure*}

\begin{figure*}[h]
    \centering
        \includegraphics[width=0.47\linewidth]{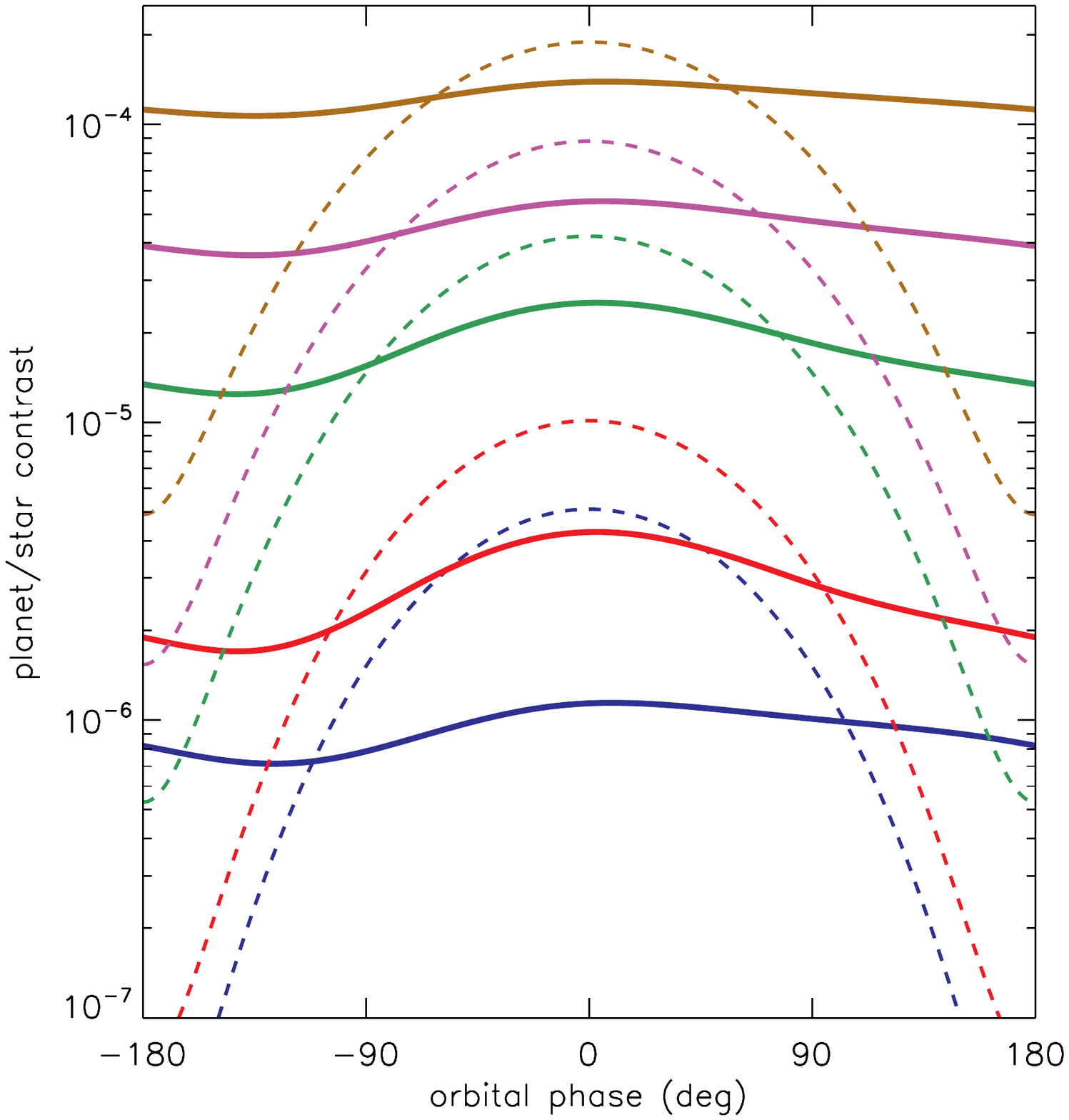}
        \includegraphics[width=0.47\linewidth]{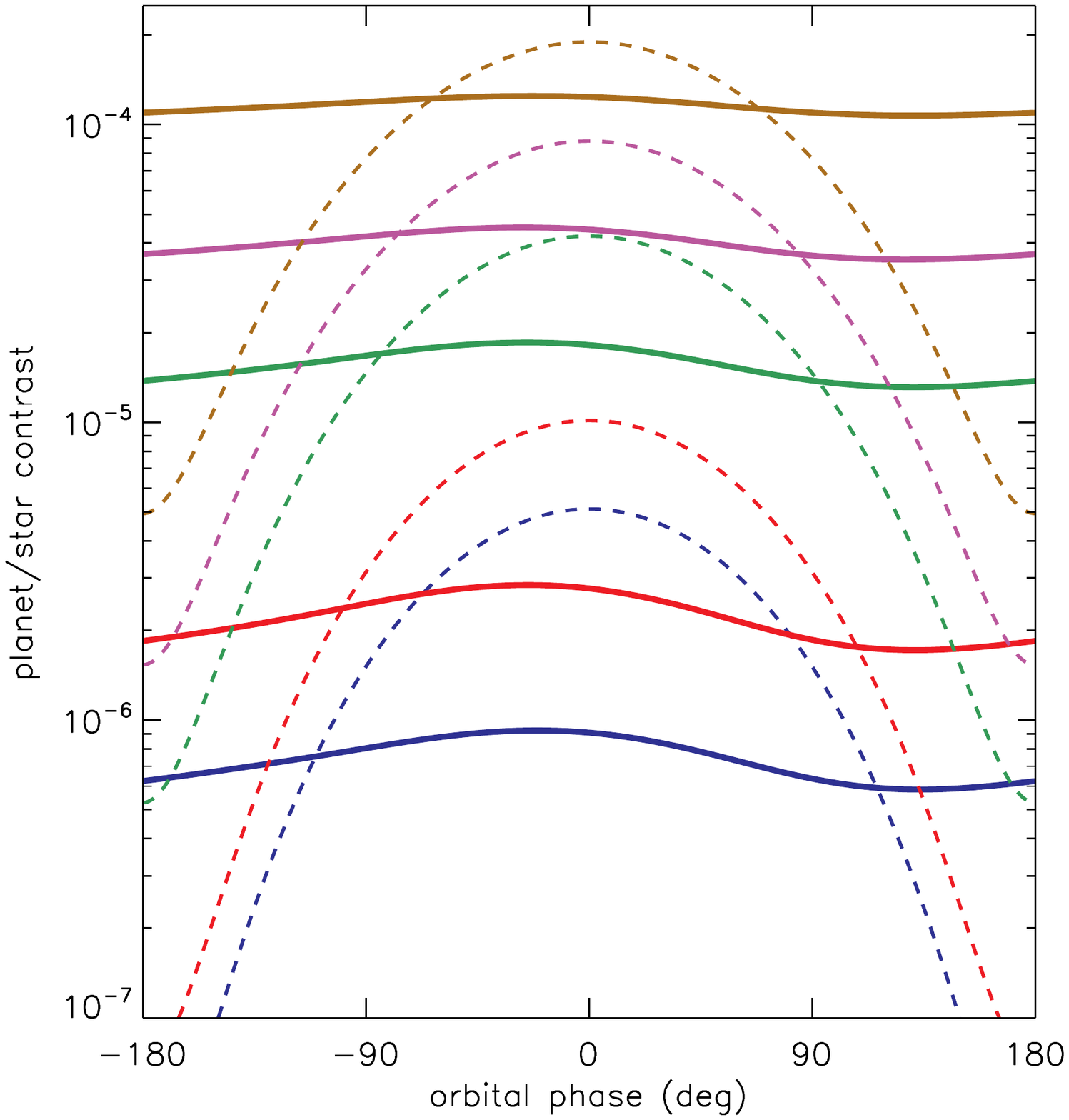}
        \caption{Emission phase curves computed for an aquaplanet with a 
10~mbar N$_2$-dominated (+~376ppm of CO$_2$) atmosphere in synchronous rotation (left) and in 3:2 spin-orbit resonance (right).}
\label{fig:IRphase:tenuousEarth}
\end{figure*}

\begin{figure*}[h]
    \centering
        \includegraphics[width=0.432\linewidth]{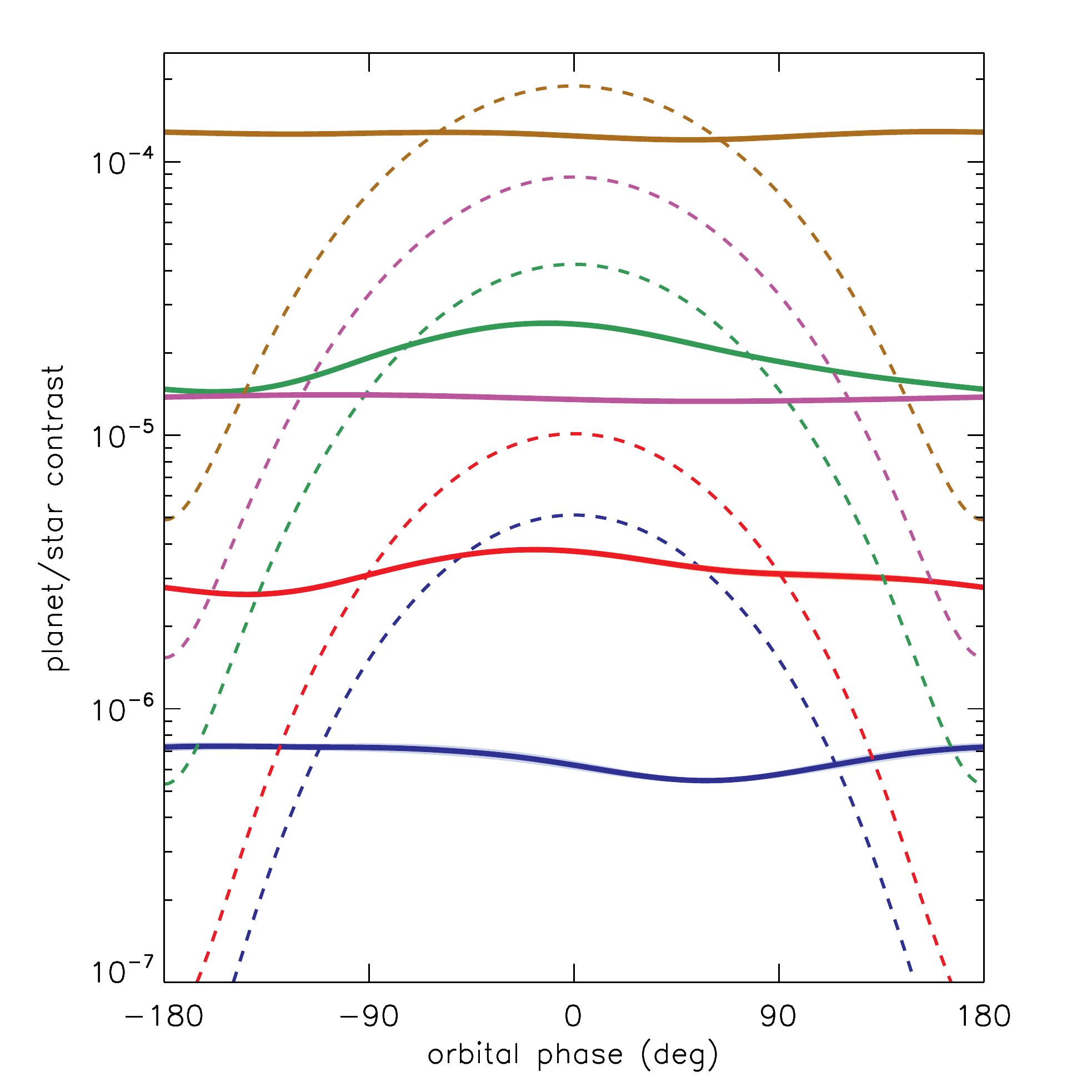}
        \includegraphics[width=0.47\linewidth]{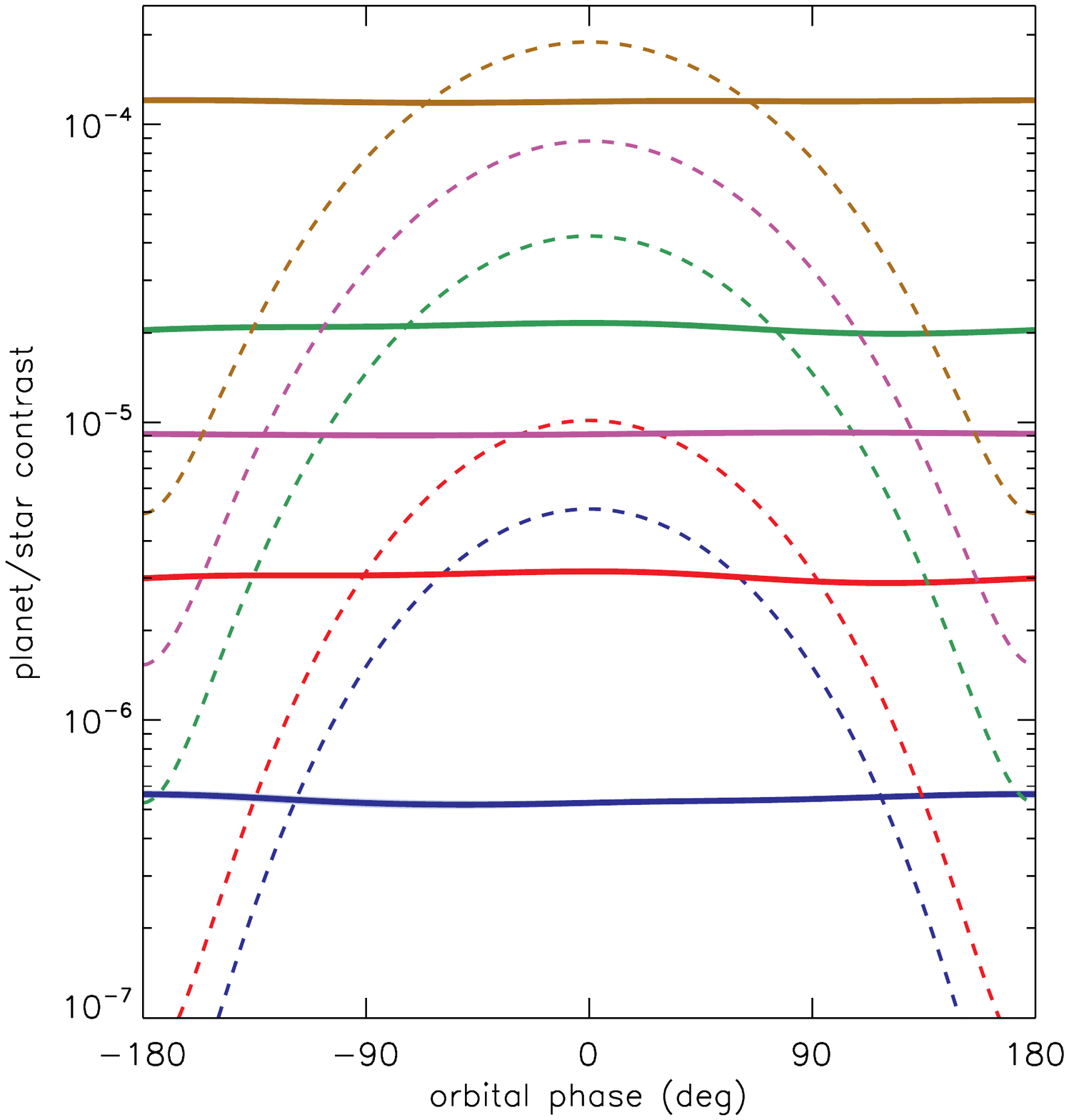}
        \caption{Emission phase curves computed for an aquaplanet with an 
Earth-like atmosphere in synchronous rotation (left) and in 3:2 spin-orbit resonance (right).}
\label{fig:IRphase:Earth}
\end{figure*}

\begin{figure*}[h]
    \centering
        \includegraphics[width=0.47\linewidth]{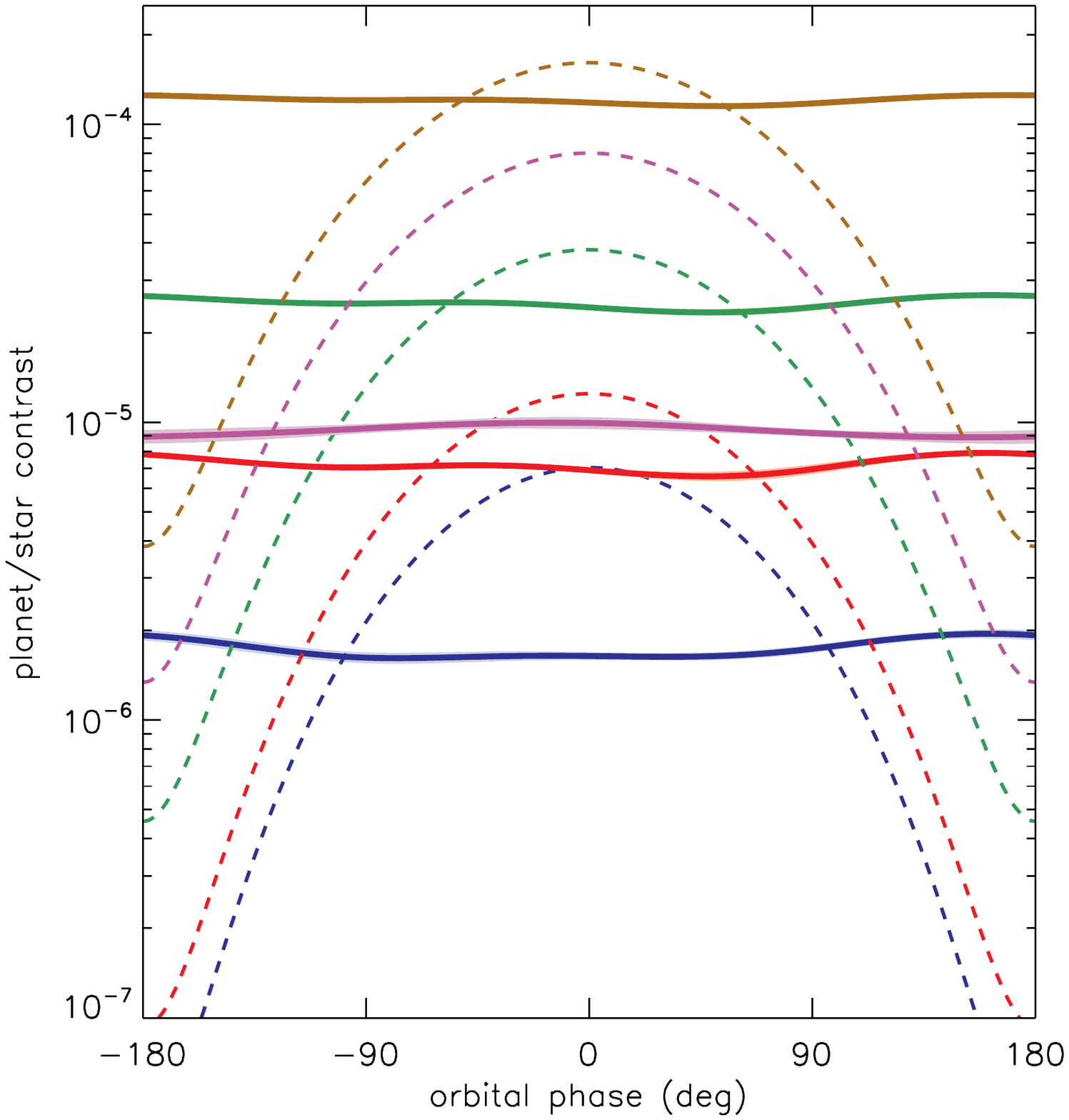}
        \includegraphics[width=0.47\linewidth]{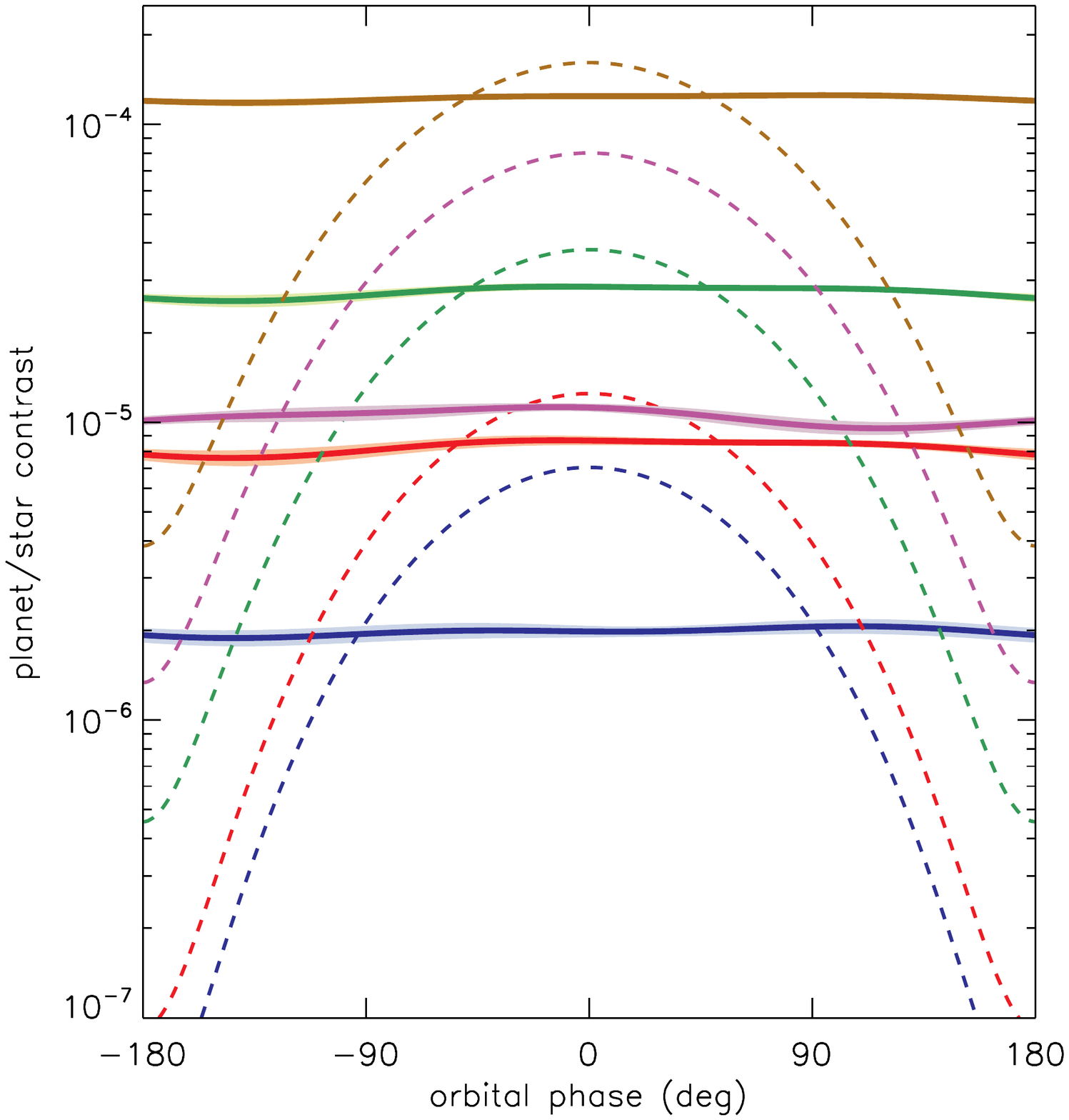}
        \caption{Emission phase curves computed for an aquaplanet with a 
1~bar CO$_2$-dominated atmosphere in synchronous rotation (left) and in 3:2 spin-orbit resonance (right).}
\label{fig:IRphase:CO2}
\end{figure*}

Emission lightcurves are controlled by the temperature longitudinal distribution at the surface if the atmosphere is transparent in the observed band or at the emitting atmospheric layer otherwise. \\
\noindent - Dry case, Earth-like atmosphere (Fig.~\ref{fig:IRphase:dryEarth}):  In both cases, the 15~$\mu$m CO$_2$ band is emitted from a uniform high altitude layer, which produces a flat phase curve. In the synchronous case, other bands exhibit phase curves that are similar to that of an airless planet except at high phase angle and long wavelengths where the warming of the night side by atmospheric circulation flattens the curves. In the 3:2 case, the flattening is more pronounced and is associated with a lag as surface temperature peaks in the afternoon. \\
\noindent - Dry case, 1 bar of CO$_2$ (Fig.~\ref{fig:IRphase:dryCO2}):  Similar to the dry Earth case except that the 6.7~$\mu$m band is flattened due CO$_2$ CIA absorption.\\
\noindent - Aquaplanet, mainly frozen, 10 mbar of N$_2$; 376 ppm of CO$_2$ (Fig.~\ref{fig:IRphase:tenuousEarth}): Here, the 15~$\mu$m CO$_2$ band is no longer opaque and exhibits the same behavior as the other bands. Compared with the previous case, and despite a more tenuous atmosphere, phase curves are more flattened due to the additional redistribution of latent heat (and the relatively high surface thermal inertia for the asynchronous case).  \\
\noindent - Aquaplanet, mainly frozen, 1 bar of N$_2$; 376 ppm of CO$_2$ (Fig.~\ref{fig:IRphase:Earth}): efficient zonal heat redistribution flattens all the phase curves in the 3:2 case. In the synchronous case, the 11 and 7.7~$\mu$m bands coming from either surface or clouds show an increase at small phase angle with some asymmetry due to the patch of clouds eastward of the substellar point (see Fig.~\ref{fig:cloudmapEarth}. The 6.7~$\mu$m band emitted in the troposphere is flat with a slight decrease also due to the clouds. The 15 and 24~$\mu$m bands are flat and not affected by clouds, emerging from higher levels. \\
\noindent - Aquaplanet, 1 bar CO$_2$ (Fig.~\ref{fig:IRphase:CO2}): There are no difference between the synchronous and 3:2 cases. All atmospheric windows are closed by either CO$_2$ or H$_2$O absorption and the emission comes from different but horizontally-uniform layers. This produces flat lightcurves at different brightness temperatures, with some shallow attenuation by the clouds at the most transparent wavelengths (11 and 7.7~$\mu$m).  \\

\begin{figure*}[h]
    \centering
        \includegraphics[width=0.47\linewidth]{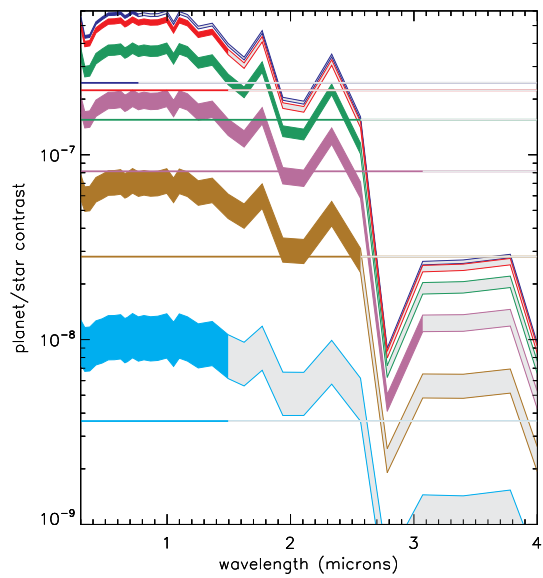}
        \caption{Reflection spectra computed for a dry planet with a Venus-like atmosphere (including aerosols).}
        \label{fig:venus1}
\end{figure*}

\begin{figure*}[h]
    \centering
        \includegraphics[width=0.47\linewidth]{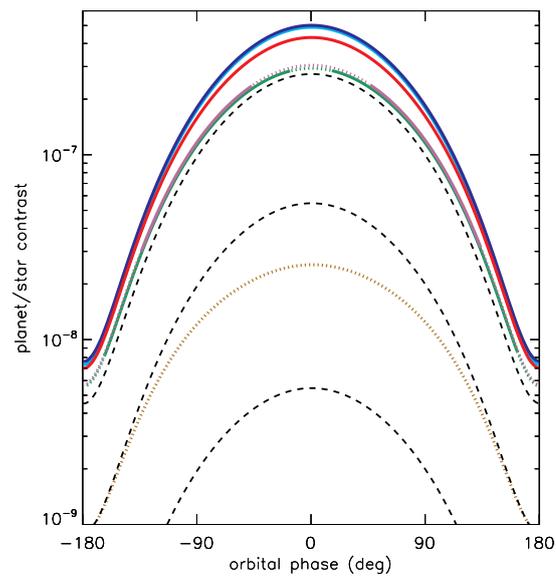}
        \caption{Reflection phase curves computed for a dry planet with a Venus-like atmosphere (including aerosols).}
        \label{fig:venus2}
\end{figure*}

\clearpage

\end{appendix}
\end{document}